\documentclass[amsmath,amssymb,aps,10pt,prd,letterpaper,nofootinbib,balancelastpage,notitlepage,superscriptaddress,twocolumn,floatfix]{revtex4-2}
\usepackage{graphicx}	
\usepackage{dcolumn}	
\usepackage{ragged2e}	
\usepackage{bm}		    
\usepackage{dsfont}
\usepackage[sort&compress]{natbib}	 	
	\setcitestyle{square,numbers,comma}	
\usepackage[colorlinks=true,urlcolor=blue,linkcolor=black,citecolor=red]{hyperref}	
\newcommand{\up}[1]{\textsuperscript{#1}}				
\newcommand{\tabref}[2][]{Tab{#1}.~\ref{tab:#2}}		
\newcommand{\figref}[2][]{Fig{#1}.~\ref{fig:#2}}		
\newcommand{\secref}[2][]{Sec{#1}.~\ref{sec:#2}}		
\newcommand{\appref}[2][x]{Appendi{#1}~\ref{app:#2}}	
\renewcommand{\eqref}[2][]{Eq{#1}.~(\ref{eq:#2})}		
\newcommand{\eqrefRange}[2]{Eqs.~(\ref{eq:#1})--(\ref{eq:#2})}		
\newcommand{\citeR}[2][]{Ref{#1}.~\cite{#2}}			
\newcommand{\paragraphdash}[1]{\indent\emph{#1}---\ignorespaces} 
\newcommand{\orcid}[1]{\href{https://orcid.org/#1}{\,\includegraphics[width=8px]{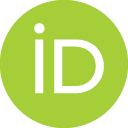}}}
\newcommand{\lb}{\ensuremath{\left}}					
\newcommand{\rb}{\ensuremath{\right}}					
\newcommand{\nl}{\nonumber \\ & \quad }					

\newcommand{\D}[1]{\frac\partial{\partial #1}}
\newcommand{\LL}{\mathcal{L}}
\DeclareMathOperator{\IM}{\text{Im}}
\DeclareMathOperator{\RE}{\text{Re}}
\newcommand{\phihat}{\bm{\hat{\phi}}}
\newcommand{\thetahat}{\bm{\hat{\theta}}}
\newcommand{\rhat}{\bm{\hat{r}}}
\newcommand{\Hz}{\,\text{Hz}}
\newcommand{\eV}{\,\text{eV}}
\newcommand{\id}[1]{\ensuremath\mathds{1}_{#1}}

\begin{document}

\title{Search for dark-photon dark matter in the SuperMAG geomagnetic field dataset}
\date{\today}
\author{Michael A.~Fedderke\orcid{0000-0002-1319-1622}}
\email{mfedderke@jhu.edu}
\affiliation{Department of Physics and Astronomy, The Johns Hopkins University, Baltimore, MD 21218, USA}
\author{Peter W.~Graham\orcid{0000-0002-1600-1601}}
\email{pwgraham@stanford.edu}
\affiliation{Stanford Institute for Theoretical Physics, Department of Physics, Stanford University, Stanford, CA 94305, USA}
\affiliation{Kavli Institute for Particle Astrophysics \& Cosmology, Stanford University, Stanford, CA 94305, USA}
\author{Derek F.~Jackson Kimball\orcid{0000-0003-2479-6034}}	
\email{derek.jacksonkimball@csueastbay.edu}
\affiliation{Department of Physics, California State University -- East Bay, Hayward, CA 94542, USA}
\author{Saarik Kalia\orcid{0000-0002-7362-6501}\,}	
\email{saarik@stanford.edu}
\affiliation{Stanford Institute for Theoretical Physics, Department of Physics, Stanford University, Stanford, CA 94305, USA}

\begin{abstract}
In our recent companion paper~\cite{Fedderke:2021rys}, we pointed out a novel signature of ultralight kinetically mixed dark-photon dark matter.
This signature is a quasi-monochromatic, time-oscillating terrestrial magnetic field that takes a particular pattern over the surface of the Earth.
In this work, we present a search for this signal in existing, unshielded magnetometer data recorded by geographically dispersed, geomagnetic stations.
The dataset comes from the SuperMAG Collaboration and consists of measurements taken with one-minute cadence since 1970, with $\mathcal{O}(500)$ stations contributing in all.
We aggregate the magnetic field measurements from all stations by projecting them onto a small set of global vector spherical harmonics (VSH) that capture the expected vectorial pattern of the signal at each station.
Within each dark-photon coherence time, we use a data-driven technique to estimate the broadband background noise in the data, and search for excess narrowband power in this set of VSH components; we stack the searches in distinct coherence times incoherently.
Following a Bayesian analysis approach that allows us to account for the stochastic nature of the dark-photon dark-matter field, we set exclusion bounds on the kinetic mixing parameter in the dark-photon dark-matter mass range $2\times10^{-18}\,\text{eV} \lesssim m_{A'} \lesssim 7\times10^{-17}\,\text{eV}$ (corresponding to frequencies~$6\times 10^{-4}\,\text{Hz}\lesssim f_{A'} \lesssim 2\times 10^{-2}\,\text{Hz}$).
These limits are complementary to various existing astrophysical constraints.
Although our main analysis also identifies a number of candidate signals in the SuperMAG dataset, these appear to either fail or be in tension with various additional robustness checks we apply to those candidates.
We report no robust and significant evidence for a dark-photon dark-matter signal in the SuperMAG dataset.
\end{abstract}
\maketitle


\section{Introduction}
\label{sec:introduction}
Over an enormous range of scales from the dwarf-galactic to the cosmological, there is overwhelming evidence for the existence of dark matter (DM) via its gravitational effects.
However, despite a broad and decades-long experimental program to detect the effects of any non-gravitational interactions which the dark matter may possess, either in the laboratory or via astrophysical probes, the identity of the dark matter remains elusive.
The difficulty of the search for the nature of dark matter stems in part from its extremely broad range of allowed masses, spanning some $\sim 80$ orders of magnitude, from ultralight fuzzy dark matter around $10^{-21}\eV$~\cite{Hu:2000ke,Hui:2016ltb,Kobayashi:2017jcf,Irsic:2017yje,Nadler:2020prv}, up to macroscopic primordial black hole dark matter around $\sim 10^{56}\eV$~\cite{Smyth:2019whb}.
Moreover, the various possible DM candidates that populate this allowed mass range give rise to a diverse array of potential phenomenological effects that cannot all be searched for using a single experimental approach.
In the past decade or so, there has in particular been a rapid growth of interest in novel experimental techniques aiming to detect bosonic DM candidates that admit a classical wave description,%
\footnote{\label{ftnt:pedantry1}%
     To be precise, we mean here that given the local galactic abundance of the DM, $\rho_{\textsc{dm}}\sim0.3\,$GeV/cm${}^{3}$, excitations of the bosonic dark-matter quantum field have expected local occupancy numbers (i.e., number of particles per cubic de Broglie wavelength) in the vicinity of the Earth that are greater than 1.
     This occurs for DM masses lighter than $\sim 10\eV$ assuming $v_{\textsc{dm}}\sim 10^{-3}$.
     } %
of which one well-motivated example is the kinetically mixed~\cite{Holdom:1985ag} dark photon~\cite{Nelson:2011sf}, sometimes also referred to as the `hidden photon'; see, e.g., \citeR[s]{Wagner:2010mi, Redondo:2010dp, Bahre:2013ywa, Graham:2014sha, Chaudhuri:2014dla, Phipps:2019cqy, TheMADMAXWorkingGroup:2016hpc, Baryakhtar:2018doz, Lawson:2019brd, Gelmini:2020kcu, Jaeckel:2013sqa,Horns:2012jf,Suzuki:2015sza,Andrianavalomahefa:2020ucg,Cantatore:2020obc,Su:2021jvk}.
Such dark-photon dark matter (DPDM) can be produced in the early Universe in a variety of model-dependent and -independent ways; e.g.,  \citeR[s]{Graham:2015rva,Ahmed:2020fhc,Kolb:2020fwh,Arias:2012az,Agrawal:2018vin,dror2019parametric,bastero2019vector,ema2019production,co2019dark,long2019dark,Nakai:2020cfw,nakayama2020gravitational,Salehian:2020asa,Moroi:2020has,Bastero-Gil:2021wsf,co2021gravitational,Nakayama:2021avl}.

In a recent companion paper~\cite{Fedderke:2021rys}, we pointed out the existence of a new signature of ultralight kinetically mixed  dark-photon dark matter: a spatially and temporally coherent, oscillating, terrestrial magnetic field signal that is narrowband in frequency, and that takes a particular vectorial field pattern over the whole surface of the Earth.
This signal arises because of the same photon--dark-photon mixing effects responsible for the generation of the signal in, e.g., DM Radio~\cite{Chaudhuri:2014dla}.
In \citeR{Fedderke:2021rys}, we provided a high-level summary and the results of an experimental search for this novel signal that we undertook using a publicly available geomagnetic field dataset maintained by the SuperMAG Collaboration~\cite{SuperMAGwebsite,Gjerloev:2009wsd,Gjerloev:2012sdg}.
This dataset, which exists primarily for geophysical metrology and solar activity research purposes, consists of time-series magnetic field measurements obtained with unshielded three-axis magnetometers located at $\mathcal{O}(500)$ ground stations that are widely dispersed over the surface of the Earth and that, collectively, have been recording data continuously since the early 1970s with a sampling rate of (at least) once per minute~\cite{SuperMAGwebsite,Gjerloev:2009wsd,Gjerloev:2012sdg}.
We reported no significant evidence for the existence of a robust dark-photon dark-matter signal in these data in the dark-photon mass range $2~\times~10^{-18}\eV \lesssim m_{A'} \lesssim 7\times10^{-17}\eV$ corresponding to frequencies $6\times 10^{-4}\,\text{Hz}\lesssim f_{A'} \lesssim 2\times 10^{-2}\,\text{Hz}$. 
We thus placed direct observational bounds on the kinetic mixing parameter $\varepsilon$ that are complementary to various existing astrophysical constraints~\cite{Dubovsky:2015cca,Bhoonah:2018gjb,McDermott:2019lch,Wadekar:2019xnf,Kovetz:2018zes}. 
In this paper, we supplement \citeR{Fedderke:2021rys} by providing a detailed technical description of this experimental search.

The remainder of this paper is structured as follows: in \secref{theory}, we briefly summarize the main features of the signal we described in detail in \citeR{Fedderke:2021rys}.
We then give a description of the SuperMAG dataset in \secref{superMAGdescription}, before giving a high-level description of our analysis strategy for this dataset in \secref{analysisOverview}.
With this high-level overview as a guidepost, we give a detailed technical description of the analysis in \secref{analysisDetails}.
The results of this analysis in the form of exclusion bounds on dark-photon dark-matter parameter space are shown at \figref{resultsExclusion} and discussed in \secref{analysisResults}.
Our analysis in \secref{analysisDetails} also identifies a number of na\"ive signal candidates in the SuperMAG data, in addition to placing bounds on parameter space; we test these candidates for robustness in \secref{analysisChecks}.
On the basis of those tests and other indicia, we find no na\"ive signal candidate for which there is robust evidence of a real signal, although a handful of these candidates would be of potential interest to examine in follow-up work.
We discuss our results and conclude in \secref{conclusion}.
There are a number of appendices that provide additional information, conventions, or details.
\appref[ces]{FTconventions} and \ref{app:vectorSphericalHarmonics} give our conventions for the Fourier transform and vector spherical harmonics, respectively.
\appref{Xk} gives some additional technical details of the signal as it appears in the SuperMAG dataset in our analysis construction.
\appref{likelihoodDetails} contains some derivations of important statistical results used in our analysis construction in \secref{analysisDetails}.
Finally, \appref{noiseValidation} contains a series of detailed validation checks on the data-driven noise estimation procedures applied in our analysis.

\section{Signal}
\label{sec:theory}
In our recent companion paper~\cite{Fedderke:2021rys}, we showed that kinetically mixed dark-photon dark matter generates a coherent magnetic field signal across the surface of the Earth of the form
\begin{align}
    \bm{B}(\Omega,t)&=\sqrt{\frac\pi3} \lb(\varepsilon m_{A'}\rb) \lb( m_{A'} R \rb) \nl
	\times  \RE\lb[ \sum_{m=-1}^1A'_m\bm{\Phi}_{1m}(\Omega)e^{-2\pi i(f_{A'}- f_d m)t} \rb],
    \label{eq:signal}
\end{align}
as measured in the rotating Earth-fixed frame, where $\Omega = (\theta,\phi)$ is the location on the surface of the Earth (in the geographic co-ordinate system referenced to True Geographic North), $\varepsilon$ is the kinetic mixing parameter (as defined in \citeR{Fedderke:2021rys}); $m_{A'} \equiv 2\pi f_{A'}$ is the DPDM mass (with $f_{A'}$ being the corresponding cycles-per-second frequency);
\footnote{\label{ftnt:hbarc1}%
    We work in natural units where $\hbar = c = 1$.
    The mass--frequency conversion is thus $f_{A'} \approx 24 \,\text{mHz} \times ( m_{A'} / 10^{-16}\,\text{eV})$.
} %
$R$ is the radius of the Earth; $A'_m$ are the (complex) amplitudes describing the (amplitude and phase of the) three different polarization modes of the dark photon in the vicinity of the Earth (as measured in a non-rotating fixed inertial frame);%
\footnote{\label{ftnt:shorthand}%
		As discussed in \citeR{Fedderke:2021rys}, the $A_m'$ technically describe the amplitudes of the polarization modes of the sterile component (in the interaction basis) of the DPDM, as measured in the vicinity of the Earth but well outside the atmosphere, and in the inertial frame.
		Our convention for the $A_m'$ is such that $A'_\pm = \mp\tfrac{1}{\sqrt{2}} \lb( A_x' \mp i A_y' \rb)$
		so that the Cartesian components (in the inertial frame) of the dark vector potential are given by 
		\begin{align*}
		\quad A'_x &= -\tfrac{1}{\sqrt{2}} \lb( A'_+ - A'_- \rb); &
		A'_y &= -\tfrac{i}{\sqrt{2}} \lb(A'_+ + A'_- \rb); &
		A'_z &= A'_0.
		\end{align*}
		We employ the shorthand $A'_{\pm} \equiv A'_{\pm1}$.
	} %
$\bm{\Phi}_{\ell m}$ are vector spherical harmonics (see \appref{vectorSphericalHarmonics} for conventions); and the additional frequency $f_d = (\text{sidereal day})^{-1}$ appears in the $m=\pm1$ modes owing to the rotation of the Earth~\cite{Fedderke:2021rys}. 

As discussed in detail in \citeR{Fedderke:2021rys}, \eqref{signal} is a good description of the signal within a single DM coherence time $T_{\text{coh}} \sim 2\pi / (m_{A'} v_{\textsc{dm}}^2) \sim 10^6 f_{A'}^{-1}$, where we have taken $v_{\textsc{dm}} \sim 10^{-3}$ to be a representative value for the galactic DM velocity dispersion. 
Within that coherence time, the complex amplitudes $A_{0,\pm}$ characterizing the local DM field remain approximately constant, but in general they evolve significantly from one coherence time to the next.%
\footnote{\label{ftnt:darkEFieldNotOnALine}%
		Generically, this implies that the `dark electric field' $\bm{E'} \sim m_A \bm{A'}$ of the dark-photon dark matter is \emph{not} simply a vector of fixed direction in 3D space with an amplitude oscillating at frequency $f_{A'}$.
		Instead, each of the components of $\bm{E'}$ executes oscillations at frequency $f_{A'}$, which implies that $\bm{E'}$ has a periodic (with period $T_{A'} = f_{A'}^{-1}$) variation of both its instantaneous amplitude \emph{and} its direction in 3D space.
		In the generic case, $|\bm{E'}|$ does not vanish instantaneously at any moment in time, and the tip of the unit vector $\bm{\hat{E}'}$ traces out a closed periodic curve with period $f_{A'}^{-1}$; that curve evolves secularly on characteristic timescales of order the coherence time.
			} %
Indeed, since the local DPDM field can be thought of as being comprised of the sum of a large number of independent plane waves with frequencies $f \sim f_{A'} \lb[ 1 + \mathcal{O}(v_\textsc{dm}^2) \rb]$, each with its own random phase, within a single coherence time each of the real and imaginary parts of the $A_m'$ can by virtue of the central limit theorem be described as a random draw from a zero-mean normal distribution with standard deviation $\sqrt{\rho_{\textsc{dm}}} / (\sqrt{3} m_{A'})$, such that together they satisfy
\begin{align}
\frac{1}{2} m_{A'}^2 \langle|A'|^2\rangle &= \rho_{\textsc{dm}},
\label{eq:Asquared}
\end{align}
where
\begin{align}
|A'|^2 = \sum_{m=-1}^1\lb(\RE\lb[A'_m\rb]^2+\IM\lb[A'_m\rb]^2\rb);
\label{eq:Amag}
\end{align}
the angle-brackets $\langle\,\cdots\rangle$ describe an average over times $\tau$ much longer than the coherence time, $\tau \gg T_{\text{coh}}$; and $\rho_{\textsc{dm}}$ is the average local (to the Earth) dark-matter mass-density, which we will take to be fixed at $\rho_{\textsc{dm}} = 0.3\,\text{GeV/cm}^3$ throughout this paper.

We pause to note that there is some discussion in the literature regarding the appropriate treatment of the DPDM polarization state; see, e.g., discussion in \citeR[s]{Arias:2012az,Caputo:2021eaa}.
We have assumed above, and will continue to do so throughout this paper, that the DPDM field is a sum of plane waves with random individual phases and randomly oriented individual polarization states; this guarantees that the overall polarization state will necessarily randomize over a coherence time. 
However, certain production mechanisms (e.g., misalignment) may possibly give rise to a polarization state that does not evolve (significantly) in time today, because every individual mode is produced in the early Universe with the same (or similar) polarization; provided that structure formation and interactions with matter do not spoil this, the DPDM field today would then consist of a sum of plane waves with individual random phases but all the same (or similar) polarization states.
Such a DPDM field would still exhibit phase-decoherence over a coherence time and thus amplitude fluctuations from one such time to the next, but the polarization state would of course not randomize significantly from one coherence time to the next.

A crucial feature of the signal is the factor of $\lb( m_{A'}R \rb)$ in \eqref{signal}, which encodes that the signal suffers a suppression in the ratio of the radius of the Earth to the (Compton) wavelength of the DM.
Na\"ively, however, one might be tempted to think that the depth of the atmosphere $L_{\text{atmos}} \ll R$ would instead be the relevant length scale governing this suppression (see, e.g., brief comments in \citeR{Dubovsky:2015cca}), which would have implied that this factor would instead be replaced by a factor of $m_{A'}L_{\text{atmos}} \sim 10^{-2} m_{A'}R$, dramatically weakening the signal prediction.
We discuss at length in \citeR{Fedderke:2021rys} why this is not in fact the case.

Numerically, the signal \eqref{signal} is expected to have an amplitude on the order of
\begin{align}
    B \sim 0.7\,\text{nG} \times \lb( \frac{\varepsilon}{10^{-5}} \rb)\times \lb( \frac{m_{A'}}{4\times 10^{-17}\,\text{eV}} \rb), \label{eq:signalAmplitude}
\end{align}
where we assumed for the purposes of this rough estimate that $|\bm{A}'| \sim \sqrt{2\rho_{\textsc{dm}}}/m_{A'}$ with $\rho_{\textsc{dm}} = 0.3\,\text{GeV/cm}^3$, and evaluated the maximum value of the field on the Earth's surface.
While this signal is very small in amplitude~(many orders of magnitude smaller than the static geomagnetic field, which is of order $B_{\oplus}\sim 0.5$\,G~\cite{magneticFieldEarth}), it is at nonzero frequency, effectively monochromatic with a long coherence time and has a very particular global field pattern over the surface of the Earth; it can thus be meaningfully distinguished from many noise sources via techniques that are tailored to search for the specific spatial and frequency structure of the signal.

The magnetic field amplitude of the signal \eqref{signalAmplitude} is also potentially \emph{much} smaller than the individual point-in-time, single-station, single-field-component digital measurement resolution of the magnetic field stations that contribute to SuperMAG, which are typically in the 10--100\,pT~$\approx$~100--1000\,nG range~\cite{Mann:2008wrb,Chi:2013sdn,Engebretson:1995gna,Yumoto:2001sfb,Tanskanen:2009wdg,Love:2013fbq,Lichtenberger:2013adg,1992NASCP3153..321H,INTERMAGNETmanual}.
However, given that the instantaneous fluctuating random noise in the detectors greatly exceeds%
\footnote{\label{ftnt:noise}%
    Of course, the single-station instantaneous noise averages down dramatically when considering the whole time series of the data over the hundreds of stations; we can thus detect a signal that is much smaller than the instantaneous single-detector noise.
    } %
this measurement resolution~\cite{Gjerloev:2009wsd,Gjerloev:2012sdg,SuperMAGwebsite} (see generally \secref{analysisNoise} and \appref{noiseValidation}), the fact that the signal amplitude is sub-readout-resolution does \emph{not} degrade the sensitivity for our signal search.
This can be understood from the following intuitive argument:
suppose the station digital measurement readout resolution is $\rho$, and the signal has amplitude $\mathcal{A}$ in some given field component. 
If $\mathcal{A}<\rho$, then on average only a fraction $\xi \sim \mathcal{A}/\rho$ of single-station, single-field-component measurements will be impacted by the signal being present.
This is because only that fraction of point-in-time noise realizations lie close enough to the break-point in the digital readout rounding for the presence of the signal to alter the reading of the magnetometer. 
However, for those fraction $\xi$ of measurements, the readout is changed by a full resolution unit $\rho > \mathcal{A}$, which is $\rho / \mathcal{A} \sim 1/\xi$ larger than the signal amplitude $\mathcal{A}$.
These two effects thus effectively cancel out.
And indeed, we have verified numerically as well in simple cognate examples that, in the presence of the readout digitization noise, a narrowband signal of sub-resolution amplitude added to super-resolution instantaneous random noise remains narrowband and visible in the finite-resolution data, provided of course that the signal amplitude is larger than the averaged-down noise level.

\section{SuperMAG data}
\label{sec:superMAGdescription}
We now turn to a general description [\secref{superMAGoverview}] of the magnetic field dataset we have analyzed in this paper to search for the signal shown at \eqref{signal}, before turning to a more extensive discussion of various salient details in \secref[s]{superMAGCoordinateSystem}--\ref{sec:superMAGtemporal}.

\subsection{Overview}
\label{sec:superMAGoverview}

\begin{figure}[t]
\includegraphics[width=\columnwidth]{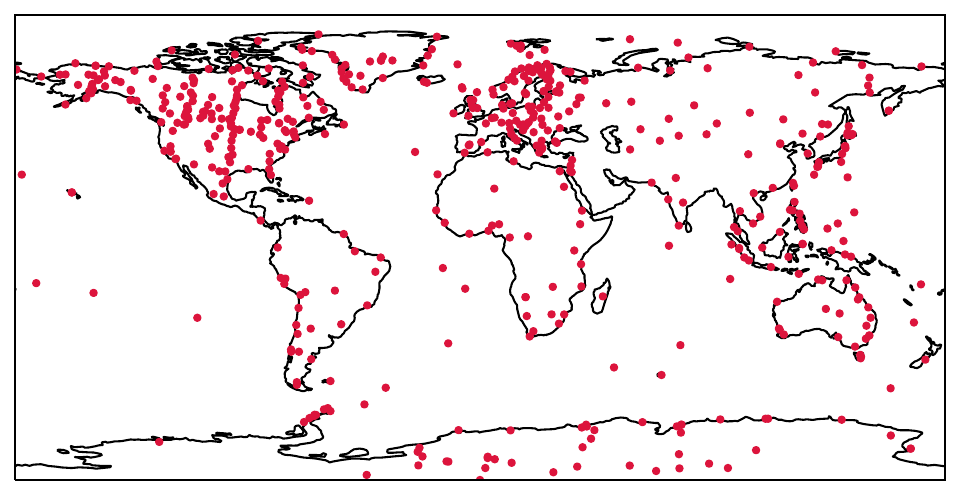}
\caption{\label{fig:SuperMAGstations}%
    Locations of SuperMAG geomagnetic observatories whose data are included in the analysis in this paper~\cite{SuperMAGwebsite,Gjerloev:2009wsd,Gjerloev:2012sdg}. 
    World map (equirectangular projection) created using \texttt{cartopy}~\cite{cartopy}.} 
\end{figure}

The SuperMAG Collaboration~\cite{Gjerloev:2009wsd,Gjerloev:2012sdg} maintains and makes available for research purposes a large archival dataset of three-axis geomagnetic field measurements sourced from around $\mathcal{O}(500)$ individual measurement stations%
\footnote{\label{ftnt:memberStations}%
		These stations are maintained either by SuperMAG member organizations or national scientific bodies; see, e.g., \citeR[s]{Love:2013fbq,Lichtenberger:2013adg,Tanskanen:2009wdg,Yumoto:2001sfb,Engebretson:1995gna,Chi:2013sdn,Mann:2008wrb} and references therein.
	} %
which are geographically dispersed across the surface of the Earth; see \figref{SuperMAGstations}.
These data are presented in a common format, in a well-defined co-ordinate system~\cite{Gjerloev:2012sdg}, with common temporal measurement resolution and synchronization, and (where relevant) are pre-processed in a common manner.

The data product with which we will be primarily interested in this paper is their `low fidelity' dataset, which encompasses measurements made with one-minute resolution, beginning in 1970~\cite{Gjerloev:2012sdg,SuperMAGwebsite}.
The SuperMAG Collaboration has also recently released a `high fidelity' dataset of measurements taken by $\mathcal{O}(100)$ stations with one-second temporal resolution in the time-frame 2012--2020~\cite{SuperMAGwebsite}.
We defer analysis of the one-second resolution data to future work.

\begin{figure}[t]
\includegraphics[width=\columnwidth]{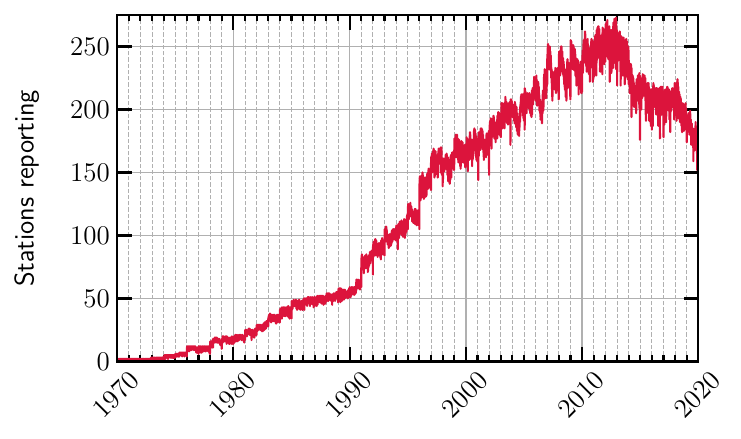}
\caption{\label{fig:station_count}%
    	Number of stations in the SuperMAG dataset~\cite{SuperMAGwebsite,Gjerloev:2009wsd,Gjerloev:2012sdg} reporting as a function of the date.
	We discuss the manifest annual trends in \secref{superMAGtemporal}.
	} 
\end{figure}

While various individual stations contributing to this dataset have come online and/or gone offline since 1970, and even otherwise operational stations do not have 100\% uptime (so that data from individual stations are unavailable during certain periods of time), the cumulative number of stations in this dataset for which at least some amount of data are available and analyzed in this work is 494, and the number of stations operational in recent years has fluctuated between around 150 and 250 at any given time~\cite{Gjerloev:2012sdg,SuperMAGwebsite}; see \figref{station_count}.
Note however that for technical reasons,%
\footnote{\label{ftnt:toolittledataearly1}%
    Insufficiently many stations are operative in 1970 and 1971 for our analysis to be applied; see also footnote \ref{ftnt:tooLittleDataEarly} for a similar point.
    } %
we restrict our attention to the 48 years of data taken starting at the beginning of 1972 and concluding at the end of 2019.

\subsection{SuperMAG co-ordinate system}
\label{sec:superMAGCoordinateSystem}
The three-axis magnetic field measurements supplied by SuperMAG are reported for every station in locally well-defined co-ordinate systems whose definitions vary from station to station~\cite{Gjerloev:2012sdg}.
As described in this subsection, these local co-ordinate systems must be rotated to obtain the magnetic field components in the global geographic co-ordinate system that we will require for our analysis.

As discussed in detail in \citeR{Gjerloev:2012sdg}, SuperMAG assumes that each station has correctly reported the orientation of the vertical component (i.e., the component radially directed at the center of the Earth) $B_{\textsc{z}}$ of their field, \emph{into the ground} being positive.
However, owing to a proliferation of possible conventions in use by individual station operators to report the two orthogonal components of the field in the plane perpendicular to the vertical (`the horizontal plane'), SuperMAG undertakes a data-driven procedure to rotate the horizontal field components reported by every station into a co-ordinate system whose orthogonal basis vectors are oriented along \emph{Local} Magnetic North (LMN) and \emph{Local} Magnetic East (LME); the corresponding magnetic field components along these directions are $B_{\textsc{lmn}}$ and $B_{\textsc{lme}}$, respectively.

The instantaneous rotation angle about the vertical axis that would be required to perform this transformation is obtained in an unambiguous way from the field data themselves by demanding that the `typical value' (as defined in \citeR{Gjerloev:2012sdg}) of $B_{\textsc{lme}}$ in a sliding 17-day window period centered on the observation time is zero by definition~\cite{Gjerloev:2012sdg}; the rotation angles that are actually applied to the data are smoothed versions of these instantaneous rotation angles, where the smoothing is performed over the same 17-day period~\cite{Gjerloev:2012sdg,GjerloevPrivate}.

For the purposes of the discussion here, it is only relevant to note that the 17-day ($\sim 1.5\times 10^{6}\,\text{s}$) time windows used in these procedures are much longer than the intrinsic timescales associated with the oscillation of the DPDM in our mass range of interest: $6\times 10^{-4}\Hz \lesssim f_{A'} \lesssim 2\times 10^{-2}\Hz$, corresponding roughly to $2\times 10^{-18}\eV \lesssim m_{A'} \lesssim 7\times 10^{-17}\eV$.
Since our analysis procedure will construct an observable linear in the magnetic field, the principle of superposition implies that the rotation procedure cannot induce or mask signals that are in-band.%
\footnote{\label{ftnt:coherenceIsNotDestroyed}%
		One might however also na\"ively be concerned that, even if this procedure might not induce or mask in-band signals, it might somehow impact the coherence of the DPDM signal over timescales longer that 17 days (provided that $T_{\text{coh}} > 17\,\text{days}$).
		This is however not the case: we remind the reader the statement that the coherence time $T_{\text{coh}} \sim 10^{6} T_{A'}$ [where $T_{A'}=1/f_{A'}$] is simply the statement that the width of the DPDM signal peak in Fourier space is $\sim 10^{-6}$ of the carrier frequency $f_{A'}$ set by the DPDM mass.
		For our range of interest, $6\times 10^{-4}\Hz\lesssim f_{A'} \lesssim 2\times10^{-2}$\Hz, and the above considerations imply that all the information regarding the signal coherence is similarly restricted to (approximately) the same frequency range.
		On the other hand, the 17-day smoothing and rotation effects discussed here will only impact frequencies at or below $(17\,\text{days})^{-1} \sim 7\times 10^{-7}\Hz$.
		As such, these modifications do not impact the coherence of the signal in the data.
	} %

While the local co-ordinate systems described above have the advantage of being unambiguous, their definitions are inherently local and time-dependent.
In order to obtain the field components in the rigid, time-independent, global geographic co-ordinate system required for our analysis, the field components in these local co-ordinate systems must be rotated to the global frame using the per-station time-dependent true magnetic declination angles%
\footnote{\label{ftnt:declinationConvention}%
		Conventionally $\delta$ is defined as the angle between True Geographic North (TGN) and LMN, with the sign convention chosen such that $\delta$ is positive when the direction of LMN lies eastward (i.e., clockwise on a compass vane) of TGN, and is negative when LMN lies westward (i.e., counterclockwise on a compass vane) of TGN.
	} %
$\delta(\Omega,t)$.
The SuperMAG data products include the time-dependent declination angles $\delta(\Omega,t)$ for each station, assuming the International Geomagnetic Reference Field model of the appropriate epoch~\cite{Gjerloev:2012sdg,GjerloevPrivate}; these declination angles again vary only on timescales that are out-of-band for our DPDM mass range of interest.
A simple 2D rotation about the vertical axis can thus be applied to yield the magnetic field components $B_{\textsc{tgn}}$ and $B_{\textsc{tge}}$ reported along the directions of True Geographic North (TGN) and True Geographic East (TGE), respectively:
\begin{align}
\begin{pmatrix}
	B_{\textsc{tgn}} \\
	B_{\textsc{tge}}
\end{pmatrix}
	&= 
\begin{pmatrix}
	\cos \delta	& - \sin\delta \\
	\sin \delta	& \cos\delta
\end{pmatrix}
\begin{pmatrix}
	B_{\textsc{lmn}} \\
	B_{\textsc{lme}}
\end{pmatrix}.
\label{eq:declRot}
\end{align}

Finally, note that the field components in the global geographic (i.e., Earth-fixed) co-ordinate system  $(\rhat,\thetahat,\phihat)$ are related to those discussed above by $B^\theta = -B_{\textsc{tgn}}$ (recall, $\thetahat$ points to the geographic South Pole), $B^{\phi} = B_{\textsc{tge}}$, and $B^r = - B_{\textsc{z}}$ ($\rhat$ points locally out of the ground,%
\footnote{\label{eq:sphericalEarthSociety}%
    We assume an exactly spherical surface for the Earth through this paper; this approximation is accurate at the 0.3\% level~\cite{WGS84}.
    } %
whereas the SuperMAG convention is to measure $B_{\textsc{z}}$ positive when pointing down).

\subsection{Post-processing by SuperMAG}
\label{sec:superMAGpostProcessing}
In addition to the 17-day windowing procedure outlined in \secref{superMAGCoordinateSystem} that is used to obtain field components in the SuperMAG co-ordinate system, the default SuperMAG data product magnetic field measurements have also been post-processed to remove a `baseline' field component consisting of a sum of time-varying diurnal and slower annual components (as well as a constant offset irrelevant for the purposes of searching for a time-dependent signal, as here).

The procedure used to perform this subtraction is detailed in \citeR{Gjerloev:2012sdg}; for the present purposes we simply note that the procedure utilized to remove the diurnal component involves examining magnetic field data in discrete coarse-grained intervals of 30 minutes in length.
While this diurnal baseline subtraction can result in the removal of even quite monochromatic signals with periods longer than 30 minutes (see, e.g., Fig.~6 of \citeR{Gjerloev:2012sdg}, where a strong six-hour signal is removed from the $B_{\textsc{lmn}}$ data from a single station), the coarse-graining of the data into 30 minute intervals implies that this procedure should not significantly impact any frequencies somewhat higher than $(30\,\text{min})^{-1} \sim 5\times 10^{-4}\Hz$, although it could impact frequencies around or below this.
Since the lower end of our frequency range of interest is $f_{A'}\sim 6\times10^{-4}\Hz$, the effects of this diurnal baseline subtraction are mostly out-of-band for the DPDM mass range of greatest interest to us.
We have explicitly re-run our analysis pipeline on the non-baseline-subtracted dataset that is also available from SuperMAG~\cite{SuperMAGwebsite} and verified that the diurnal subtraction is not observed to remove any DM-likeline features in the results near our frequency range of interest.

The annual baseline subtraction procedure makes use of data which are aggregated using the same 17-day sliding window that was employed to determine the co-ordinate system rotations, and is thus also well out-of-band.
Therefore, it would be consistent to use the data either with or without these two time-dependent baseline subtractions for the purposes of our analysis.
However, in order to determine the appropriate data-driven weighting to give the measurements from each station in our analysis [see \eqref[s]{weights_phi} and (\ref{eq:weights_theta}) below], it is more appropriate to utilise the data with the time-dependent baseline subtracted, as this disregards some noise below our frequency range of interest.

Later in our analysis treatment we also assume that a station that is not reporting data reports exactly zero field (instead of the mean field).
For consistency with this treatment, we must work with data whose mean DC value is zero in order to avoid introducing artificial discontinuities of order the size of the mean field.
Therefore, we will work with the fully baseline subtracted data throughout.

\subsection{Temporal features in SuperMAG data}
\label{sec:superMAGtemporal}
It is typical for the number of stations that are active to change significantly at the beginning of a calendar year; see \figref{station_count}.
This is likely due to a tendency for stations to report data associated with specific calendar years.
Importantly for us, this means that the amount of available data fluctuates relatively little within a calendar year, but may fluctuate significantly between calendar years.
In our noise analysis, we will therefore compute separate noise spectra for each calendar year.
This, of course, assumes that the noise level remains relatively constant within a calendar year.
We assess the validity of this assumption in \appref{calendarYear}.

\section{Analysis strategy}
\label{sec:analysisOverview}
We now turn to a high-level description of the analysis procedures we have utilized in order to search for the signal (described in \secref{theory}) in the SuperMAG data (described in \secref{superMAGdescription}).
Details of the implementation of this analysis follow in \secref{analysisDetails}.

Station $i$, located at geographic co-ordinates $\Omega_i=(\theta_i,\phi_i)$, reports a time series of three-axis magnetic field measurements; we denote the field measured at time $t_j$ as $\bm{B}_i(t_j)$.
We denote the set of sampling times at which station $i$ reports valid measurements as $\mathcal{T}_i$.
As already described, the $\mathcal{T}_i$ vary among stations, making a straightforward analysis of the individual stations fairly complicated.

Two possible analysis approaches suggest themselves: 
(A) from the predicted signal \eqref{signal}, one could construct the expected per-station signals $\bm{B}^{\text{signal}}_i(t_j)$ at every time $t_j \in \mathcal{T}_i$. 
One could then perform a simultaneous joint search in the observed $\bm{B}^{\text{observed}}_i(t_j)$ over all stations for the correlated signal predictions in all the stations, and thereby extract a signal or limits on the value of $\varepsilon$ as a function of $m_{A'}$; or
(B) one could exploit the observation that our signal \eqref{signal} is predicted to be in only a small number of global vector spherical harmonics (VSH). 
Therefore, one could instead first extract from all the individual station measurements $\bm{B}^{\text{observed}}_i(t_j)$ a small number of time series that give the projections of the entire collection of station measurements onto the field components of the independent global vector spherical harmonic modes of interest. 
One could then perform a search on these distilled VSH component time series for the signal \eqref{signal} and thereby extract a signal or limits on the value of $\varepsilon$ as a function of $m_{A'}$.
For technical reasons, we find it simpler to utilize approach (B).

Our analysis strategy will thus be to first identify the appropriate global VSH components of interest to find a signal of the form \eqref{signal}; we find that there are five such components of interest, which we denote $X^{(1)},\ldots,X^{(5)}$.
We will then combine all station measurements $\bm{B}_i(t_j)$ available at a given time $t_j$ to extract the values of $X^{(1)}(t_j),\ldots,X^{(5)}(t_j)$, which are the distilled time series of the VSH components previously mentioned.
As noted, the signal \eqref{signal} is very narrow in frequency, so it is most appropriate to search for the signal in the frequency domain; however, since the total duration of the data-taking for the available SuperMAG data is in many cases significantly longer than the signal coherence time, a straightforward Fourier transform of the full time series would in general result in a non-monochromatic peak in frequency space if a DM signal were present; extracting a rigorous limit or signal amplitude estimate would then require knowledge of the exact shape of this peak, which is a fairly nontrivial problem that relies on detailed knowledge of the velocity dispersion of the DM (see, e.g., \citeR[s]{Turner:1990qx,Derevianko:2016vpm,Foster:2017hbq,OHare:2017yze,Centers:2019dyn,Lisanti:2021vij,Gramolin:2021mqv} on this and related points applicable to dark-photon and axion-like dark matter).
In order to avoid this issue, whenever the coherence time of the signal is shorter than the available data-taking duration, we instead break up our full time series $X^{(1)}(t_j),\ldots,X^{(5)}(t_j)$ into a number of shorter subseries, each of which has a duration of (approximately; see next section) a single coherence time for the frequency of interest.
We then Fourier transform each of these subseries to the frequency domain, and because the signal is then coherent in each of the subseries by construction, we can conduct independent searches for a monochromatic (i.e., single frequency bin) signal in the frequency domain subseries.
As a second step, we then incoherently stack the results obtained from the searches on subseries into a single result for the frequency of interest, which we do taking into account that the signal phase and DM amplitude vary stochastically from one coherence time to the next.
The requisite estimates for the noise in the VSH time series are obtained in a data-driven fashion within each coherence time, as detailed in the next section.

We utilize a Bayesian analysis framework: assuming a reparametrization-invariant (Jeffreys) prior on $\varepsilon$ at each mass $m_{A'}$, we use the SuperMAG data to extract the fully marginalized Bayesian posterior distribution on $\varepsilon$, from which we extract upper limits on $\varepsilon$ at each mass $m_{A'}$; see~Refs.~\cite{Centers:2019dyn,roussy2021experimental} for a similar approach.

In the cases where our search indicates the possible existence of a signal at some threshold significance in the full dataset, we first verify that it has the appropriate frequency-domain width for a DM signal.
If the candidate peak passes this test, we then perform subsampling checks to test whether or not the full-dataset signal is consistent with being the expected global DM signal, or whether the degree either of geographical variation in the signal between different random selections of stations, or of temporal variation in the signal between different disjoint subsets of the data broken up over time, are too great to be consistent with the DM interpretation, perhaps indicating that the signal is being driven by some large noise fluctuation in a small number of stations, or for some finite duration of time.

\section{Analysis details}
\label{sec:analysisDetails}
In the previous section, we provided a high-level overview of our analysis strategy; in this section we will provide a detailed description of the analysis.
We begin in \secref{analysisTimeSeries}, with a discussion of the selection of the five time series $X^{(1)}(t_j),\ldots,X^{(5)}(t_j)$ and the procedure by which we combine the different station measurements into these five time series. 
Additionally, we discuss the breaking of these time-series data into single-coherence-time subseries.
In \secref{analysisSignal}, we discuss how the same procedure affects a hypothetical signal \eqref{signal}, as this will determine the expectation values of the $X^{(n)}(t_j)$ that enter in the likelihood function we utilize to construct the posterior distribution on $\varepsilon$.
A noise estimate is also required to construct the likelihood function; in \secref{analysisNoise}, we outline our data-driven noise estimation procedure (with validation checks discussed in \appref{noiseValidation}).
In \secref{analysisBayesian}, we construct the likelihood function, and then derive the posterior distribution on $\varepsilon$ given the observed SuperMAG data.
Finally, in \secref{analysisFreqChoice}, we address a technical point related to the choice of frequencies in our analysis and the way in which they relate to an approximation we make for the coherence time of the signal of interest.

For the convenience and reference of the reader, we collect in \tabref{variables} a variety of the analysis variables we will define in this section, a cross-reference to where they are defined, and a brief description of each.

\begin{table}[t]
\caption{\label{tab:variables}%
    A selection of analysis variables we define in \secref{analysisDetails}, a cross-reference to the location in the text where they are defined, and a brief summary description.
    This table is provided for the reference and convenience of the reader.%
    }
\begin{ruledtabular}
\begin{tabular}{lp{1.85cm}p{5cm}}
Variable        &   Cross Ref.    &   Description     \\ \hline
$X_i^{(n)}(t_j)$  &   \eqrefRange{X1i}{X5i}   &   $n$-th type of projection of the magnetic field measured at station $i$ onto a specific VSH component at time $t_j$ \\[1ex]
$X^{(n)}(t_j)$  &   \eqref{Xn}   &   Weighted sum of the $X_i^{(n)}(t_j)$ over all stations $i$ \\[1ex]
$X_k^{(n)}(t)$    &   \secref{analysisTimeSeriesSubsets} & $X^{(n)}(t)$ restricted to times during the $k$-th coherence time \\[1ex]
$\vec{X}_k$   &   \eqref{vecXdefn}    & 15-dimensional analysis vector consisting of the Fourier transforms of the $X_k^{(n)}(t)$ at frequencies $f_{A'},f_{A'}\pm\hat{f}_d$\\[1ex]
$\langle\vec{X}_k\rangle$   &   \secref{analysisSignal},\newline \appref{Xk}     & Expected value of $\vec{X}_k$ in the presence of the signal \eqref{signal}\\[1ex]
$x^{(m)}(t_j)$  &   \secref{analysisNoise}  &   Hypothetical realization of the data $X^{(m)}(t_j)$ over a duration $\tau$ contained entirely within a calendar year $a$
\end{tabular}
\end{ruledtabular}
\end{table}

\subsection{Time series}
\label{sec:analysisTimeSeries}

\subsubsection{Selection of VSH components}
\label{sec:analysisTimeSeriesSelection}
The first point to address is the selection of the appropriate time-series VSH components on which to perform our analysis.
We would like to extract a set of variables defined on the available magnetic field measurements which keep only the information in the measured fields which could have a spatial overlap%
\footnote{\label{ftnt:TemporalOverlap}%
		The temporal overlap is considered by going to the frequency domain later.
	} %
with the signal; that is, the dot-product $\bm{B}(\Omega_i,t_j)\cdot \bm{B}_{i}(t_j)$ of the expected signal $\bm{B}(\Omega_i,t_j)$ and the observed fields $\bm{B}_i(t_j)$ should form the basis of the information we wish to extract.
The signal \eqref{signal} is proportional to the expression $\bm{B}(\Omega_i,t_j)\propto \RE \lb[ \sum_m a'_m(t_j) \bm{\Phi}_{1m}(\Omega_i) \rb]$ for some complex $a'_m(t_j)$ that are related to $A'_m$.
Now, noting that the $\bm{B}_{i}(t_j)$ are real, it is easy to see that
\begin{align}
\bm{B}(\Omega_i,t_j)\cdot \bm{B}_{i}(t_j)
&\propto \RE \lb[ \sum_m a'_m(t_j) \bm{\Phi}_{1m}(\Omega_i) \cdot \bm{B}_i(t_j)\rb]; \label{eq:dotProductSum}
\end{align}
using the explicit expressions for the $\bm{\Phi}_{1m}(\Omega_i)$ VSH that are displayed at \eqrefRange{Phi1m1}{Phi1p1}, it can be shown that this sum can be written as a linear combination of the variables
\begin{align}
X_i^{(1)}(t_j) &\equiv \sin\phi_i \cdot B_i^\theta(t_j) \nonumber\\
              &\propto \RE[\Phi_{11}^\theta(\Omega_i)\cdot B_i^\theta(t_j)], \label{eq:X1i} \\
X_i^{(2)}(t_j) &\equiv \cos\phi_i \cdot B_i^\theta(t_j) \nonumber\\
                &\propto \IM[\Phi_{11}^\theta(\Omega_i)\cdot B_i^\theta(t_j)], \label{eq:X2i}\\
X_i^{(3)}(t_j) &\equiv \cos\phi_i\cos\theta_i \cdot B_i^\phi(t_j)  \nonumber\\
                &\propto \RE[\Phi_{11}^\phi(\Omega_i)\cdot B_i^\phi(t_j)], \label{eq:X3i} \\
X_i^{(4)}(t_j) &\equiv -\sin\phi_i\cos\theta_i \cdot B_i^\phi(t_j)  \nonumber\\
                &\propto \IM[\Phi_{11}^\phi(\Omega_i)\cdot B_i^\phi(t_j)], \label{eq:X4i}\\
X_i^{(5)}(t_j) &\equiv \sin\theta_i \cdot B_i^\phi(t_j)  \nonumber\\
                &\propto \RE[\Phi_{10}^\phi(\Omega_i)\cdot B_i^\phi(t_j)]. \label{eq:X5i}
\end{align}
These $X^{(n)}_i(t_j)$ hold all the information about the observed fields at station $i$ that could possibly spatially overlap with the expected signal we wish to constrain or observe, and we can thus structure our analysis around these variables.

\subsubsection{Combination of stations}
\label{sec:analysisTimeSeriesCombination}
To combine the results from all the stations into a small number of time series, we simply take the weighted averages over all stations of these projections on the VSH components:
\begin{align}
X^{(n)}(t_j) &= \frac1{W^{(n)}(t_j)}\sum_{\{i|t_j\in\mathcal T_i\}}w_i^{(n)}(t_j) X_i^{(n)}(t_j),
\label{eq:Xn}
\end{align}
where the notation `$\{i|t_j\in\mathcal T_i\}$' indicates that the sum is over the set of stations $i$ such that there is a valid field measurement from station $i$ at time $t_j$.%
\footnote{\label{ftnt:equivalentFormulation}%
	    We could equivalently formulate this as saying that the sum is over all stations $i$, but that the weights are zero-ed out at all times when a station is not reporting valid data: $w^{(n)}_i(t_j\notin T_i) = 0$.
	} %
The weights $w_i^{(n)}(t_j)$ we choose for the station at location $\Omega_i$ will be taken to be constant within the time span over which we will assume stationarity of the noise distributions [one calendar year; see \secref[s]{superMAGtemporal} and \ref{sec:analysisNoise}], so that the noise distributions we estimate from the data are informed entirely from the magnetic field noise and from the fluctuations in which stations are reporting, and we do not inject additional temporal variation via explicitly time-dependent weights.

The choice of weights for each period of assumed noise stationarity (i.e., one calendar year) could in principle be arbitrary; however, a reasonable assumption is to take the weights to be informed by the per-station white noise levels, assuming the noise between stations is uncorrelated (see discussion in \secref{analysisNoise} below).
That is, we will take $w_i^{(n)}$ for $n=1,2$ to be the inverse of the station-$i$ white noise level in $B_i^\phi$ over a given year, while $w_i^{(n)}$ for $n=3,4,5$ will be taken to be the inverse of the station-$i$ white noise level in $B_i^\theta$ over a given year.
More specifically, for all $t$ within year $a$, we take
\begin{align}
    \label{eq:weights_phi}
    w_i^{(n)}(t)&=\lb[\frac1{N_i^a}\sum_{t_j\in\mathcal T_i^a } \lb[B_i^\phi(t_j)\rb]^2\rb]^{-1} & [n&= 1,2], \\
    w_i^{(n)}(t)&=\lb[\frac1{N_i^a}\sum_{t_j\in\mathcal T_i^a}\lb[B_i^\theta(t_j)\rb]^2\rb]^{-1} & [n&=3,4,5],
    \label{eq:weights_theta}
\end{align}
where $\mathcal{T}_i^a$ is the subset of $\mathcal{T}_i$ [see \secref{analysisOverview} and below \eqref{Xn}] contained entirely within year $a$, and $N_i^a$ is the corresponding number of samples in $\mathcal{T}_i^a$.
The normalizing total weight is then simply defined by
\begin{align}
W^{(n)}(t_j)=\sum_{\{i|t_j\in\mathcal T_i\}}w_i^{(n)}(t_j);
\label{eq:W}
\end{align}
note that even though in our analysis all the $w_i^{(n)}(t_j)$ are themselves constant within a year, $W^{(n)}(t_j)$ may still change on more rapid timescales because the number of stations reporting generically changes over time.

\subsubsection{Coherent-signal data subsets}
\label{sec:analysisTimeSeriesSubsets}
The $X^{(n)}(t_j)$ time series contain all the relevant information we need to proceed with our data analysis in \secref{analysisDetails}. 

However, as we have already explained in \secref{analysisOverview}, the coherence time of the signal can be shorter than the full duration of available SuperMAG data, and we wish to avoid having to search for signals that have a resolvable width in frequency space, as this complicates the analysis significantly (and depends in part on the exact DPDM lineshape). 
Instead, we will perform our search over total durations that are longer than the intrinsic signal coherence time by first analyzing the data coherently within each separate, disjoint coherence time, and then incoherently combining the results from these single-coherence-time searches.
By way of concrete example, we mean that if we are confronted with searching for a signal with a six-year coherence time ($f_{A'}\sim 5.3$\,mHz for $v_{\textsc{dm}}\sim 10^{-3}$), we would separate the total 48 years of available SuperMAG data into eight separate consecutive data subsets. 
We then perform eight independent fully coherent searches for the signal, one in each of these subsets; finally, we stack these search results incoherently to obtain a final search result.

To this end, let us denote by $X_k^{(n)}$ the subseries of the time series $X^{(n)}$ that contains only the data from the ${k\text{-th}}$ disjoint interval $[k=1,\ldots,K]$ of duration $T$ within the full SuperMAG data taking duration.
As we discuss below in some detail in \secref{analysisFreqChoice}, we will take $T$ to be \emph{approximately} equal to the coherence time for the signal: $T\approx T_{\text{coh}} \sim 10^{6} f_{A'}^{-1} \sim 2\pi /(m_{A'}v_{\textsc{dm}}^2)$, assuming $v_{\textsc{dm}}\sim 10^{-3}$.
\footnote{\label{ftnt:imprecisionExcused}%
		Unless more precision is required to avoid confusion, in order to avoid the repetitive incantation of `approximate coherence time' and other such caveats, we will hereinafter simply refer to the duration of time $T$ as `the coherence time', and to any such time period of duration $T$ as a `coherence time', with this approximation implicitly understood.
	} %

We will analyze each of the subseries $X_k^{(n)}$ independently in the frequency domain; we denote the Fourier transform (FT) of the subseries $X_k^{(n)}(t)$ by $\tilde{X}_k^{(n)}(f)$.
We can see from \eqref{signal} that a signal in the data would contribute power not only at the cycles-per-second frequency corresponding to the DPDM mass, $f_{A'} = m_{A'} / ( 2\pi )$; to the extent that the data contain a signal spatially oriented such that there is some $m=\pm1$ contribution, there will also be power at $f = f_{A'}\pm f_d$ where, as before, $f_d = (\text{sidereal day})^{-1}$.
Therefore it will be relevant to consider the Fourier transforms $\tilde{X}_k^{(n)}(f)$ at $f=f_{A'}$ and at $f=f_{A'}\pm f_d$.%
\footnote{\label{ftnt:someMightBeZero}%
		Note however that it is possible that some of the $\tilde{X}_k^{(n)}(f)$ we thus consider are identically zero for a signal; we compute the expected $\tilde{X}_k^{(n)}(f)$ assuming a signal is present in \secref{analysisSignal}.
	} %

Actually, since we obtain the FT of a time-domain signal of total duration $T$ and discrete sampling cadence $\Delta t$ (for a total of $N=T/\Delta t$ samples) via the Discrete Fourier Transform (DFT) [or, more precisely, by the Fast Fourier Transform (FFT) implementation of the DFT], we only obtain independent frequency information at a discrete set of predetermined frequencies $f_k = k \Delta f$ where $\Delta f = 1/T$ and $k=0,\ldots,N-1$.
We are thus not generically able to obtain (at least not within the context of the FFT) the FT value $\tilde{X}_k^{(n)}(f)$ at exactly all the frequencies $f_{A'}$ and $f_{A'}\pm f_d$.
Instead, we will consider the FT at $f_{A'}$, which we will always by construction take to be an exact DFT frequency ($f_{A'} = m \Delta f$ for some integer $m$; see \secref{analysisFreqChoice}); and at $f_{A'}\pm\hat f_d$, where $\hat f_d$ is the closest multiple of $\Delta f$ to $f_d$ (i.e., $\hat{f}_d \equiv n \Delta f$ for the integer value of $n$ such that $| n\Delta f - f_d |$ is minimized).
We note that a refinement of this approach, in particular one that considers the full lineshape in the Fourier domain, would likely be possible at additional computational expense.

With this in mind, we define a new 15-dimensional `analysis vector' $\vec X_k$ that contains the values of the FT $\tilde{X}_k^{(n)}(f)$ for $n=1,\ldots,5$ at frequencies $f=f_{A'}$ and $f = f_{A'} \pm \hat{f}_d$:%
\interfootnotelinepenalty=10000
\footnote{\label{ftnt:15vs3}%
		Here, and throughout, we use $\vec {x}$ to denote a vector $x$ with 15 components, and $\bm{y}$ to indicate a vector $y$ with three (usually spatial) components.
	} %
\interfootnotelinepenalty=100
\begin{align}
\vec X_k = \begin{pmatrix}
			\tilde X_k^{(1)}(f_{A'}-\hat f_d)\\
			\tilde X_k^{(2)}(f_{A'}-\hat f_d)\\
			\tilde X_k^{(3)}(f_{A'}-\hat f_d)\\
			\tilde X_k^{(4)}(f_{A'}-\hat f_d)\\
			\tilde X_k^{(5)}(f_{A'}-\hat f_d)\\
			\tilde X_k^{(1)}(f_{A'})\\
			\tilde X_k^{(2)}(f_{A'})\\
			\tilde X_k^{(3)}(f_{A'})\\
			\tilde X_k^{(4)}(f_{A'})\\
			\tilde X_k^{(5)}(f_{A'})\\
			\tilde X_k^{(1)}(f_{A'}+\hat f_d)\\
			\tilde X_k^{(2)}(f_{A'}+\hat f_d)\\
			\tilde X_k^{(3)}(f_{A'}+\hat f_d)\\
			\tilde X_k^{(4)}(f_{A'}+\hat f_d)\\
			\tilde X_k^{(5)}(f_{A'}+\hat f_d)
		\end{pmatrix}.
		\label{eq:vecXdefn}
\end{align}

These analysis vectors $\vec X_k$ are the central objects we use in \secref{analysisBayesian} to construct the likelihood function on which our analysis is based; we will consider them to be multivariate Gaussian variables, an assumption we validate in \appref{gaussianity}.  
As such we will need to know their expectation values given an injected signal [\secref{analysisSignal}] and their covariance matrices [\secref{analysisNoise}].

\subsection{Signal}
\label{sec:analysisSignal}
Having described the general procedures we utilize to search for signals appearing in the relevant VSH components of the SuperMAG data in the previous subsection, in this subsection we will derive the expected values of the $\vec X_k$ variables which arise under the dark-photon dark-matter signal model, \eqref{signal}; we denote the expected $\vec X_k$ under the signal hypothesis with parameter $\varepsilon$ by $\langle \vec X_k \rangle$.

In principle, this derivation amounts to simply substituting \eqref{signal} into the definitions of the time series in \eqrefRange{X1i}{X5i}; however, it is useful to examine intermediate results here, so we will develop this section pedagogically.

As a first step, it is useful to more explicitly understand the signal \eqref{signal} in the case where the DPDM is polarized along any of the three inertial Cartesian axes [i.e., the set of rigid, mutually orthogonal axes fixed in space with respect to the (average) positions of a field of distant stars, not the body-fixed axes rigidly attached to the rotating Earth].
To this end, we define variables $c_{i}$ which define the orientation of the DPDM field along the inertial $i$-axis for $i=x,y,z$:
\begin{align}
c_i &\equiv\frac{\sqrt2\pi f_{A'}A'_i}{\sqrt{\rho_{\textsc{dm}}}};
\end{align}
or, in terms of the $A_{\pm,0}$, we have
\begin{align}
c_x &= -\frac{\pi f_{A'}(A'_+-A'_-)}{\sqrt{\rho_{\textsc{dm}}}}, \label{eq:cx}\\
c_y &= -\frac{i\pi f_{A'}(A'_++A'_-)}{\sqrt{\rho_{\textsc{dm}}}}, \label{eq:cy} \\
c_z &= \frac{\sqrt2\pi f_{A'}A'_0}{\sqrt{\rho_{\textsc{dm}}}} \label{eq:cz}.
\end{align}
{\indent}These are normalized such that for a (hypothetical) linearly polarized DPDM signal that is oriented along the unit vector $\bm{\hat{n}}$ (in the inertial frame) and that satisfies $\frac{1}{2} m_{A'}^2|\bm{A'}|^2=\rho_{\textsc{dm}}$, we would have $c_i = \hat{n}^i$ where $i=x,y,z$ and $\hat{n}^i$ denotes the $i$-th Cartesian component of $\bm{\hat{n}}$.
Expanding \eqref{signal} in terms of these variables, we can write 
\begin{align}
\bm{B} &\equiv \hspace{0.87em} \RE[c_x]\cdot\bm{B}^{x}_{R}+\IM[c_x]\cdot\bm{B}^{x}_{I}, \nl
		 +\RE[c_y]\cdot\bm{B}^{y}_{R}+\IM[c_y]\cdot\bm{B}^{y}_{I}, \nl
		+\RE[c_z]\cdot\bm{B}^{z}_{R}+\IM[c_z]\cdot\bm{B}^{z}_{I},
\end{align}
where we have defined
\begin{widetext}
\begin{align}
\bm{B}^{x}_{R}(\Omega,t) &= \pi\varepsilon f_{A'}R\sqrt{2\rho_{\textsc{dm}}}\lb(\sin(2\pi f_dt+\phi)\thetahat+\cos(2\pi f_dt+\phi)\cos\theta\phihat\rb)\cos(2\pi f_{A'}t),\\
\bm{B}^{x}_{I}(\Omega,t) &= \pi\varepsilon f_{A'}R\sqrt{2\rho_{\textsc{dm}}}\lb(\sin(2\pi f_dt+\phi)\thetahat+\cos(2\pi f_dt+\phi)\cos\theta\phihat\rb)\sin(2\pi f_{A'}t),\\
\bm{B}^{y}_{R}(\Omega,t) &= \pi\varepsilon f_{A'}R\sqrt{2\rho_{\textsc{dm}}}\lb(-\cos(2\pi f_dt+\phi)\thetahat+\sin(2\pi f_dt+\phi)\cos\theta\phihat\rb)\cos(2\pi f_{A'}t),\\
\bm{B}^{y}_{I}(\Omega,t) &= \pi\varepsilon f_{A'}R\sqrt{2\rho_{\textsc{dm}}}\lb(-\cos(2\pi f_dt+\phi)\thetahat+\sin(2\pi f_dt+\phi)\cos\theta\phihat\rb)\sin(2\pi f_{A'}t),\\
\bm{B}^{z}_{R}(\Omega,t) &= -\pi\varepsilon f_{A'}R\sqrt{2\rho_{\textsc{dm}}}\sin\theta\cos(2\pi f_{A'}t)\phihat,\\
\bm{B}^{z}_{I}(\Omega,t) &= -\pi\varepsilon f_{A'}R\sqrt{2\rho_{\textsc{dm}}}\sin\theta\sin(2\pi f_{A'}t)\phihat.
\end{align}
\end{widetext}
Here $\Omega\equiv(\theta,\phi)$ denotes the spherical co-ordinates of the observation point on the surface of the Earth, and $\thetahat$ and $\phihat$ are the associated unit vectors (recall: $\theta$, $\phi$, $\thetahat$, and $\phihat$ are all defined in the body-fixed frame that an observer co-rotating with the surface of the Earth would naturally use).

Substituting these expressions for $\bm{B}$ into \eqrefRange{X1i}{X5i} and Fourier transforming yields the contribution of each polarization to $\langle \vec X_k\rangle$.

We define several auxiliary time-dependent functions $H^{(n)}(t_j)$ [$n=1,\ldots,7$] which will allow us to more compactly express our results provided that the per-station weights are taken to be%
\footnote{\label{ftnt:weightChoice}%
    From \eqrefRange{X1i}{X5i}, we see that the $X_i^{(n)}$ for $n=1,2$ depend only on $B_i^{\theta}$, while the $X_i^{(n)}$ for $n=3,4,5$ depend only on $B_i^{\phi}$.
    Therefore, since we use per-station data-driven estimates for the station weights (see \secref{analysisTimeSeriesCombination}), it stands to reason that the weights applied for $X_i^{(n)}$ for $n=1,2$ should be common, while those for $X_i^{(n)}$ for $n=3,4,5$ should also be common, but with the latter common value distinct from the former.} %
$w_i^{(n)}\equiv w_i^{(\theta)}$ for $n=1,2$ and $w_i^{(n)}\equiv w_i^{(\phi)}$ for $n=3,4,5$:
\begin{align}
H^{(1)}(t_j)&=\frac1{W^{(\theta)}(t_j)}\sum_{i:t_j\in\mathcal T_i}w_i^{(\theta)}(t_j)\cos^2\phi_i,\label{eq:H1i}\\
H^{(2)}(t_j)&=\frac1{W^{(\theta)}(t_j)}\sum_{i:t_j\in\mathcal T_i}w_i^{(\theta)}(t_j)\sin\phi_i\cos\phi_i,\\
H^{(3)}(t_j)&=\frac1{W^{(\phi)}(t_j)}\sum_{i:t_j\in\mathcal T_i}w_i^{(\phi)}(t_j)\cos^2\theta_i,
\end{align}
\linebreak
\begin{align}
H^{(4)}(t_j)&=\frac1{W^{(\phi)}(t_j)}\sum_{i:t_j\in\mathcal T_i}w_i^{(\phi)}(t_j)\cos^2\phi_i\cos^2\theta_i,\\
H^{(5)}(t_j)&=\frac1{W^{(\phi)}(t_j)}\sum_{i:t_j\in\mathcal T_i}w_i^{(\phi)}(t_j)\sin\phi\cos\phi\cos^2\theta_i,\\
H^{(6)}(t_j)&=\frac1{W^{(\phi)}(t_j)}\sum_{i:t_j\in\mathcal T_i}w_i^{(\phi)}(t_j)\cos\phi_i\sin\theta_i\cos\theta_i,\\
H^{(7)}(t_j)&=\frac1{W^{(\phi)}(t_j)}\sum_{i:t_j\in\mathcal T_i}w_i^{(\phi)}(t_j)\sin\phi_i\sin\theta_i\cos\theta_i,\label{eq:H7i}
\end{align}
where $W^{(\theta)}$ and $W^{(\phi)}$ are defined as in \eqref{W}. 
Armed with these functions, we are in a position to write down the contributions to $\langle \vec X_k\rangle$. 
For instance, in the case that the signal is a dark photon oriented entirely in the $z$-direction (and has a phase such that $c_z\in \mathbb{R}$), we have\linebreak
\begin{widetext}
\begin{align}
\langle\vec X_k\rangle_{\bm{B}=\bm{B}^{z}_{R}} 
	&=-\pi\varepsilon f_{A'}R\sqrt{\frac{\rho_{\textsc{dm}}}2}
			\begin{pmatrix}
				0\\
				0\\
				\tilde H_k^{(6)}(-\hat f_d)+\tilde H_k^{(6)}(2f_{A'}-\hat f_d)\\
				-\tilde H_k^{(7)}(-\hat f_d)-\tilde H_k^{(7)}(2f_{A'}-\hat f_d)\\
				\tilde{\mathbf1}_k(-\hat f_d)-\tilde H_k^{(3)}(-\hat f_d)+\tilde{\mathbf1}_k(2f_{A'}-\hat f_d)-\tilde H_k^{(3)}(2f_{A'}-\hat f_d)\\
				0\\
				0\\
				\tilde H_k^{(6)}(0)+\tilde H_k^{(6)}(2f_{A'})\\
				-\tilde H_k^{(7)}(0)-\tilde H_k^{(7)}(2f_{A'})\\
				\tilde{\mathbf1}_k(0)-\tilde H_k^{(3)}(0)+\tilde{\mathbf1}_k(2f_{A'})-\tilde H_k^{(3)}(2f_{A'})\\
				0\\
				0\\
				\tilde H_k^{(6)}(\hat f_d)+\tilde H_k^{(6)}(2f_{A'}+\hat f_d)\\
				-\tilde H_k^{(7)}(\hat f_d)-\tilde H_k^{(7)}(2f_{A'}+\hat f_d)\\
				\tilde{\mathbf1}_k(\hat f_d)-\tilde H_k^{(3)}(\hat f_d)+\tilde{\mathbf1}_k(2f_{A'}+\hat f_d)-\tilde H_k^{(3)}(2f_{A'}+\hat f_d)
			\end{pmatrix}
			\label{eq:XexpzExact}\\
	&\approx-\pi\varepsilon f_{A'}R\sqrt{\frac{\rho_{\textsc{dm}}}2}
			\begin{pmatrix}
				0\\
				0\\
				\tilde H_k^{(6)}(-\hat f_d)\\
				-\tilde H_k^{(7)}(-\hat f_d)\\
				\tilde{\mathbf1}_k(-\hat f_d)-\tilde H_k^{(3)}(-\hat f_d)\\
				0\\
				0\\
				\tilde H_k^{(6)}(0)\\
				-\tilde H_k^{(7)}(0)\\
				\tilde{\mathbf1}_k(0)-\tilde H_k^{(3)}(0)\\
				0\\
				0\\
				\tilde H_k^{(6)}(\hat f_d)\\
				-\tilde H_k^{(7)}(\hat f_d)\\
				\tilde{\mathbf1}_k(\hat f_d)-\tilde H_k^{(3)}(\hat f_d)
			\end{pmatrix} \equiv \varepsilon\vec\mu_{zk}. \label{eq:XexpzApprox}
\end{align}
\end{widetext}
Here, $H_k^{(m)}$ represents the subseries of $H^{(m)}$ consisting of the same sampling times as the subseries $X_k^{(m)}$ of $X^{(m)}$, and $\tilde H_k^{(m)}$ is its Fourier transform.
The series $\mathbf 1_k$ is a series of 1's at these same sampling times, and $\tilde{\mathbf 1}_k$ its Fourier transform.%
\footnote{\label{ftnt:lengthMismatch}%
		Therefore typically $\tilde{\mathbf 1}_k(0)=T$. 
		However, we do adjust this to account for the situation in which no stations report valid measurements for some subset of times within the $k$-th coherence time, or the analysis duration for any one $k$ happens to be shorter than $T$ (e.g., for the last interval).
	} %

Note that we have made the approximation in moving from \eqref{XexpzExact} to \eqref{XexpzApprox} that $\tilde H_k^{(m)}$ decays rapidly with increasing frequency, so that we may discard high frequency contributions (i.e., those at $f=2f_{A'}$, and $f=2f_{A'} \pm \hat{f}_d$).
We verified that, with our choices of weightings, this is a valid approximation: for instance, $\tilde H_k^{(m)}(2f_{A'}\pm\hat f_d)/\tilde H_k^{(m)}(\pm\hat f_d)$ is typically of order a few percent, and $\tilde H_k^{(m)}(2f_{A'})/\tilde H_k^{(m)}(0)$ is typically $\sim 10^{-5}$.
This approximation has the computational advantage that we need only know the Fourier transforms of the $H^{(m)}_k(t_j)$ at three frequencies, and so we can avoid performing an FFT.

Under this same assumption, it is not difficult to see that if the signal is oriented along the $z$-direction but with $i c_z \in \mathbb{R}$, we have
\begin{align}
\langle\vec X_k\rangle_{\bm{B}=\bm{B}^{z}_{I}}\approx-i\varepsilon\vec\mu_{zk}. 
\label{eq:imag}
\end{align}
We may likewise define $\vec\mu_{xk}$ and $\vec\mu_{yk}$ as the contributions to $\langle\vec X_k\rangle$ coming from the $x$- and $y$-polarizations, and an exact analog of \eqref{imag} holds for these too; the full expressions for $\vec\mu_{xk}$ and $\vec\mu_{yk}$ are shown in \appref{Xk}. 

It follows immediately that the full expression for the expectation value of the $\vec X_k$ under the signal hypothesis \eqref{signal} can be written in terms of the $\vec\mu_{ik}$ ($i=x,y,z$) as
\begin{align}
\langle\vec X_k\rangle=\varepsilon\cdot(c_{xk}^*\vec\mu_{xk}+c_{yk}^*\vec\mu_{yk}+c_{zk}^*\vec\mu_{zk}),
\label{eq:combinedXkSignalExpectation}
\end{align}
where the $c_{ik}$ ($i=x,y,z$) encode the inertial-frame polarization state of DPDM during the $k$-th coherence time, and ${}^*$ denotes complex conjugation.

\subsubsection{Signal in other VSH modes}
The signal \eqref{signal} was derived in \citeR{Fedderke:2021rys} under the assumption of an exactly spherical geometry; i.e., assuming that the spherical ionosphere acts as the outer boundary for the lower atmospheric cavity in which the dark-photon signal is sourced.
In this case, the dark photon sources only a $\bm\Phi_{1m}$ component of the magnetic field.
However, as we discussed at length in Sec.~II.B of \citeR{Fedderke:2021rys}, details of the ionosphere call this assumption into question, and suggest that the aspherical magnetopause may instead act as the outer boundary of the geometry.
We showed in Sec.~III.C of \citeR{Fedderke:2021rys} that when the spherical ionospheric outer boundary assumption is relaxed, the signal \eqref{signal} in general receives additional contributions from other VSH (e.g., $\bm\Psi_{\ell m}$ and $\bm Y_{\ell m}$ contributions; see \appref{vectorSphericalHarmonics} for definitions); however the $\bm\Phi_{1m}$ component shown at \eqref{signal} remains correct to leading order in an $m_{A'}R(\ll1)$ expansion.

In principle, these additional field contributions are distinguishable from the $\bm\Phi_{1m}$ component at \eqref{signal} due to the global orthogonality of the VSHs; see \eqref{orthogonality}.
If the station locations $\Omega_i$ were uniformly distributed over the Earth's surface and the weights were taken to be $w_i^{(n)}(t_j)=1$ at all stations $i$ and times $t_j$, then the definition of $X^{(n)}$ in \eqref{Xn} would approximate a uniform integral over the sphere in the limit of many stations.
This would project out any $\bm\Psi_{\ell m}$ or $\bm Y_{\ell m}$ contributions to the observed magnetic field $\bm B_i(t_j)$, leaving only the contributions from \eqref{signal}.
Following the analysis through, this would imply that \eqref{combinedXkSignalExpectation} would give the exact signal expectation for the $\vec{X}_k$.

However, due to the nonuniformity of the station distribution, differing noise levels among stations, and variations in the number of stations reporting at a given time, \eqref{Xn} for $X^{(n)}$ deviates from approximating a uniform integral over the sphere.
This implies that field contributions from other VSH modes arising from the magnetospheric asphericity could give unsuppressed contributions to the time series $X^{(n)}$; we estimate that this `leakage' of other VSH components into $X^{(n)}$ could be at the level of tens of percent.
However, while such contributions in principle enter the $X^{(n)}$ in such a way that the expected $\langle \vec{X}_k \rangle$ in the presence of the full signal that includes these other VSH modes would deviate from \eqref{combinedXkSignalExpectation} at the level of an $\mathcal{O}(1)$ factor, it would require a highly unlikely environmental fine-tuning for these modifications to completely cancel the signal contribution \eqref{combinedXkSignalExpectation} that we search for.
For instance, the asphericity in the magnetopause is variable with Solar activity as its shape is strongly sculpted by the radial outflow of the variable Solar wind, and other Solar activity (Coronal Mass Ejection events, etc.); the Earth also rotates inside of it.
It would be exceedingly surprising for some conspiracy between the stochastically varying DPDM field and the evolving magnetopause shape in which the Earth rotates to somehow engineer cancellation of all three components of the vectorial signal \eqref{signal} as it enters the $X^{(n)}$ at \eqref{Xn}, and for that cancellation to be maintained precisely for $\mathcal{O}(50)$ years when considered over all $\mathcal{O}(500)$ stations that switch on and off over time and have varying noise levels completely uncorrelated with the DPDM signal.

While a more refined future analysis may hope to deal with these signal additional contributions more precisely, we are satisfied that these considerations imply that our search is still accurate at the level of (at worst) $\mathcal{O}(1)$ factors even when they are present and not explicitly accounted for.

\subsection{Noise spectra}
\label{sec:analysisNoise}

The statistical analysis of the SuperMAG magnetic field dataset---as expressed in terms of the variables $\vec X_k$ [see \secref{analysisTimeSeries}]---in order to search for a signal of the form $\langle \vec X_k \rangle$ [see \secref{analysisSignal}] requires a quantitative estimate of the noise; we utilize a data-driven noise estimation procedure, which we detail in this subsection.

Our analysis is constructed around the assumptions that the noise in the data time series $\vec X_k$ is (1) Gaussian, and (2) statistically stationary within each calendar year.
We quantify the extent to which (1) and (2) are acceptable assumptions in detail in \appref{noiseValidation}.

Let $x^{(m)}(t_j)$ $[m=1,\ldots,5]$ represent a single hypothetical realization of the data time series which we have denoted as $X^{(m)}(t_j)$, taken over some time span of duration $\tau$ contained entirely within a single calendar year $a$.
Assume the data are taken with a measurement cadence $\Delta t$, such that $\tau \equiv N \Delta t$ with $N \in \mathbb{Z}$ and, here, $t_j = j \Delta t$ for $j=0,\ldots,N-1$; additionally, we consider $x^{(m)}(t_j)$ to be obtained under the assumption that no DPDM signal is present in the data.%
\footnote{\label{ftnt:noiseNoSignal}%
		Note that even if any true dark-photon signal were present in the data, it would have to be very large to invalidate this approach. 
		The DPDM signal line has a width of order $\sigma_f \sim 10^{-6}f_{A'}$. 
		However, the spacing of the DFT frequencies in \eqref{noise} is approximately $(\Delta f)' \sim 1\times10^{-6}\Hz$ if $\tau = 16834\,$min.
		Because our frequency range of interest is $ 6\times 10^{-4}\Hz\lesssim f_{A'} \lesssim 2\times 10^{-2}\Hz$,
		this means that $(\Delta f)'$ lies in the range $1700\gtrsim (\Delta f)'/\sigma_f\gtrsim50$.
		A true DPDM signal would thus have to be huge, at least 50 times larger than the noise level in neighboring bins, to make even an $\mathcal{O}(1)$ impact on the noise estimate.
		For signals smaller than this, the estimate we have outlined here is acceptably accurate.
		For a large signal, the noise estimate outlined here would be formally incorrect; however, we would still see an obvious signal candidate in this case, but further analysis would be required to extract an accurate noise estimate; see, for instance, our signal injection analysis in \secref{analysisChecksValidation} and \figref{resultsInjected}.
	} %
Then, $x^{(m)}(t_j)$ is simply a single hypothetical duration-$\tau$ realization of the noise in the data time series $X^{(m)}(t)$ in year $a$.
We define the two-sided cross-power spectral density of the noise for year $a$ by
\begin{align}
\langle\tilde x^{(m)}(f_p')\, \tilde x^{(n)}(f_q')^*\rangle_{\varepsilon=0} \equiv \tau S_{mn}^a(f_p')\, \delta_{pq},
\label{eq:noise}
\end{align}
where $\langle\, \cdots \rangle_{\varepsilon=0}$ denotes the expectation taken over all possible noise realizations [i.e., with no signal, $\varepsilon=0$], ${\tilde x}^{(m)}(f)$ is the DFT of $x^{(m)}(t_j)$ evaluated at one of the set of DFT frequencies $f_{p,q}'$ 
(see below for discussion and definition of $f'_p$), and $\delta_{pq}$ is the Kronecker delta.

Our data-driven noise estimate of year $a$ is constructed from $S_{mn}^a(f)$, which we wish to estimate from our single realization of the actual data time series, $X^{(m)}(t)$.
One of our fundamental analysis assumptions is that the noise properties of the data are statistically stationary within each calendar year period; see \appref{calendarYear} for validation of this assumption.
Therefore, we divide each calendar year of data $X^{(m)}(t)$ (with $t$ entirely within year $a$) into many temporal `chunks', each of duration $\tau$, and treat each chunk as an independent noise realization [that is, we convert the ensemble average in \eqref{noise} over hypothetical noise realizations to a straight average over chunks of the actual data, under the assumption of noise stationarity].

Since the length of calendar years varies between leap and non-leap years and we wish to use as much of our data as possible, we do not fix the length of $\tau$ universally, but instead choose a universal minimum value $\tau_\text{min}$, and divide each individual calendar year evenly into chunks whose durations exceed $\tau_\text{min}$.
We choose the shortest such duration that allows us to evenly divide the entire year.
Namely for a year of length $T^a$, we use $N_{\text{chunks}}$ chunks of length $\tau$, where
\begin{align}
    N_{\text{chunks}} &\equiv \lb\lfloor\frac{T^a}{\tau_\text{min}}\rb\rfloor, &
    \tau &= \lb\lfloor\frac{T^a}{N_{\text{chunks}}}\rb\rfloor,
\end{align}
and where the second expression assumes a unit of time measurement of minutes (i.e., the `floor function' notation in the second expression is abused to mean `round this result to the nearest minute').

Generically, computing the DFT of a time series of duration $\tau$ can be computationally difficult if the number of sample points in the duration $\tau$ is not a power of 2 (since the measurement cadence of SuperMAG data is $\Delta t = 1\,$min, this means that $\tau$ itself should be a power of 2 when measured in minutes).
We therefore pad our time series $x^{(m)}$ with zeros to extend the number of data points in the chunk to the next power of 2 (i.e., we add additional values of $x^{(m)}(t_j)=0$ at assumed sample times $t_j = j \Delta t$ with $j=N,\ldots,2^p-1$ for some $p\in\mathbb{Z}$).%
\footnote{\label{ftnt:ensembleAverageOK}%
    With an appropriate re-scaling of the normalization of the power spectral density (PSD) computed from the padded data (see footnote \ref{ftnt:paddingRescaling}), the ensemble average of the re-normalized PSD from the padded data and the ensemble average of the PSD from the unpadded data agree statistically with their respective standard deviations of the mean.
    This step is purely for computational advantage.
    }
We therefore find it convenient to choose $\tau_\text{min}$ to be a power of 2, and thus take the extended, padded chunk duration to be $2\tau_\text{min}$.%
\footnote{\label{ftnt:paddingRescaling}%
    Naive application of the definition of the PSD at \eqref{PSDdefn} taking the padded duration and padded number of data points yields the incorrect normalization for the desired PSD in this case because of the dead time associated with the padding. 
    However, since we pad in such a way as to maintain the same $\Delta t$ in both the padded and un-padded data, the normalization of the FFT given at \eqref{DFTdefn} is correct, and the only modification we must make is to re-scale the PSD computed per \eqref{PSDdefn} by a factor of $( 2\tau_{\text{min}}) / \tau$; cf.~\eqref{singleNoiseEstimate} and the comments in footnote \ref{ftnt:proportionalRescaling}. 
    }
The frequencies $f_p'$ at which the DFT $\tilde x^{(m)}$ is computed will thus be multiples of $(\Delta f)' = 1/(2\tau_\text{min})$.
We find $\tau_\text{min}=16384\,\text{min}=2^{14}$\,min to be an adequate choice.
[This implies $\tau=16425$\,min for non-leap years and $\tau=16470$\,min for leap years. 
Additionally, the DFT frequencies $f_p'$ will be multiples of $(\Delta f)' = 1/(32768\,\text{min}) \sim 5 \times 10^{-7}\,$Hz.]
We justify this choice in \appref{tauMin}, and show that our results do not depend strongly on the specific choice we have made.

For the $i$-th chunk of actual data $X^{(m)}$ in year $a$, we compute the quantity%
\footnote{\label{ftnt:proportionalRescaling}%
    The value of $\tau$ appearing in the denominator of \eqref{singleNoiseEstimate} is actually taken to be $\tau \equiv N^i_{\text{data}} \Delta t$, where $N^i_{\text{data}}$ is the number of data sampling points within chunk $i$ for which at least one station has a valid measurement to allow the construction of $x^{(m)}$ (which necessarily is none of the points that have been padded with zeros). 
    Generically, there is at least one station reporting at every time throughout the $i$-th chunk, and this procedure has no effect, yielding a value for $\tau$ that matches the value discussed in the main text; however, for the small number of cases where no stations happen to report data for some duration of the $i$-th chunk, $\tau$ as appearing in \eqref{singleNoiseEstimate} is proportionally re-scaled to a smaller value.
    } %
\begin{align}
S_{mn}^{a,i}(f_p') \equiv \frac{\tilde x^{(m)}(f_p')\tilde x^{(n)}(f_p')^*}\tau,
\label{eq:singleNoiseEstimate}
\end{align}
and average over all $M$ chunks within year $a$ in order to estimate $S_{mn}^a(f)$:
\begin{align}
S_{mn}^{a}(f_p') \approx \frac{1}{M} \sum_{i=1}^M S_{mn}^{a,i}(f_p').
\label{eq:noiseEstimate}
\end{align}
This process allows us to estimate $S_{mn}^{a}(f_p')$ at the discrete frequencies $f_p' = p (\Delta f)'$ for $p\in\mathbb{Z}$.
However, in the course of analyzing the data over durations longer than $2\tau_\text{min}$, we will have access to a finer frequency spacing than $(\Delta f)'$, and so we really need access to $S_{mn}^{a}(f)$ sampled over this finer frequency range; since it is not possible to directly estimate $S_{mn}^{a}(f)$ on that finer grid with only our single data realization, our analysis interpolates the $S_{mn}^{a}(f_p')$ estimated as at \eqref{noiseEstimate} to intermediate frequencies.
Although this is approximate, there is no obvious superior approach.

Armed with the estimate \eqref{noiseEstimate} for the noise cross-power spectra $S_{mn}^{a}(f_p')$, which yields the covariances between $\tilde X^{(m)}$ within a given year, we then compute the covariances of the analysis variables $\vec X_k$ as defined at \eqref{vecXdefn}.
Suppose that $N_k^a$ is the number of data points in the subseries $X_k^{(m)}$ which were obtained in year $a$, so that $\sum_a N_k^a \equiv \aleph$ where $\aleph$ is the number of data points in the subseries $X_k^{(m)}$, then we have
\begin{align}
\langle\tilde X_k^{(m)}(f)\tilde X_k^{(n)}(f)^*\rangle_{\varepsilon=0}=\sum_a T_k^a \cdot S_{mn}^a(f),
\end{align}
where $T_k^a = N_k^a \Delta t$ is the duration of time corresponding to the number of data samples in the subseries $X_k^{(m)}$ in year $a$, assuming a measurement cadence of $\Delta t$, such that in turn we have $\sum_a T_k^a = T$, the total duration of the $k$-th coherence time (except for the situations already noted in footnote \ref{ftnt:lengthMismatch}, which are also handled appropriately here); see also \secref{analysisTimeSeriesSubsets} and the more detailed discussion in \secref{analysisFreqChoice}.

We may then write the covariance matrix for the $\vec X_k$ schematically as
\begin{widetext}
\begin{align}
\Sigma_k &\equiv\text{Cov}(\vec X_k,\vec X_k)
=\begin{pmatrix}\sum_aT_k^a\cdot S_{mn}^a(f_{A'}-\hat f_d)\\&\sum_aT_k^a\cdot S_{mn}^a(f_{A'})\\&&\sum_aT_k^a\cdot S_{mn}^a(f_{A'}+\hat f_d)\end{pmatrix},
\end{align}
\end{widetext}
for the appropriate values of $m$ and $n$ in the relevant locations;
this matrix takes a block diagonal form because we assume the DFT results at distinct frequencies are uncorrelated variables, and $\vec X_k$ is constructed in such a way that the successive blocks of entries all refer to the same frequency.\\

\subsection{Bayesian statistical analysis}
\label{sec:analysisBayesian}
In the previous two subsections, we computed the expected $\vec X_k$ under the signal hypothesis \eqref{signal}, and discussed our data-driven noise estimation procedure.
We can now synthesize these developments to construct a likelihood function for our model in terms of the expected signal vectors $\vec\mu_{ik}$ and the estimated covariance matrix $\Sigma_k$.
We can then use that likelihood function to construct the marginalized Bayesian posterior for $\varepsilon$ given the data.

\subsubsection{Likelihood function}
\label{sec:analysisBayesianLikelihood}
Up to normalization, the likelihood function for the $k$-th coherence time given the signal hypothesis \eqref{signal} with a kinetic-mixing parameter $\varepsilon$ is (we set the normalization factor for the likelihood to 1 arbitrarily)%
\footnote{\label{ftnt:why3a}%
        The normalization of the RHS of this equation (that is the normalization of $\ln\LL_k$, not $\LL_k$) cannot be chosen arbitrarily.
        It is set by demanding that $\langle \vec{X}_k \vec{X}_k^\dagger\rangle = \Sigma$, or equivalently that $\vec{Y}_k$ as defined by \eqref{YkDefn} satisfies $\langle \vec{Y}_k \vec{Y}_k^\dagger \rangle = 1$.
	} %
\begin{widetext}
\begin{align}
- \ln \LL_k\lb(\varepsilon,\bm{c}_k\big|\vec X_k\rb) = \lb(\vec X_k-\varepsilon\sum_ic_{ik}^*\vec\mu_{ik}\rb)^\dagger\Sigma_k^{-1}\lb(\vec X_k-\varepsilon\sum_ic_{ik}^*\vec\mu_{ik}\rb),
\label{eq:LLk}
\end{align}
\end{widetext}
where $\bm{c}_k$ is the 3-vector with entries $c_{ik}$ for $i=x,y,z$ [i.e., the variables defined in \eqref{combinedXkSignalExpectation} which specify the arbitrary phase and spatial orientation of the DPDM polarization vector in the inertial frame for the $k$-th coherence time].  
Assuming that all coherence times are treated as independent `experiments', the full likelihood function $\LL$ over all the available data will be taken to be the product of these $\LL_k$ over all $k$, or 
\begin{align}
\LL \equiv \prod_{k} \LL_k.
\label{eq:LL}
\end{align}

Before proceeding to utilize this likelihood to construct a Bayesian posterior on $\varepsilon$, it will be advantageous and simplifying to make some changes of notation.
Since $\Sigma_k$ is by construction a Hermitian, positive-definite matrix, it is possible to decompose it as $\Sigma_k=A_kA_k^\dagger$, for some invertible $A_k$.
If we then define
\begin{align}
\vec Y_k &=A_k^{-1}\vec X_k, \label{eq:YkDefn}\\
\vec\nu_{ik}&=A_k^{-1}\vec\mu_{ik}\quad [i=x,y,z],
\end{align}
it can be shown that \eqref{LLk} can be expressed as
\begin{align}
-\ln \LL_k\lb(\varepsilon,\vec c_k\big|\vec Y_k\rb)= \lb|\vec Y_k-\varepsilon\sum_ic_{ik}^*\vec\nu_{ik}\rb|^2.
\label{eq:Yk}
\end{align}

Now, let $N_k$ be the $15\times 3$ matrix whose first, second, and third columns take entries equal to the corresponding components of $\vec\nu_{ik}$ for $i=x,y,z$, respectively. 
\eqref{Yk} can then be rewritten as
\begin{align}
-\ln \LL_k\lb(\varepsilon,\bm{c}_k\big|\vec Y_k\rb)
&= \lb|\vec Y_k-\varepsilon N_k \bm{c_{k}}^{*} \rb|^2.
\label{eq:LLYk2}
\end{align}
The singular value decomposition of $N_k$ can be written as%
\footnote{\label{ftnt:SVDconventions}%
        Our convention is that of \citeR{Press:1992sdg}; an alternative convention would take $U_k$ to be a square unitary matrix (here, $15\times 15$), and $S_k$ to be rectangular diagonal (here, $15\times 3$).
	} %
\begin{align}
N_k=U_kS_kV_k^\dagger,
\end{align}
where $U_k$ is a $15\times3$ matrix with orthonormal columns [so that, specifically, $U_k^\dagger U_k = \id{3}$], $S_k$ is a real $3\times3$ diagonal matrix, and $V_k$ is a $3\times3$ unitary matrix. 
We also define the 3-vector variables
\begin{align}
\bm{d}_k&=V_k^\dagger\bm{c}_k^{*}, &
\bm{Z}_k&=U_k^\dagger\vec Y_k.
\end{align}
We can then re-write \eqref{LLYk2} as
\begin{align}
-\ln \LL_k\lb(\varepsilon,\bm{c}_k\big|\vec Y_k\rb)
&= \lb|\vec Y_k-\varepsilon U_kS_kV_k^\dagger \bm{c}_{k}^{*} \rb|^2\\
&= \lb| \bm{Z}_k -\varepsilon S_k \bm{d}_{k} \rb|^2 + \lb( | \vec Y_k |^{2} - \lb| \bm{Z}_k \rb|^{2} \rb),
\label{eq:likelihoodref1}
\end{align}
where to obtain the second expression we have expanded out, used $U_k^\dagger U_k = \id{3}$, added and subtracted $\lb| \bm{Z}_k \rb|^2$, and simplified.

Our immediate goal now is to use \eqref{likelihoodref1} to define a likelihood function in terms of the variables $\bm{Z}_k$, which will be central to our analysis going forward.

To this end, consider the following preparatory argument. 
The matrix $P_k=U_k U_k^\dagger$ is an orthogonal projection operator: $P_k^2=P_k=P_k^\dagger$, so let us write $\vec{Y}_k \equiv \vec{A}_k + \vec{B}_k$, where we define $\vec{A}_k \equiv P_k \vec{Y}_k$ and $\vec{B}_k \equiv (\id{15} - P_k )\vec{Y}_k$.
It follows that $|\vec{Y}_k|^2 = |\vec{A}_k|^2 + |\vec{B}_k|^2$.
Now, we also have $U_k^\dagger P_k = U_k^\dagger$ since $U_k^\dagger U_k = \id{3}$, so it also follows that $U_k^\dagger (\id{15}-P_k) = 0$, and so $U_k^\dagger \vec{B}_k = 0$.
Therefore, $\bm{Z}_k \equiv U_k^\dagger \vec{Y}_k = U_k^\dagger \vec{A}_k$; i.e., $\bm{Z}_k$ depends on $\vec{A}_k$, but is independent of $\vec{B}_k$. 
Moreover, it is easy to show that $|\bm{Z}_k|^2 = |\vec{A}_k|^2$.

Armed with that knowledge, consider now the term in $(\,\cdots)$-brackets in \eqref{likelihoodref1}.
This term is (\emph{a}) independent of the parameters $\varepsilon$ and $\bm{c}_k$, and (\emph{b}) equal to $|\vec{Y}_k|^2 - |\bm{Z}_k|^2 = |\vec{Y}_k|^2 - |\vec{A}_k|^2 = |\vec{B}_k|^2 \lb(= | \vec Y_k - U_k U_k^\dagger\vec Y_k |^2\rb)$; it thus does not depend on $\bm{Z}_k$.
These observations imply, respectively, that (\emph{a}${}^\prime$) the ${(\,\cdots)\text{-term}}$ can simply be dropped from \eqref{likelihoodref1} in constructing a likelihood for $\varepsilon$ and $\bm{c}_k$ in terms of the $\bm{Z}_k$:
\begin{align}
- \ln \LL_k\lb(\varepsilon,\bm{d}_k\big|\bm{Z}_k\rb)
\equiv \big|\bm{Z}_k-\varepsilon S_k\bm{d}_k\big|^2,
\label{eq:LLZk}
\end{align}
where we dropped an additional irrelevant constant offset;
and (\emph{b}${}^\prime$) the resulting likelihood at \eqref{LLZk} is still also interpretable in the usual way (again up to a constant offset) as the probability density for the $\bm{Z}_k$ given the parameters, which we will see is necessary for our arguments in \secref{analysisChecks}.%
\footnote{\label{ftnt:otherArgument}%
    Indeed, for the purposes of \secref{analysisDetails} only, observation (\emph{a}) would have sufficed.
    This is because \eqref{likelihoodref1} gives a likelihood, which is a function of parameters for fixed data, and we only use this in \secref{analysisDetails} to construct a marginalized posterior on $\varepsilon$ in \eqref{posterior} below.
    Any term in \eqref{likelihoodref1} that is a function of the data only and independent of the parameters gives no useful information about those parameters, and constitutes a piece of the parameter-independent normalization constant for that marginalized posterior; but the structure of that parameter-independent normalization constant is irrelevant, since the posterior gets re-normalized to a give a unit integral.
    This leads to conclusion (\emph{a}${}^\prime$).
    The reason that this argument is insufficient is that in \secref{analysisChecks} we again interpret the likelihood \eqref{LLZk} [or, really, the marginalized likelihood \eqref{likelihood}] expressed in terms of the $\bm{Z}_k$ as the probability density for $\bm{Z}_k$ to be observed given theb parameters.
    Naturally, this is usually exactly what a likelihood like \eqref{LLk} is, by definition: $\LL_k(\varepsilon,\bm{c}_k\big|\vec{X}_k) \equiv \alpha\cdot p(\vec{X}_k|\varepsilon,\bm{c}_k)$  with $\alpha$ a numerical constant.
    However, had we dropped a parameter-independent but $\bm{Z}_k$-\emph{dependent} term in \eqref{likelihoodref1}, we could no longer make the cognate identification for \eqref[s]{LLZk} or (\ref{eq:likelihood}).
    As such, it is important for the arguments in \secref{analysisChecks} that observation (\emph{b}) is true.
    } %

Physically, what has happened here is that the full 15-dimensional analysis vectors $\vec{X}_k$ that we constructed at \eqref{vecXdefn} hold much more information about the measured magnetic fields than just the pieces necessary to find the signal \eqref{signal}, as is clear from the fact that the signal expectations $\langle X_k \rangle$ are expressible as a sum over only three linearly independent vectors in the 15-dimensional space; see \eqref{combinedXkSignalExpectation}.
What the foregoing mathematical manipulations have succeeded in identifying is the relevant part of the data $\vec{X}_k$ to keep in the likelihood, $\bm{Z_k}$; and the part that is superfluous to the signal search, $\vec{B}_k$.

The cognate full likelihood combining all the coherence times (assuming they are independent `experiments') is given by
\begin{align}
\LL\lb(\varepsilon,\{ \bm{d}_k \}\big| \{ \bm{Z}_k \} \rb) \equiv \prod_k \LL_k\lb(\varepsilon,\bm{d}_k \big| \bm{Z}_k \rb).
\label{eq:LLZ}
\end{align}

\subsubsection{Marginalized likelihood function}
\label{sec:analysisBayesianMarginalizedLikelihood}
Our goal is to construct the posterior distribution for $\varepsilon$ in a Bayesian analysis framework.
In constructing this posterior however, we must account for the fact that our model for the DPDM field is such that we may not treat the $\bm{d}_k$ simply as arbitrary model parameters which can be specified by us: instead, the statistical behavior of the DPDM field that emerges from the field being the sum of a large number of interfering plane waves (see discussion in \secref{theory} and \citeR{Fedderke:2021rys}) dictates that the individual $\bm{d}_k$ should themselves be treated as random variables that must be drawn from the appropriate distribution; see, e.g., \citeR[s]{Derevianko:2016vpm,Foster:2017hbq,Centers:2019dyn,roussy2021experimental,Lisanti:2021vij}.
Within the Bayesian framework, the appropriate procedure to fold that information into the $\varepsilon$ posterior is to marginalize the combined likelihood \eqref{LLZ} over the $\bm{d}_k$.

In this subsection we discuss the appropriate likelihood that describes the distribution of the $\bm{d}_k$, and then construct the marginalized combined likelihood.

Given the discussion in \secref{theory} [in particular \eqref{Asquared}], and the definitions of the $\bm{c}_k$ in \eqrefRange{cx}{cz}, both the real and imaginary parts of $\bm{c}_k$ are independent normally distributed variables with mean zero which satisfy $\langle|\bm{c}_k|^2\rangle=1$.
Since $V_k$ is a unitary matrix, the same is true for the derived $\bm{d}_k$: $\langle|\bm{d}_k|^2\rangle=1$.
Therefore, the appropriate auxiliary likelihoods for the $\bm{d}_k$ should be taken to be
\begin{align}
\LL_k\lb(\bm{d}_k\rb)=\exp\lb(-3|\bm{d}_k|^2\rb),
\label{eq:LLdk}
\end{align}
up to an irrelevant overall normalization.%
\footnote{\label{ftnt:why3b}%
        The numerical factor of 3 in the exponent arises from assuming that the probability density function for each of the $\RE d_k^i$ and $\IM d_k^i$ for $i=1,2,3$ takes the (common) form of a zero-mean normal distribution with unknown width, $f(x) \propto \exp[-\alpha x^2]$ for $x=\RE d_k^1, \IM d_k^1, \hdots, \IM d_k^3$, and then finding the value of $\alpha$ such that the normalization condition $\langle|\bm{d}_k|^2\rangle=1$ is satisfied.
        See also footnote \ref{ftnt:why3a}.
	} %

The combined auxiliary likelihood for the $\bm{d}_k$ is thus
\begin{align}
\LL\lb( \{\bm{d}_k\}\rb)=\prod_k \LL_k\lb(\bm{d}_k\rb),
\label{eq:LLd}
\end{align}
again assuming that the polarization vectors in distinct coherence times are independent random draws.

The marginalized combined likelihood, defined as 
\begin{align}
\LL\lb( \varepsilon \big| \{ \bm{Z}_k \} \rb) &\equiv \int \bigg[ \prod_{i,k} d\!\lb( \RE d_k^i \rb) \cdot d\!\lb( \IM d_k^i \rb) \bigg] \nl\qquad
\times \LL\lb(\varepsilon,\{ \bm{d}_k \}\big|\{ \bm{Z}_k \} \rb) \LL\lb( \{\bm{d}_k\}\rb),
\label{eq:marginalLikelihood}
\end{align}
is thus given by (see \appref{marginalizedLikelihood} for a detailed derivation)
\begin{align}
\LL(\varepsilon\big|\{\bm{Z}_{k}\})\propto\prod_{i,k}\frac1{3+\varepsilon^2s_{ik}^2}\exp\lb(-\frac{3|z_{ik}|^2}{3+\varepsilon^2s_{ik}^2}\rb),
\label{eq:likelihood}
\end{align}
where $z_{ik}$ is the $i$-th component of $\bm{Z}_k$, and $s_{ik}$ is the diagonal $(i,i)$-element of the matrix $S_k$ (i.e., the $i$-th singular value of $N_k$); in both cases, $i=1,2,3$.

\subsubsection{Priors and posteriors}
\label{sec:analysisBayesianPriorsAndPosteriors}
Bayes' theorem constructs the marginalized posterior for $\varepsilon$, denoted by $p(\varepsilon|\{\bm{Z}_{k}\})$, from the marginalized likelihood given by \eqref{likelihood}, and the prior on $\varepsilon$, denoted by $p(\varepsilon)$:
\begin{align}
p(\varepsilon|\{\bm{Z}_{k}\}) & \propto\LL\lb(\varepsilon\big|\{\bm{Z}_{k}\}\rb)\cdot p(\varepsilon).
\end{align}

We must thus specify a choice of prior on $\varepsilon$.
Following \citeR{Centers:2019dyn}, we will take the (reparametrization-invariant) objective Jeffreys prior~\cite{Cowan:2018swr} for $\varepsilon$; in a similar context, this choice of prior has the additional feature that it yields limits from a Bayesian analysis which are broadly in agreement with an alternative, frequentist approach~\cite{Centers:2019dyn}.

The Jeffreys prior is defined formally in terms of the Fisher information matrix~\cite{Cowan:2018swr}; applying the formal definition, we show in \appref{jeffreysPrior} that, for our analysis, this prior takes the form
\begin{align}
p(\varepsilon)\propto\sqrt{\sum_{i,k}\frac{4\varepsilon^2s_{ik}^4}{\lb(3+\varepsilon ^2s_{ik}^2\rb)^2}}.
\label{eq:prior}
\end{align}

The posterior for $\varepsilon$ is thus
\begin{align}
p(\varepsilon|\{\bm{Z}_{k}\})
\equiv \mathcal{N}& \times \lb[\sum_{i,k}\frac{4\varepsilon^2s_{ik}^4}{\lb(3+\varepsilon ^2s_{ik}^2\rb)^2}\rb]^{\frac{1}{2}} \nonumber \\
	&\times 
	\prod_{i,k}\frac1{3+\varepsilon^2s_{ik}^2}\exp\lb(-\frac{3|z_{ik}|^2}{3+\varepsilon^2s_{ik}^2}\rb),
\label{eq:posterior}
\end{align}
where $\mathcal{N}$ is a normalization factor.
We can without loss of generality%
\footnote{\label{ftnt:fieldRedef}%
    Since the kinetic mixing term is the only term in the Lagrangian (see \citeR{Fedderke:2021rys}) that is odd in $A'$ (in the interaction basis), a trivial field definition $A' \rightarrow - A'$ maps $\varepsilon \rightarrow -\varepsilon$.
    Moreover, both the prior and posterior are even in $\varepsilon$.
    } %
restrict $\varepsilon\geq0$, and demand that $\mathcal{N}$ is set such that $\int_{0}^\infty d\varepsilon\, p(\varepsilon|\{\bm{Z}_{k}\}) =1$.%
\footnote{\label{ftnt:maxVarepsilon}%
    Note that in the \emph{kinetically mixed basis} in which $\LL \supset -\frac{1}{4} F^2 - \frac{1}{4} (F')^2 -\frac{1}{2}\epsilon F F'$,
    there is a bound on $|\epsilon|<1$ for the physical region of parameter space that is smoothly connected to $\epsilon=0$; at $\epsilon = \pm 1$, one or other of the two linear combinations $F\pm F'$ becomes a non-propagating degree of freedom (i.e., the kinetic term vanishes).
    However, in the \emph{interaction basis} we use in this work, we have $\varepsilon = \epsilon/\sqrt{1-\epsilon^2}$, so $\varepsilon$ is unbounded above in the physical region of parameter space.
    Note however that our computation of the signal \eqref{signal} is only valid for $\varepsilon \ll 1$ [i.e., we have neglected terms at $\mathcal{O}( \varepsilon^2 )$]~\cite{Fedderke:2021rys}.
    However, our posterior distributions have little support for $\varepsilon \gtrsim 1$; our results are thus self-consistent. } %
With the appropriately normalized posterior, we can then set upper bounds on $\varepsilon$; for instance, the 95\% credible upper limit (local significance) $\hat\varepsilon$ will be given by solving
\begin{align}
\int_0^{\hat\varepsilon}d\varepsilon~p(\varepsilon|\{\bm{Z}_{k}\})=0.95.
\label{eq:95credibleUpper}
\end{align}

\subsection{Coherence time approximation and choice of frequencies}
\label{sec:analysisFreqChoice}
Our analysis to this point has been constructed to obtain a bound at a single frequency $f_{A'}$, but we have not yet specified how this frequency was chosen.
One would ideally simply scan this frequency in some range.
However, computationally we require the use of an FFT which can only evaluate bounds at discrete frequencies, and the specific set of frequencies depends on the duration of data we choose to analyze coherently.
We discuss these issues further in this subsection.

In this subsection, we will be more precise about our usage of the term `coherence time'; cf.~footnote \ref{ftnt:imprecisionExcused}.
Let $T$ refer to the length of the data subseries analyzed in a coherent fashion under the analysis procedures thus far outlined in \secref{analysisDetails}, and denote by
\begin{align}
    T_\text{coh}(f_{A'})=\min\lb[ \frac{1}{ f_{A'}v_{\textsc{dm}}^2},\, T_\text{tot} \rb]; \quad v_{\textsc{dm}} \sim 10^{-3},
    \label{eq:TcohDefn}
\end{align}
the shorter of the actual DPDM signal coherence time, and the total duration $T_{\text{tot}}=48\,\text{yrs}$ of SuperMAG data available for analysis.

For the following reasons, it has been implicit in our analysis construction to this point that $T\approx T_\text{coh}(f_{A'})$:
\indent (1) beginning in \secref{analysisTimeSeriesSubsets} we split the SuperMAG data into subseries of length $T$, and assumed for the purposes of constructing the likelihood in \secref{analysisBayesian} that the polarization of the signal was constant for the duration of each [i.e., that the $\bm{d}_k$ for each $k$ were a single random  draw from the expected distribution, \eqref{LLdk}]. 
For that to be a consistent assumption, each signal subseries must not extend beyond a single actual DPDM coherence time, because the polarization wanders randomly on the latter timescale: $T\lesssim T_\text{coh}(f_{A'})$; and  \\
\indent (2) in \secref{analysisBayesian}, we explicitly constructed the joint likelihood over all the duration-$T$ intervals by treating the polarization in each interval as having a distinct random orientation uncorrelated with that in the neighboring intervals, if any (i.e., each of the $\bm{d}_k$ is a distinct random draw from the distribution defined by \eqref{LLdk}, uncorrelated with previous or future draws). 
Because we additionally analyse our data in \emph{contiguous} blocks of duration $T$, this is only a good assumption if the duration of each block is long enough that, at the start of the subsequent block, the polarization has effectively been randomised by phase drifts.
That latter time period is however again simply the definition of the DPDM coherence time, so we have: $T\gtrsim T_\text{coh}(f_{A'})$.%
\footnote{\label{ftnt:spacing}%
    We note that if our analysis were over non-contiguous blocks, the criterion is simply that the start times of consecutive subseries are spaced by at least $T_\text{coh}(f_{A'})$, and not that the subseries' durations themselves must be at least $T_\text{coh}(f_{A'})$ long. 
    However, we have mandated that there is no gap between consecutive subseries in order to maximize data usage, so the criterion for our analysis as constructed is indeed as stated in the text.} %

Since both $T\lesssim T_\text{coh}(f_{A'})$ and $T\gtrsim T_\text{coh}(f_{A'})$ are needed, we have to take $T\approx T_\text{coh}(f_{A'})$ for consistency.%
\footnote{\label{ftnt:specialCaseTDefn}%
   Technically, without additional assumptions, only point (1) holds for the case where $1/(f_{A'} v_{\textsc{dm}}^2) \gtrsim T_{\text{tot}}$, such that $T_{\text{coh}}(f_{A'}) = T_{\text{tot}}$ per \eqref{TcohDefn}.
   In that case, for point (2) to hold, the necessary assumption is simply that we wish to analyze \emph{all} the available data to maximize the statistical power of the search; we implement this assumption by analyzing the whole dataset coherently with $T=T_{\text{tot}}$ in this case.
   Since we define $T_{\text{coh}}(f)$ such that $\max\lb[T_{\text{coh}}(f)\rb] = T_{\text{tot}}$, this case is automatically handled correctly by the discussion in the main text.
   } %
Temporarily setting aside that $T_\text{coh}(f_{A'})$ itself is only known approximately (because the DM velocity profile is not known exactly, and the entire concept of the DPDM coherence time arises precisely because of the velocity dispersion in the interfering constituent plane waves), there is a computational problem in assuming $T=T_\text{coh}(f_{A'})$ exactly: it makes the duration of the signal to be analyzed an explicit function of the frequency at which the analysis is being performed [at least for all frequencies such that $1/(f_{A'} v_{\textsc{dm}}^2) < T_{\text{tot}}$].
That would preclude the application, necessary here owing to the multi-gigabyte volume of the full SuperMAG dataset, of the FFT algorithm to process the analysis of many frequencies simultaneously, because the FFT relies on having a fixed-duration signal to transform.
Having to either perform the slow DFT for each frequency, or indeed having to re-perform the FFT for every frequency of interest, would be computationally prohibitive given available resources.

At a high-level, our solution to this computational issue seeks a trade-off between implementing the condition $T \approx T_\text{coh}$,%
\footnote{\label{ftnt:approxSolnTcoh}%
    Note that, in some sense, this solution exploits the existing inherent uncertainty in the exact length of the DPDM coherence time to our advantage: we are not honor-bound to implement an inefficient analysis strategy to obtain $T=T_\text{coh}(f_{A'})$ exactly when the latter is only approximately known.
    We have some freedom to instead design an efficient analysis strategy that obtains $T \approx T_\text{coh}(f_{A'})$.
    } %
and still being able to exploit the computational speedup of the FFT algorithm to process many frequencies simultaneously.
We will break up our full range of frequencies of interest into some small number of narrower frequency ranges, indexed by $n$, and perform our analysis for the frequencies in each such range $n$ using a fixed duration of the data subseries, $T=T_n$.
We chose $T_n$ to be independent of frequency within each individual frequency range $n$ (but varying for different $n$) such that $T_n \approx T_\text{coh}$ is satisfied, up to some fixed tolerance (which we take to be 3\%), for all of the frequencies that lie within range $n$.
Since $T_n$ is fixed within each frequency range $n$, we can then utilize the FFT algorithm to obtain results simultaneously for the whole set of FFT frequencies that lie within range $n$, $\{f_{ni}\}$ [where the $f_{ni}$ are multiples of $1/T_n$; see below].
As we will not actually require too many different $n$ (indeed we need only 56 such ranges) to cover our whole frequency range of interest in this way, this strategy allows us to construct results under the assumption that $T \approx T_\text{coh}(f)$ up to some known controllable tolerance, while also exploiting the FFT computational speedup, at only the modest cost of having to run the FFT algorithm 56 times.

More precisely, we choose $T_n$ to be%
\footnote{\label{ftnt:FFTefficiency}%
        Let $M$ be the number of data points corresponding to the time interval $T_n$. 
		For computational purposes in the FFT, it is preferable for all the prime factors of $M$ to be small.
		Therefore, we actually choose $T_n$ such that $M$ is the integer within 10 of the estimate implied by \eqref{Tn} that has the minimal largest prime factor.
	} %
\begin{align}
T_n\approx\frac{T_\text{tot}}{(1+q)^{2n}},
\label{eq:Tn}
\end{align}
where $q = 0.03$ fixes the aforementioned 3\% tolerance, and the consecutive set of integers $n=0,\ldots,55$ is chosen such that $T_n$ ranges from $T_\text{tot}$ down to approximately $10^6$~minutes [i.e., the coherence time, assuming $v_{\textsc{dm}} = 10^{-3}$, corresponding to the sampling rate of the SuperMAG data, which is $1/(1\,\text{min})$].

The set of frequencies $\{f_{ni}\}$ that we will consider to fall within range $n$ will be $f_{ni}=i/T_n$ for $i= i_{n}^{\text{min}},\ldots,i_{n}^{\text{max}}$.
For $n\neq0$, we take $i_{n}^{\text{min}} = \lfloor 10^6/(1+q) \rfloor$, while for the  special case $n=0$ (i.e., when the entire dataset is treated coherently), we have $i_{n}^{\text{min}} = 0$.
The value of $i_{n}^{\text{max}}$ is defined iteratively starting with the highest-frequency set, and for each $n$ is taken to be the largest integer such that $\max\lb[ \{ f_{ni} \} \rb] < \min\lb[\{ f_{n+1,i} \} \rb]$; this means that, approximately, $i_{n}^{\text{max}} \approx \lfloor 10^6(1+q) \rfloor$.%
\footnote{\label{ftnt:discardFreqs}%
    While the FFT algorithm run on each duration-$T_n$ dataset will also generally yield results for frequencies $f_{ni}$ with $i$ outside the range shown in the text, for those frequencies the coherence time approximation tolerance will not be satisfied.
    We thus discard those results and utilize a different $n$ for the construction of the results at the corresponding frequencies.
    } %
Because $i_{n}^{\text{max}}$ must be iteratively constructed beginning with the set of frequencies containing the highest frequency, we must specify the highest frequency in the construction: this is taken to be one DFT frequency bin below the SuperMAG sampling frequency (i.e., twice the Nyquist frequency), such that $i_{55}^{\text{max}} = T_{55}/(1\,\text{min}) - 1$.

Defined this way, the individual sets of frequencies $\{f_{ni}\}$ cover non-overlapping ranges of frequencies.
Moreover, for frequencies $f_{ni}$ such that $T_{\text{coh}}(f_{ni}) < T_{\text{tot}}$, we have
\begin{align}
\lb|\frac{T_n-T_\text{coh}(f_{ni})}{T_\text{coh}(f_{ni})}\rb|
&= \lb|\frac{i}{10^6} - 1 \rb|\\
& \leq \lb|\frac{i_{n}^{\text{max}}}{10^6} - 1 \rb|\\
& \leq q.
\end{align}
Meanwhile it is easy to show that frequencies with $T_{\text{coh}}(f_{ni}) = T_{\text{tot}}$ necessarily have $n=0$ and so $T_\text{coh}(f_{ni})=T_n$ trivially by \eqref{Tn}.
Therefore, we indeed approximate the coherence time (or total data duration) to within a fixed percentage for every frequency $f_{ni}$ within every range $n$.

\begin{figure}[t]
\includegraphics[width=0.95\columnwidth]{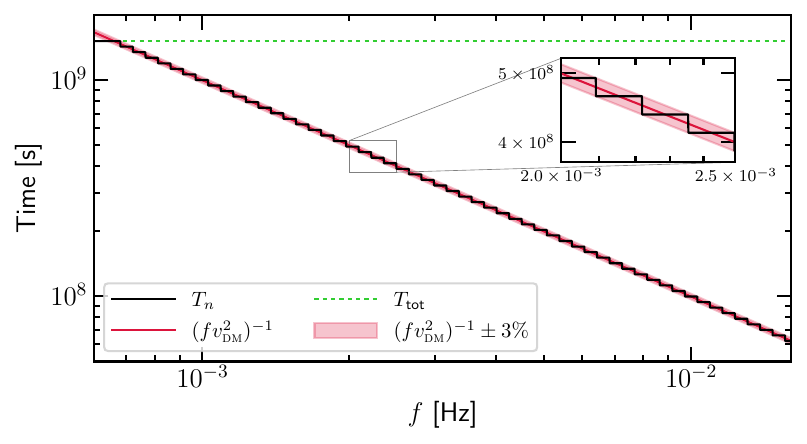}
\caption{\label{fig:TcohApprox}%
 	Graphical representation of the scheme used to approximate $T_{\text{coh}}(f)$ as outlined in \secref{analysisFreqChoice}.
	The solid black line shows the values of $T_n$ employed in the analysis, as a function of frequency.
	The solid red line shows the approximate coherence time, $(f v_{\textsc{dm}}^2)^{-1}$, for the DPDM signal, while the red shaded band gives a 3\% tolerance around this approximate value; note importantly that for all $f$ such that $(f v_{\textsc{dm}}^2)^{-1}<T_{\text{tot}}$, $T_n$ lies within this tolerance of $(f v_{\textsc{dm}}^2)^{-1}$.
	The dotted green line shows the total duration of the dataset: note that once $(f v_{\textsc{dm}}^2)^{-1}>T_{\text{tot}}$, $T_n = T_{\text{tot}}$ is assumed (i.e., the data are all analyzed in a single coherent block).
	}
\end{figure}

We show a graphical representation of this approximation scheme in \figref{TcohApprox}.

\subsection{Correction for finite signal width}
\label{sec:analysisLimitDegradation}
Our analysis construction to this point has operated on the assumption that the dark-photon signal is exactly monochromatic within a coherence time, so that the entirety of the signal power appears in a single DFT frequency bin; in order words, we assumed exact coherence of the signal for a full coherence time $T_{\text{coh}}\sim (f_{A'}v_{\textsc{dm}}^2)^{-1}$.
Indeed, in the preceding subsection we matched the DFT frequency bin width to the coherence time to within 3\% over the entire frequency range we consider in order to preserve this property.%
\footnote{\label{ftnt:correciotnBreak}%
    For $f_{A'}\lesssim 6.4\times 10^{-4}\Hz$, preservation of this property begins to fail because the signal coherence time begins to exceed (3\% more than) the available data duration; see left edge of \figref{TcohApprox}.
    As the frequency is decreased further, the coherence time further exceeds the data duration, and the signal therefore begins to become much narrower than a single DFT bin.
    This concentration of signal power in a single bin more closely matches our analysis construction, which implies that the degradation factor should be smoothly tapered to 1 (i.e., no degradation) for $f_{A'}\ll 6.4\times 10^{-4}\Hz$.
    However, the lowest frequency that we explicitly present limits for in this work (see \figref{resultsExclusion}) is $f_{A'} = 6\times 10^{-4}\,$Hz; at this frequency, the coherence time is still within 10\% of the available data duration, and so we find it unnecessary to implement any such tapering of the degradation factor in presenting our results.
} %
However, this is a slight oversimplification of the situation: the DPDM signal is actually $\sigma_f \sim 1/T_{\text{coh}}$ wide in frequency space, so while we do expect the majority of the signal power to appear in the DFT bin corresponding to $f_{A'}$, some power will appear in the neighboring (few) bins as well.
Given the way our analysis is constructed, if we did not account for this, we would set limits that are too aggressive.

While a more sophisticated approach to this analysis would have considered this spreading of the signal power from the beginning of the analysis construction, we leave such an improvement to future work.
Instead, precisely because we have matched the DFT bin-width to the signal width to high accuracy over the whole frequency range, we can apply a simple frequency-independent rescaling factor to approximately correct for this in a \emph{post hoc} fashion.
That is, we proceed by simply degrading the limit on the kinetic mixing parameter from \eqref{95credibleUpper}:
\begin{align}
\hat{\varepsilon} \rightarrow \hat{\varepsilon}' \equiv \zeta \cdot \hat{\varepsilon},
\label{eq:degradeFactor}
\end{align}
with $\zeta>1$. 
In all of our results to follow, we present the degraded limits $\hat{\varepsilon}'$ unless otherwise explicitly noted.

It remains to estimate $\zeta$.
For the purposes of this estimate, we ignore the vectorial nature of the DPDM field, and focus only on the frequency-space spreading (this is equivalent to considering each vectorial component of the DPDM field independently); see \secref{analysisChecksValidation} for the cognate signal injection that accounts for the vectorial nature of the signal and that validates this approach.

Assume that the DPDM field (component) is a sum of a large number of plane waves (see, e.g., Sec.~II.A of \citeR{Fedderke:2021rys}):
\begin{align}
    A'(t) &\sim \frac{\sqrt{2\rho_{\textsc{dm}}}}{m_{A'}} \frac{1}{\sqrt{2\mathcal{N}+1}}\nl \times \sum_{n=-\mathcal{N}}^{\mathcal{N}} \exp\lb[ i m_{A'} t \sqrt{ 1 + \bm{v}_n^2 } + i \phi_n \rb],
    \label{eq:Azeta1}
\end{align}
where $\bm{v}_n$ are samples drawn from an assumed galactic-frame Maxwellian velocity distribution with an rms speed $v_0 \sim 10^{-3}$, and $\phi_n$ is a random phase.
Now analyse this field in the Fourier domain via the DFT, and let the largest single-bin value of the resulting PSD be $S_*$.%
\footnote{\label{ftnt:shiftMax}%
    Note that the average frequency of the DPDM field constructed in this fashion is $2\pi f_{A'} = m_{A'} \langle \sqrt{ 1 + \bm{v}^2} \rangle \approx m_{A'} ( 1 + v_0^2 / 2)$, which differs from the standard relationship we have employed to this point, $2\pi f_{A'} = m_{A'}$, by a frequency shift of order (half) the DFT bin spacing. 
    The correct way to interpret this shift is to identify the physical mean frequency of the DPDM field with the frequency at which we set limits, and consider this shift to be a modification to the relationship between $f_{A'}$ and $m_{A'}$; the correction is however negligible everywhere except for the frequency--mass identification.
    This procedure guarantees that the highest-power DFT bin is (except for fluctuations) the bin centered on $f_{A'}$.
    See the discussion in \secref{analysisChecksValidation}.
    } %
Let the sum over the whole PSD of this field (i.e., the total signal power) be $\Sigma_S$.
Because our analysis is very roughly constructed so as to compare single-bin signal power to single-bin noise power, and because the PSD of the resulting magnetic field signal \eqref{signal} arising from the DPDM is proportional to $\varepsilon^2$, the appropriate degradation factor would then be $\zeta \sim \sqrt{ \Sigma_S / S_* }$.
Averaging over 100 distinct random realizations of DPDM fields of this type constructed from sums of $2\mathcal{N}+1=5001$ plane waves, we estimate numerically that the degradation factor would be $\zeta \sim 1.24(1)$.
We therefore set $\zeta = 1.25$ as the degradation factor.

This 25\% degradation factor is comparable to the uncertainties on many of our noise properties (see \appref{noiseValidation}), and so proceeding in this way is consistent with the overall accuracy of our full analysis.

We discuss this degradation factor further in \secref{analysisChecksValidation}, where we verify that an injected signal would be correctly reconstructed.

\subsection{Results}
\label{sec:analysisResults}

\begin{figure*}[p]
\includegraphics[width=\textwidth]{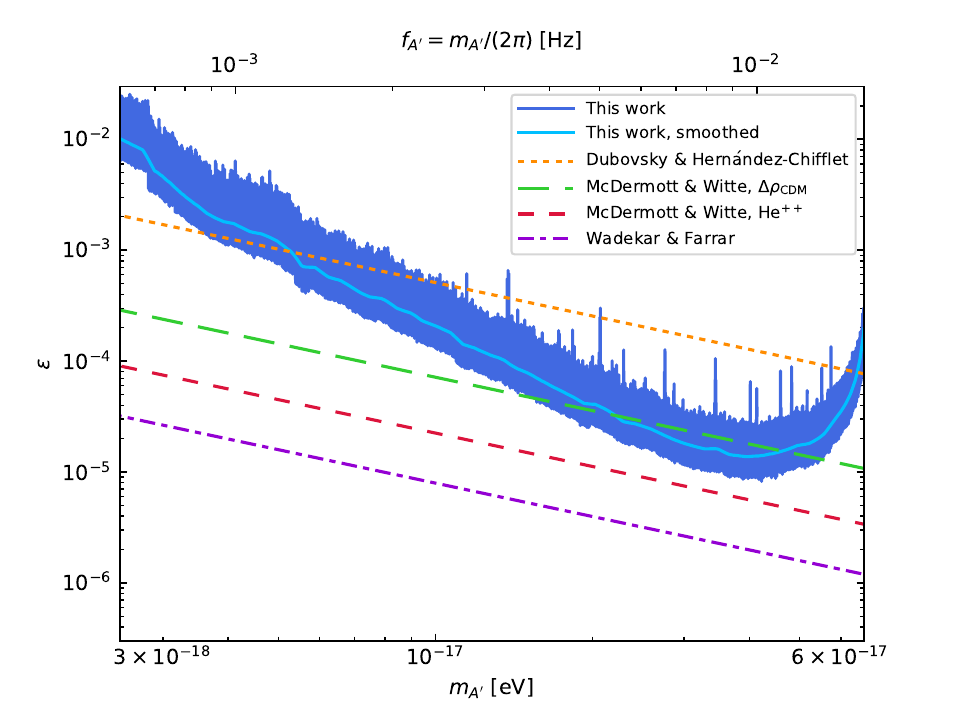}
\caption{\label{fig:resultsExclusion}%
        Exclusion bounds on the kinetic mixing parameter $\varepsilon$ of the dark-photon dark matter as a function of the dark-matter mass $m_{A'}$ (frequency $f_{A'}$).
        The darker blue line (appearing as a band owing to frequency-to-frequency limit fluctuations) shows our 95\% credible upper limit (local significance) [cf.~\eqref[s]{95credibleUpper} and (\ref{eq:degradeFactor})] on the kinetic mixing parameter $\varepsilon$ as a function of the dark-photon dark-matter mass (corresponding Compton frequency noted on upper axis), assuming that the dark photon constitutes all of the local dark-matter density, $\rho_{\textsc{dm}} = 0.3\,\text{GeV/cm}^3$, but taking into account the stochastic variations expected for classical-field dark matter (see, e.g., \citeR[s]{Derevianko:2016vpm,Foster:2017hbq,Centers:2019dyn,roussy2021experimental,Lisanti:2021vij}).
        These limits include the effect of the 25\% degradation factor discussed in \secref{analysisLimitDegradation}.
        To guide the eye and give a sense of the relative density of stronger vs.~weaker limits in narrow frequency bands, we also show as the lighter blue solid line the sliding average of the limit taken over the neighboring $\pm 25000$ frequencies. 
        Various sharply rising narrow spikes in our limits provide a variety of potential candidate signals; we examine these in detail in \secref{analysisChecks}, where we conclude that none constitute robust evidence for a real signal.
        The various other lines show a variety of existing astrophysical limits arising from dark-photon dark-matter heating of gas in a number of astrophysical environments: the ionized interstellar medium in the Milky Way (dotted orange)~\cite{Dubovsky:2015cca}; the intergalactic medium around helium reionization (short-dashed red, labeled `$\text{He}^{++}$')~\cite{McDermott:2019lch}; and gas in the Leo T dwarf galaxy (dot-dashed purple)~\cite{Wadekar:2019xnf}.
        A DM-depletion limit from nonresonant dark-photon--photon conversion~\cite{McDermott:2019lch} is also shown (long-dashed green, labeled `$\Delta \rho_{\textsc{cdm}}$').
        Our limits are complementary to these existing bounds as they arise from terrestrial experimental data (analogous to `direct detection'), and are thus subject to completely different sources of systematic uncertainty as compared to the other bounds shown (which are analogous to `indirect detection').
	} 
\end{figure*}

We now have in place all the relevant tools to set upper bounds on the kinetic mixing parameter $\varepsilon$; the results of our analysis are shown as the blue band in \figref{resultsExclusion} as 95\% credible upper limits (local significance) on $\varepsilon$ as a function of the dark-photon mass $m_{A'}$.
These constraints are complementary to the existing astrophysical limits also presented in \figref{resultsExclusion}, which arise from dark-photon heating of gas in the interstellar medium in the Milky Way (dotted orange)~\cite{Dubovsky:2015cca}, the intergalactic medium around the time of helium reionization (short-dashed red)~\cite{McDermott:2019lch}, and in the Leo T dwarf galaxy (dash-dotted purple)~\cite{Wadekar:2019xnf};%
\footnote{\label{WadekarPrivate}%
    Per \citeR{WadekarPrivate2021}, the limits in \citeR{Wadekar:2019xnf} are mildly weaker than those in the \texttt{arXiv v1} and \texttt{v2} preprints of that paper, on account of \emph{inter alia} updated gas metallicity measurements of Leo T that were incorporated in the published version of \citeR{Wadekar:2019xnf}.
    }%
\up{,}%
%
\footnote{\label{ftnt:otherLimits}%
 Limits similar to those in \citeR{Wadekar:2019xnf} appear also in \citeR{Bhoonah:2018gjb}.
 The latter reference also gives a stronger preliminary bound based on a gas cloud of anomalously low (and disputed) temperature, which we do not show here; see the discussion in \citeR[s]{Bhoonah:2018wmw,Farrar:2019qrv,Bhoonah:2018gjb} and our comments in \citeR{Fedderke:2021rys}.
} %
or from dark-photon--photon conversion depleting dark matter (long-dashed green)~\cite{McDermott:2019lch}.
Future bounds based on 21\,cm observations are expected to become strong in this mass range~\cite{Kovetz:2018zes}, but we do not show current limits or projections here in light of the EDGES global 21\,cm anomaly~\cite{Bowman:2018sdg}.

We note the existence of some clearly visible sharp peaks in the exclusion bounds shown in \figref{resultsExclusion}.
These peaks are among some 30 na\"ive signal candidates that we identify in the data.
We discuss these na\"ive signal candidates in detail in the next section, where we conclude that none of them clearly survive robustness checks on their consistency with the expected signal properties.
Because we dismiss all these signal candidates, we can reasonably also plot in \figref{resultsExclusion} as a guide to the eye a smoothed version of our limits (light blue solid line) that is obtained by averaging our limits over the $\pm 25000$ neighboring frequency bins.

Finally, we note that for $m_{A'} \lesssim 3\times 10^{-17}\eV$ our limits scale with increasing mass $m_{A'}$ faster than $m_{A'}^{-1}$ [cf.~\eqref{signal}], which is a manifestation of the decreasing noise at higher frequency in the SuperMAG magnetic field data.
This trend is only terminated at the upper end of the plotted mass range owing to decreased sensitivity around and above the Nyquist frequency ($f_{\text{Nyq}}$ corresponds to a mass $m_{A'}\sim 3\times 10^{-17}\eV$).
This observation is highly encouraging because, assuming that this noise trend were to be maintained in the higher-cadence (i.e., one-second) SuperMAG data currently being released, it is plausible that this search method would allow access to kinetic-mixing parameter space at higher frequency that is currently unconstrained by astrophysical observations; we have not however undertaken any analysis of the higher-cadence data to check whether this is the case---this is deferred to future work.
In any event, we note that even our existing limits are subject to completely independent systematics as compared to the existing astrophysical constraints in the mass range where we have presented limits in \figref{resultsExclusion}, and are already therefore complementary.

\section{Candidates, validation, and rejection}
\label{sec:analysisChecks}
Our results in \secref{analysisResults} are phrased as exclusions (upper bounds) on the value of the parameter $\varepsilon$ as a function of the DPDM mass.
However, our analysis would be incomplete without also considering whether there are any indicia in the data of nonzero DPDM signals; indeed, this is the logical prior step.
Even casual observation of \figref{resultsExclusion} indicates the existence of multiple `peaks' in the limits: frequencies at which the bounds are considerably weaker than those at neighboring frequencies.
While this behavior may be the product of statistical fluctuations or other real non-DM-signal features in the data, it would also be expected behavior for the upper limit on $\varepsilon$ to fluctuate upward for any specific frequency or frequencies at which a real DPDM signal(s) were present in the data (with a `true' value of $\varepsilon$ somewhat smaller than the value of the limit we have placed on $\varepsilon$ for the respective frequencies).

In this section, we therefore complete our analysis by first developing in \secref{analysisChecksCandidates} formal criteria for identifying what we call `na\"ive signal candidates': we find 30 such candidates in the data.
Then, in \secref{analysisChecksTests}, we develop and apply tests to analyze whether or not the identified candidates are fully consistent with the expected properties of a DPDM signal, \eqref{signal}: on the basis of the discussion there, we conclude that none of the 30 na\"ive signal candidates can be considered robust evidence for a real DPDM signal in the SuperMAG data.
Finally, we show in \secref{analysisChecksValidation} that a mock signal of the form \eqref{signal} injected into the (partially processed) SuperMAG data would be identified by our analysis, and not rejected by the robustness tests we develop, which validates our analysis approach.
We offer discussion in \secref{analysisChecksDiscussion}.

\subsection{\texorpdfstring{Na\"ive signal candidates}{Naive signal candidates}}
\label{sec:analysisChecksCandidates}
We begin by developing the formal criterion for declaring a feature in the data to be a na\"ive signal candidate.

As a first step, we must determine the statistical significance of any such feature under the zero-signal, null hypothesis: $\varepsilon=0$.
We work with the quantities $z_{ik}$, defined in \secref{analysisBayesianMarginalizedLikelihood}, whose (marginalized) likelihood for a given $\varepsilon$ is given by \eqref{likelihood}.

\begin{figure*}[t]
\includegraphics[width=\textwidth]{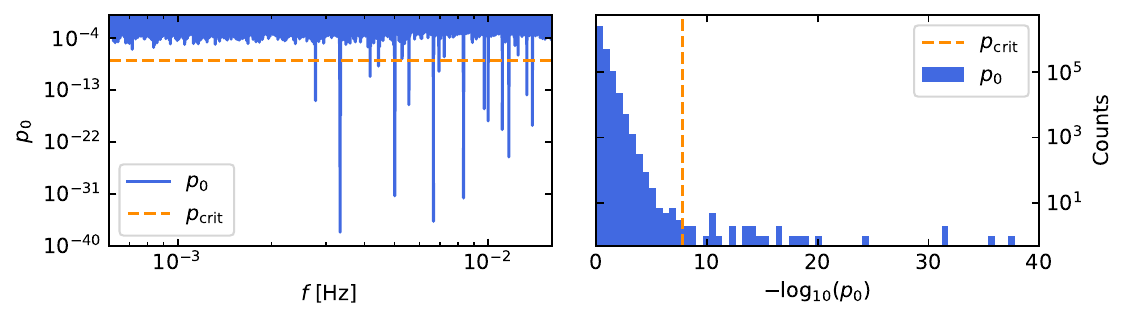}
\caption{\label{fig:pvals}%
        \textsc{Left panel:}
    	The (local) $p_0$-values (blue) for every frequency bin analyzed in our range of interest, computed per \eqref{pvalueCandidate}.
    	The threshold value for declaring a na\"ive candidate signal at 95\% confidence, $p_\text{crit}\approx 1.6\times10^{-8}$ [see \eqref{threshold}], is shown by the horizontal dashed orange line; this threshold takes into account a trials factor (i.e., the significance is global).
    	We identify 30 na\"ive signal candidates in the range $6\times10^{-4}\Hz<f_{A'}<\lb[ (1\,\text{min})^{-1} - 6\times10^{-4}\rb]\Hz$ (see text).
    	We investigate these na\"ive signal candidates in \secref{analysisChecksTests}.
    	\textsc{Right panel:} Histogram of all of the $p_0$ values that are shown in the left panel, showing the expected smoothly falling distribution with the identified signal candidates as clear outliers above the threshold $p_{\text{crit}}$, which is shown by the vertical dashed orange line.
	} 
\end{figure*}

From \eqref{likelihood}, we can see that under the null hypothesis (zero-signal; $\varepsilon=0$), the real and imaginary parts of the quantities $z_{ik}$ are described by a zero-mean multivariate normal distribution.
Therefore to determine which frequencies in our original analysis are inconsistent with the absence of a signal, we can simply compute the $\chi^2$ statistic%
\footnote{\label{ftnt:why4}%
    Arguments similar to those advanced in footnotes \ref{ftnt:why3a} and \ref{ftnt:why3b} dictate the inclusion of the numerical factor of 2 here.%
    } %
\begin{align}
Q_0=2\sum_{i,k}|z_{ik}|^2,
\end{align}
and its $p$-value
\begin{align}
p_0=1-F_{\chi^2(6K_0)}[Q_0],
\label{eq:pvalueCandidate}
\end{align}
where $F_{\chi^2(\nu)}$ is the cumulative distribution function (CDF) of the $\chi^2$-distribution with $\nu$ degrees of freedom, and $K_0$ is the number of duration-$T$ subseries into which we partitioned the full time series (i.e., the number of values through which the index $k$ ranges; see \secref{analysisTimeSeriesSubsets}).
The relevant number of degrees of freedom is $6K_0$ since each of the three Cartesian components of $\bm{Z}_k$ has independent real and imaginary parts; note that we have made no correction for parameter estimation to the na\"ive number of degrees of freedom.
\figref{pvals} shows the value of $p_0$ for each frequency in the range of interest, as well as a histogram of all $p_0$ values taken over the whole range of frequencies we analyse.

We consider a data feature at a certain frequency to be a na\"ive signal candidate (with 95$\%$ confidence) if $p_0$ is below the threshold $p_\text{crit}$ defined by
\begin{align}
(1-p_\text{crit})^{N_f}=0.95,
\label{eq:threshold}
\end{align}
where $N_f$ is the number of frequencies we consider in the range of interest; i.e., this threshold takes into account a trials factor, so the 95\% confidence is global.
For the frequency range of interest, $6\times10^{-4}\Hz<f_{A'}<\lb[ (1\text{ min})^{-1} - 6\times10^{-4}\rb]\Hz$,%
\footnote{\label{ftnt:explainRange}%
    The lower limit here is the lower end of our frequency range of interest.
    The upper limit is its reflection across the Nyquist frequency.
    } %
we have $N_f\sim 3.3\times10^6$, and the corresponding threshold is $p_\text{crit} = 1.6\times10^{-8}$; this threshold is shown as the horizontal (respectively, vertical) orange line in the left (right) panel of \figref{pvals}.

Using the criterion \eqref{threshold}, we identify 30 na\"ive DPDM candidates in our frequency range of interest; see \tabref{resampling}.
All of these na\"ive candidates are sufficiently narrow (i.e., they are only one-to-two frequency bins wide, consistent with $\Delta f/f_{A'}\sim v_{\textsc{dm}}^2 \sim 10^{-6}$) to be a potential DPDM signal.
However, \emph{we cannot yet declare any of these na\"ive candidates to be a DPDM signal}, as we must first verify that they pass further checks on their spatial and/or temporal characteristics.

\begin{table*}[t]
    \centering
    \begin{ruledtabular}
    \begin{tabular}{r|lll||llll|l||llll|l||l}
    No. & $f$ [mHz] &   $p_0$   &   $\sigma(p_0)$ &   $p_1$   &   $p_2$   &   $p_3$   &   $p_4$   &   $p_\text{time}$ &   $p_5$   &   $p_6$   &   $p_7$   &   $p_8$   &   $p_\text{geo}$  &   $p_\text{full}$ \\\hline
1   &   $2.777776$   &   $1.5\times10^{-15}$   &   $5.7$   &   $0.96$   &   $0.94$   &   $0.10$   &   $0.00$   &   $7.5\times10^{-4}$   &   $0.90$   &   $0.39$   &   $0.12$   &   $1.00$   &   $9.1\times10^{-4}$   &   $6.9\times10^{-6}$   \\
2   &   $2.777779$   &   $2.6\times10^{-11}$   &   $3.8$   &   $0.94$   &   $0.90$   &   $0.21$   &   $0.00$   &   $0.015$   &   $0.99$   &   $0.06$   &   $0.18$   &   $0.99$   &   $2.7\times10^{-3}$   &   $3.2\times10^{-4}$   \\
3   &   $3.321727$   &   $7.0\times10^{-12}$   &   $4.1$   &   $0.97$   &   $0.62$   &   $0.01$   &   $0.01$   &   $2.3\times10^{-3}$   &   $0.59$   &   $0.84$   &   $0.49$   &   $0.18$   &   $0.79$   &   $0.026$   \\
4   &   $3.333330$   &   $1.2\times10^{-9}$   &   $2.7$   &   $0.98$   &   $0.78$   &   $0.11$   &   $0.00$   &   $3.8\times10^{-3}$   &   $0.03$   &   $0.09$   &   $0.01$   &   $0.44$   &   $0.023$   &   $6.7\times10^{-4}$   \\
5   &   $3.333333$   &   $2.6\times10^{-38}$   &   $11.7$   &   $0.97$   &   $0.57$   &   $0.01$   &   $0.00$   &   $4.2\times10^{-5}$   &   $0.07$   &   $0.00$   &   $0.00$   &   $0.79$   &   $3.3\times10^{-5}$   &   $1.9\times10^{-8}$   \\
6   &   $3.344939$   &   $3.9\times10^{-18}$   &   $6.7$   &   $0.99$   &   $0.78$   &   $0.46$   &   $0.01$   &   $0.034$   &   $0.61$   &   $0.51$   &   $0.36$   &   $0.28$   &   $0.97$   &   $0.27$   \\
7   &   $4.166664$   &   $2.4\times10^{-11}$   &   $3.8$   &   $0.39$   &   $0.35$   &   $0.57$   &   $0.05$   &   $0.62$   &   $1.00$   &   $0.01$   &   $0.17$   &   $0.38$   &   $1.1\times10^{-3}$   &   $9.7\times10^{-3}$   \\
8   &   $4.432841$   &   $1.5\times10^{-9}$   &   $2.6$   &   $1.00$   &   $0.84$   &   $0.56$   &   $0.14$   &   $0.051$   &   $0.47$   &   $0.73$   &   $0.81$   &   $1.00$   &   $0.091$   &   $0.023$   \\
9   &   $4.999999$   &   $4.9\times10^{-32}$   &   $10.4$  &   $0.99$   &   $0.71$   &   $0.08$   &   $0.00$   &   $9.0\times10^{-5}$   &   $0.10$   &   $0.12$   &   $0.00$   &   $0.98$   &   $3.6\times10^{-3}$   &   $3.7\times10^{-6}$   \\
10   &   $5.011607$   &   $2.2\times10^{-17}$   &   $6.4$   &   $0.97$   &   $0.83$   &   $0.75$   &   $0.00$   &   $3.1\times10^{-3}$   &   $0.94$   &   $0.41$   &   $0.09$   &   $0.81$   &   $0.25$   &   $6.6\times10^{-3}$   \\
11   &   $5.555552$   &   $5.2\times10^{-9}$   &   $2.1$   &   $0.17$   &   $0.98$   &   $0.27$   &   $0.01$   &   $0.031$   &   $0.25$   &   $0.88$   &   $0.09$   &   $1.00$   &   $3.5\times10^{-4}$   &   $1.1\times10^{-4}$   \\
12   &   $5.555557$   &   $2.9\times10^{-16}$   &   $6.0$   &   $1.00$   &   $0.59$   &   $0.35$   &   $0.00$   &   $1.0\times10^{-3}$   &   $1.00$   &   $0.10$   &   $0.88$   &   $1.00$   &   $3.7\times10^{-4}$   &   $4.0\times10^{-6}$   \\
13   &   $6.655058$   &   $2.0\times10^{-11}$   &   $3.8$   &   $1.00$   &   $0.44$   &   $0.90$   &   $0.49$   &   $0.058$   &   $0.74$   &   $0.96$   &   $0.59$   &   $0.98$   &   $0.11$   &   $0.031$   \\
14   &   $6.666665$   &   $1.9\times10^{-36}$   &   $11.3$   &   $1.00$   &   $0.94$   &   $0.10$   &   $0.00$   &   $2.0\times10^{-5}$   &   $0.99$   &   $0.02$   &   $0.02$   &   $0.86$   &   $2.8\times10^{-3}$   &   $7.1\times10^{-7}$   \\
15   &   $6.944447$   &   $1.8\times10^{-10}$   &   $3.2$   &   $1.00$   &   $0.34$   &   $0.60$   &   $0.03$   &   $1.7\times10^{-3}$   &   $0.98$   &   $0.98$   &   $0.86$   &   $1.00$   &   $8.4\times10^{-7}$   &   $2.3\times10^{-8}$   \\
16   &   $8.321724$   &   $5.2\times10^{-14}$   &   $5.1$   &   $0.98$   &   $1.00$   &   $0.07$   &   $0.11$   &   $1.1\times10^{-4}$   &   $0.88$   &   $0.94$   &   $1.00$   &   $0.94$   &   $5.6\times10^{-3}$   &   $6.6\times10^{-6}$   \\
17   &   $8.333325$   &   $4.5\times10^{-11}$   &   $3.6$   &   $0.82$   &   $0.61$   &   $0.39$   &   $0.03$   &   $0.36$   &   $0.43$   &   $0.54$   &   $0.17$   &   $0.90$   &   $0.67$   &   $0.55$   \\
18   &   $8.333333$   &   $2.1\times10^{-32}$   &   $10.5$   &   $0.99$   &   $0.59$   &   $0.83$   &   $0.00$   &   $1.2\times10^{-3}$   &   $1.00$   &   $0.42$   &   $0.79$   &   $1.00$   &   $6.8\times10^{-5}$   &   $9.5\times10^{-7}$   \\
19   &   $8.333342$   &   $4.5\times10^{-11}$   &   $3.6$   &   $0.82$   &   $0.61$   &   $0.39$   &   $0.03$   &   $0.36$   &   $0.43$   &   $0.54$   &   $0.17$   &   $0.90$   &   $0.67$   &   $0.55$   \\
20   &   $8.344942$   &   $5.2\times10^{-14}$   &   $5.1$   &   $0.98$   &   $1.00$   &   $0.07$   &   $0.11$   &   $1.1\times10^{-4}$   &   $0.88$   &   $0.94$   &   $1.00$   &   $0.94$   &   $5.6\times10^{-3}$   &   $6.6\times10^{-6}$   \\
21   &   $9.722222$   &   $5.7\times10^{-17}$   &   $6.2$   &   $1.00$   &   $0.14$   &   $0.64$   &   $0.00$   &   $2.5\times10^{-5}$   &   $0.99$   &   $0.95$   &   $0.94$   &   $1.00$   &   $1.0\times10^{-6}$   &   $4.1\times10^{-10}$   \\
22   &   $9.999996$   &   $4.6\times10^{-19}$   &   $7.0$   &   $1.00$   &   $0.99$   &   $0.40$   &   $0.03$   &   $2.2\times10^{-4}$   &   $1.00$   &   $0.01$   &   $0.11$   &   $0.44$   &   $7.5\times10^{-3}$   &   $1.7\times10^{-5}$   \\
23   &   $10.00001$   &   $1.3\times10^{-14}$   &   $5.3$   &   $1.00$   &   $0.88$   &   $0.45$   &   $0.16$   &   $5.0\times10^{-5}$   &   $0.64$   &   $0.47$   &   $0.04$   &   $0.95$   &   $0.25$   &   $2.2\times10^{-4}$   \\
24   &   $10.01160$   &   $5.7\times10^{-9}$   &   $2.1$   &   $0.98$   &   $0.47$   &   $0.98$   &   $0.80$   &   $0.062$   &   $0.86$   &   $0.42$   &   $0.95$   &   $0.98$   &   $0.076$   &   $0.023$   \\
25   &   $11.11111$   &   $1.4\times10^{-20}$   &   $7.5$   &   $1.00$   &   $0.36$   &   $0.07$   &   $0.00$   &   $7.5\times10^{-6}$   &   $1.00$   &   $0.39$   &   $0.81$   &   $1.00$   &   $1.2\times10^{-4}$   &   $1.3\times10^{-8}$   \\
26   &   $11.65507$   &   $9.3\times10^{-13}$   &   $4.5$   &   $1.00$   &   $1.00$   &   $0.99$   &   $0.56$   &   $3.7\times10^{-7}$   &   $1.00$   &   $0.61$   &   $0.97$   &   $1.00$   &   $5.4\times10^{-7}$   &   $3.7\times10^{-12}$   \\
27   &   $11.66667$   &   $2.6\times10^{-25}$   &   $8.8$   &   $1.00$   &   $0.89$   &   $0.22$   &   $0.00$   &   $1.1\times10^{-4}$   &   $0.01$   &   $0.19$   &   $0.04$   &   $1.00$   &   $2.0\times10^{-3}$   &   $2.4\times10^{-6}$   \\
28   &   $11.67827$   &   $3.9\times10^{-13}$   &   $4.7$   &   $0.92$   &   $0.99$   &   $0.97$   &   $0.10$   &   $10.0\times10^{-3}$   &   $0.99$   &   $0.88$   &   $0.31$   &   $0.48$   &   $0.12$   &   $7.5\times10^{-3}$   \\
29   &   $13.33334$   &   $1.5\times10^{-14}$   &   $5.3$   &   $1.00$   &   $0.98$   &   $0.71$   &   $0.00$   &   $3.2\times10^{-4}$   &   $0.31$   &   $0.34$   &   $0.43$   &   $0.99$   &   $0.34$   &   $1.5\times10^{-3}$   \\
30   &   $13.88889$   &   $7.8\times10^{-20}$   &   $7.2$   &   $0.98$   &   $0.98$   &   $0.26$   &   $0.00$   &   $4.1\times10^{-6}$   &   $0.67$   &   $0.02$   &   $0.71$   &   $1.00$   &   $3.6\times10^{-7}$   &   $2.5\times10^{-11}$
    \end{tabular}
    \end{ruledtabular}
    \caption{%
    Na\"ive signal candidates and their various associated $p$-values.
    $p_0$ indicates the local $p$-value significance of the candidate in the original analysis under the null hypothesis of zero signal (see also \secref{analysisChecksCandidates} and \figref{pvals}); also shown is the equivalent one-sided, global Gaussian-standard-deviation significance of the signal candidate [\eqref{sigmap0}].
    The values $p_1,\ldots,p_4$ indicate the significances of the candidate in the individual subdivisions of the data on the basis of the temporal splitting outlined in \secref{analysisChecksTests}; likewise, $p_5,\ldots,p_8$ indicate the significances of the candidate in the individual geographical subdivisions discussed in the same location in the text.
    For $j=1,\ldots,8$, the $p_j$ are weighted by the posterior on $\varepsilon$ from the analysis of the full dataset; see \eqref{pjWeighted}. 
    Moreover, $p_j$ for $j=1,\ldots,8$ that lie either close to 0 or close to 1 indicate disagreement with the original analysis.
    $p_\text{time}$ (respectively, $p_\text{geo}$) is the combined significance of the temporal (geographical) split of the data in the resampling analysis.
    $p_\text{full}$ is the overall combined significance in the resampling analyses.
    We discuss these results at length in the text.
    }
    \label{tab:resampling}
\end{table*}

\subsection{Tests of candidates}
\label{sec:analysisChecksTests}
Having identified 30 na\"ive signal candidates (see \tabref{resampling}) on the basis of the criterion specified in \secref{analysisChecksCandidates}, it is important to develop tests for the robustness of those na\"ive candidates.
Candidates which fail these robustness checks can be rejected as being inconsistent with a real DPDM signal.

In particular, although the na\"ive signal candidates are indeed real magnetic field signal patterns in the data that have strong overlap with the VSH field pattern expected from a DPDM signal, it must also be the case that a DPDM signal should be present in all stations, and at all times.
Therefore, re-analysis of any subdivision of the SuperMAG dataset should, for a real signal, yield parameter determinations consistent with the analysis of the full dataset.
If, on the other hand, analysis of subdivisions of the full dataset yield inconsistent DPDM parameter determinations, that is strong evidence that the na\"ive signal candidate is not a real DM candidate; instead, this could be evidence for strong in-band local (in time or space) fluctuations driving the identification of the na\"ive candidate.

In this subsection, we develop these resampling tests and apply them to the 30 signal candidates we have identified. 

Our first task is to identify the appropriate subdivisions of the full dataset to analyze independently for this resampling analysis.
We perform two types of divisions of the data: by geographical location of the station, and by epoch of data acquisition.
For the geographical division, we randomly partition the stations into four disjoint subsets.%
\footnote{\label{ftnt:tooLittleDataEarly}%
    Because we require each disjoint subset to have at least three active stations at all times in order to construct five linearly independent time series $X^{(n)}$ $[j=1,\ldots,5]$ for each subdivision, we are forced for this part of the analysis only to ignore the first six years of available data: insufficiently many stations are continuously operative during this time. 
    Thus, for the geographical subdivisions in this resampling analysis, we analyse only the last 42 years of data available.} %
For the temporal division, we divide the full dataset into four consecutive, non-overlapping temporal intervals, each with a duration of 12 years.

For each subset of data, we re-performed the analysis described in \secref{analysisDetails} for the frequencies corresponding to each of the 30 na\"ive signal candidates only.
In performing this analysis, our choice of the time interval $T$ described at length in \secref{analysisFreqChoice} is however inherited from the analysis of the full dataset instead of being re-adjusted on the basis of the subdivided data.
Note however that this does mean that the number of duration-$T$ intervals, $K_j$, in each subdivision $j$ of the data may be different from the number of such intervals in the analysis of the full dataset, $K_0$.

For the analysis of each division of the data ${j=1,\ldots,8}$ (with ${j=1,\ldots,4}$ being the temporal splits and ${j=5,\ldots,8}$ being the geographical splits), this re-analysis procedure yields the quantities $z_{ik,j}$ and $s_{ik,j}$; these have the same definitions as the $z_{ik}$ and $s_{ik}$ in \secref{analysisDetails}, with the exception that they are evaluated only on the $j$-th subdivision of the data.
For a kinetic mixing parameter of size $\varepsilon$, \eqref{likelihood} indicates that the relevant test statistic to compute on the data in each subdivision $j$ is the following $\chi^2$ statistic:
\begin{align}
Q_j(\varepsilon)=\sum_{i,k}\frac{6|z_{ik,j}|^2}{3+\varepsilon^2s_{ik,j}^2}.
\label{eq:QjResampled}
\end{align}
Given a fixed value of $\varepsilon$, it is easy to compute the $p$-value for this test statistic: it is simply found from the CDF of the $\chi^2$ distribution as%
\footnote{\label{ftnt:no1}%
    In contrast to the definition of $p_0$ in \eqref{pvalueCandidate}, we define $p_j(\varepsilon)$ without the `$1-$'; in light of the usage of $p_j(\varepsilon)$ in \eqref[s]{pjWeighted} and (\ref{eq:Qfull}), a definition of $p_j(\varepsilon)$ with the same `$1-'$ as in \eqref{pvalueCandidate} would be equivalent.
    } %
\begin{align}
p_j(\varepsilon) = F_{\chi^2(6K_j)}[Q_j(\varepsilon)],
\end{align}
where $F_{\chi^2(6K_j)}$ is again the CDF for a $\chi^2$ distribution with $6K_j$ degrees of freedom [see below \eqref{pvalueCandidate}]; we have once again made no correction for parameter estimation to the na\"ive number of degrees of freedom.

As we wish to test for consistency of the subdivisions of the data with analysis of the full dataset, we then weight these $p$-values by the posterior on $\varepsilon$ computed in the analysis of the full dataset, $p(\varepsilon|\{z_{ik}\})$ as defined at \eqref{posterior}.
That is, we assign to subdivision $j$ the $p$-value
\begin{align}
p_j=\int d\varepsilon~p(\varepsilon|\{z_{ik}\})\cdot p_j(\varepsilon).
\label{eq:pjWeighted}
\end{align}

We utilize Fisher's method~\cite{Fisher:1958iqe,10.2307/2681650,10.2307/2529826,KOST2002183} to combine the $p_j$ into a single $p$-value for the resampling checks. 
That is, we construct a joint test statistic over all the data subsamples by summing of the logarithms of the $p_j$, and then compare it to a $\chi^2$-distribution.
Our tests here must however be two-tailed as both large and small $p_j$ indicate disagreement with the original analysis.
Therefore, the appropriate quantity whose logarithm must be summed is the minimum of $p_j$ and $1-p_j$, rather than just $p_j$.  
In other words, we will combine these $p_j$ into the single test statistic
\begin{align}
Q_\text{full}=-2\sum_{j=1}^{8}\ln\left(2\cdot\min\{p_j,1-p_j\}\right),
\label{eq:Qfull}
\end{align}
which has the corresponding joint $p$-value for $n=8$ tests:
\begin{align}
p_\text{full}=1-F_{\chi^2(2n)}(Q_\text{full}) \qquad [n=8],
\label{eq:pfull}
\end{align}
with the relevant number of degrees of freedom in Fisher's method being $2n$~\cite{Fisher:1958iqe,10.2307/2681650,10.2307/2529826,KOST2002183}.
We additionally examine the $p$-values that arise from considering the temporal-only and geographical-only splits, $p_{\text{time}}$ and $p_{\text{geo}}$, respectively; these are defined in the same fashion as $p_{\text{full}}$, but with the appropriate restriction on the sum over the data subsets $j$ in \eqref{Qfull} in each case, and the number of $\chi^2$ degrees of freedom in \eqref{pfull} set to be $2n=8$ and not $2n=16$ since there are only $n=4$ tests in each case.
All of these $p$-values for each of the 30 na\"ive signal candidates are shown in \tabref{resampling}, along with the $p_0$ value for each candidate [see \eqref{pvalueCandidate}], and its equivalent one-sided global Gaussian-standard-deviation significance, 
\begin{align} 
    \sigma(p_0) \equiv \sqrt{2}\, \text{erfc}^{-1}\lb[ 2\lb( 1 - \lb( 1-p_0\rb)^{N_{f}} \rb)\rb],
\label{eq:sigmap0}
\end{align}
where $N_{f} \approx 3.3\times 10^6$; see discussion below \eqref{threshold}.

Before proceeding to the interpretation of these $p$-values, we note an important caveat.
The procedure for the construction of $p_{\text{full}}$ (but not for $p_{\text{time}}$ or $p_{\text{geo}}$) given above is approximate in the following sense.
Each data subset $j=1,\ldots,4$ from the temporal split of the data is fully independent of each of the other distinct data subsets from the temporal split, and the same is true of the subsets $j=5,\ldots,8$ from the geographic split.
However, the temporal-split data subsets $j=1,\ldots,4$ are not fully independent of the geographical-split data subsets $j=5,\ldots,8$.
For instance, the $j=1$ data subset consists of data at all available stations for a fixed temporal duration, while the $j=5$ data subset consists of data at all available times over a fixed subset of stations; this obviously implies that data from some station that is considered in the set of stations in $j=5$ and that were taken during a time that is included in the temporal duration considered for $j=1$, will appear in both the $j=1,5$ data subsets.
If the data were exactly evenly distributed among data subsets, this would mean that there is an approximately $1/16$-th overlap between each temporal-spatial pair of data subsets; of course, the temporal split yields unevenly distributed data subsets (see \figref{station_count}), so this is only a rough estimate.
Naturally, the overlap of data in the different data subsets correlates their $p_j$ values in some complicated way, although given the relatively modest data overlap that we estimate, we do not expect this effect to be large.
A detailed accounting for this effect would, given the complexity of the SuperMAG dataset and the nontrivial analysis manipulations we apply to it, however likely require significant Monte Carlo modeling, which we consider to be beyond the scope of this work.

The preceding caveat notwithstanding, we proceed as follows.
In the absence of the correlation in the data subsets, we would reject any na\"ive signal candidate with $p_\text{full}<0.05$ (95\% confidence); because we expect that the correlation effect is mild, we will continue to automatically reject any na\"ive signal candidate with $p_\text{full}<0.01\ll 0.05$.
This automatically rejects 23 of the 30 candidates in \tabref{resampling}.
For the remaining seven candidates that are not automatically rejected, we find that four have $0.01 < p_{\text{full}} < 0.05$ (candidates numbered 3, 8, 13, and 24 in \tabref{resampling}).
We consider these to be in strong tension with the robustness checks; they are likely ruled out, but we cannot give a definitive statement absent a quantitative accounting for the correlation caveat noted above.
The remaining three candidates have $p_{\text{full}}>0.05$ (candidates numbered 6, 17, and 19 in \tabref{resampling}); these are not formally excluded, but they too have issues that put them in tension with an interpretation as a robust DPDM signal.

We discuss all seven of the candidates that are not automatically excluded in detail:

\paragraphdash{Candidate 3}
This candidate is in strong tension with the combined temporal and geographical robustness test ($p_\text{full}=0.026$), but cannot be ruled out definitively on these grounds.
It is however severely excluded by the temporal robustness test alone: $p_\text{time}=2.3\times10^{-3}$.
We reject this candidate.

\paragraphdash{Candidates 8, 13, and 24}
All three of these candidates are also in strong tension with the combined temporal and geographical robustness test, but again cannot be definitively ruled out: $0.01<p_\text{full}<0.05$.
Examining the temporal and geographical robustness tests separately, none of these candidates is clearly rejected by either alone: $0.05<p_\text{time},p_\text{geo}<0.11$.
We note however that these candidates only exhibit moderate global significances: $\sigma(p_0)=2.6$, $3.8$, and $2.1$, respectively.
We consider these to be weak candidates which are heavily disfavored by our resampling analysis.

\paragraphdash{Candidate 6}
This candidate is perhaps the most interesting of the seven.
It has a strong global significance, with $\sigma(p_0)=6.7$, and is in good agreement both with the combined robustness test ($p_\text{full}=0.27$), and the geographical robustness test ($p_\text{geo}=0.97$).
It is however in tension (but not definitively) with the temporal robustness test: $p_\text{time}=0.034$.
Because of this tension, we do not believe that there is clear or robust evidence that this candidate constitutes a signal; however it may warrant follow-up, for instance, in an analysis of the higher-cadence SuperMAG data.

\paragraphdash{Candidates 17 and 19}
While these candidates are in good agreement with the combined robustness test, as well as the separate temporal and geographical tests, we note that they lie exactly one DFT frequency bin above and below the Nyquist frequency for a one-minute data cadence.
Although not dispositive, this constitutes good reason to believe that these peaks are systematic artefacts of the analysis.
Future analysis of the higher-cadence data would settle this point definitively, as the Nyquist frequency for the one-second data would differ.

We conclude that none of the 30 na\"ive signal candidates should be considered to be a robust dark-photon dark-matter candidate signal, but that definitive exclusion of candidate 6 would require follow-up work.

Having already reached this conclusion, we have terminated our checks at this point; however, additional checks would have been possible had any candidate survived without demonstrating tension or inconsistency with the existing checks.
For instance, we could also check whether the signal shows evidence of \emph{enough} variation in the DPDM field from one coherence time to the next. 
The idea here would be to first perform parameter estimation to fix $\varepsilon$ and $m_{A'}$ for the signal using the full dataset.
Then, we would examine subsets of the data with duration equal to the signal coherence time, and for each such data subset, we would estimate: (1) the DPDM polarization state, and (2) the DPDM field amplitude.
We would then test whether the polarization state indeed randomizes on coherence-time timescales (i.e., whether the polarization state is too highly correlated between coherence times), and whether the DPDM field amplitude shows the appropriate distribution for a true DPDM field consisting of a sum of plane-waves with random phases; see, e.g., \eqref{cxSum} below.
Of course, because of the Earth's rotation, a re-orientation of the DPDM field polarization state in the galactic frame not only causes a change to the vectorial orientation of the magnetic field to be expected at each station, but it also shifts the relative power of the signal between the three frequencies $f=f_{A'},f_{A'}\pm f_d$; both effects are of course captured in \eqref{signal}, but one could additionally specifically test to ensure that the variation in the extracted DPDM polarization state is occasioned by the expected shift in the relative amount of the signal power at each of these frequencies, holding $m_{A'}$ fixed.

\subsection{Validation of analysis pipeline}
\label{sec:analysisChecksValidation}
In order to confirm that our analysis would identify and not reject a true DPDM signal in the data, we injected a mock physical DPDM signal with the expected spatial and temporal coherence properties into the SuperMAG dataset, and tested whether our analysis pipeline would correctly reconstruct it.
We confirm that (a) the extracted limit curve on the kinetic mixing parameter $\varepsilon$ is unaffected at frequencies other than in the expected vicinity of the injected signal frequency, (b) the limit on $\varepsilon$ at the injection frequency is close to the value used to construct the mock signal (up to expected deviations owing to idealized assumptions in our pipeline), and (c) our resampling analysis that tests for signal robustness on the basis of spatial and temporal coherence of the signal correctly does \emph{not} reject the injected mock signal candidates that appear at the injection frequency.

As we discussed in \citeR{Fedderke:2021rys}, the dark photon vector potential $\bm{A}'$ can be considered to be a sum over a collection of plane waves with random vectorial orientations (although see also \citeR{Caputo:2021eaa} for discussion on this point) and phase offsets, and frequencies centered on $f_{A'} = m_{A'}/(2\pi)$, but with a linewidth $\sim f_{A'} v_0^2$, where $v_0\sim 10^{-3}$ is the DM velocity dispersion in the Milky Way (MW).
Because the exact lineshape of this quasi-monochromatic DPDM signal in the Fourier domain is actually not known as it depends on the detailed properties of the collection of plane waves being summed over (e.g., the exact velocity distribution in the MW), we choose to construct a straw-man artificial injected DPDM signal as follows.

Let $c_j(t)\; [j=x,y,z]$ denote the time-dependent orientation of the DPDM signal as defined in \eqrefRange{cx}{cz} but with $A'_m=A'_m(t)$ now time-dependent.
Write $\bm{c}(t)$ as the 3-vector those $j$-th component is $c_j(t)$.
We construct $\bm{c}(t)$ as follows:
\begin{align}
    \bm{c}(t) &= \frac{1}{\sqrt{2\mathcal{N}+1}} \sum_{n=-\mathcal{N}}^{\mathcal{N}} \frac{ \bm{\hat{a}}_n + i \bm{\hat{b}}_n}{\sqrt{2}} e^{i\phi_n} \nl 
    \qquad\qquad\qquad\quad\times\exp\lb[ 2\pi i \xi f_{A'} t \sqrt{ 1 + \bm{v}_n^2 }  \rb],
    \label{eq:cxSum}
\end{align}
where $\mathcal{N} \gg1$ is a large number of random plane waves (see \tabref{injectedParameters}); $\bm{\hat{a}}_n$ and $\bm{\hat{b}}_n$ are randomly selected real unit 3-vectors specifying the polarization state of each such plane wave; $\phi_n$ is a random phase; $f_{A'}$ is the physical average frequency of the signal; $\bm{v}_n$ is a 3-velocity selected from an isotropic Gaussian velocity distribution with a single-velocity-component standard deviation $\sigma_v = v_{\textsc{dm}}/\sqrt{3}$ such that the root-mean-square DM speed is $v_{\textsc{dm}} = 10^{-3}$ (note: we neglect the difference between the Earth/Solar System and galactic frames for this purposes of this construction); and $\xi \approx 1 - v_{\textsc{dm}}^2/2 + \mathcal{O}(v_{\textsc{dm}}^4)$ is an otherwise-negligible correction factor to the mass--frequency relationship chosen such that $\langle \xi \sqrt{ 1 + \bm{v}^2 } \rangle \equiv 1$, with the average taken over the velocity distribution (i.e., we inject a signal with a slightly corrected mass $m_{A'} = 2\pi \xi f_{A'}$ such that $f_{A'}$ is the average signal frequency; this is done for the technical reason that we wish to inject the signal on average at an exact DFT frequency, and not shifted upward by order of a DFT bin half-width---see also footnote \ref{ftnt:shiftMax}).
These $\bm{c}(t)$ have the appropriate normalization (i.e., $\langle |\bm{c}(t)|^2 \rangle = 1$ with the average being an ensemble average over random phases, orientations, and velocities; or, equivalently, a temporal average over times much longer than the coherence time) and coherence properties for a realistic DPDM signal (technically, in the galactic rest-frame), assuming a non-truncated Standard Halo Model (see, e.g., \citeR{Evans:2018bqy}) for the DM velocity distribution.
Note that the injected signal \eqref{cxSum} has a frequency-space `width' of order $\sigma_f \equiv f_{A'}v_{\textsc{dm}}^2$.

\begin{table}[t]
\caption{\label{tab:injectedParameters}%
    Parameters used to generate the mock signal injected into the SuperMAG data for the purposes of analysis pipeline validation; parameters are defined in detail in the text (see \secref{analysisChecksValidation}).
    Also shown for comparison is the value of $\Delta f$, the DFT frequency spacing corresponding to the value of the approximate coherence time used in the vicinity of $f_*$ (see discussion in \secref{analysisFreqChoice}).
    }
    \begin{ruledtabular}
    \begin{tabular}{lll}
    Parameter & Symbol & Value \\ \hline
    Central frequency & $f_*$ & $7.5\times10^{-3}\,$Hz\\
    Signal `width' ($\equiv f_* v_{\textsc{dm}}^2$)     & $\sigma_f$ & $7.5\times10^{-9}\,$Hz \\
    Kinetic mixing parameter   & $\varepsilon_*$ & $10^{-3}$ \\
    Number of plane waves summed & $2\mathcal{N}+1$ & 1001  \\ \hline
    DFT frequency spacing at $f=f_*$ & $\Delta f$ & $7.45\times10^{-9}\,$Hz
    \end{tabular}
    \end{ruledtabular}
\end{table}

As in \secref{analysisSignal}, we can express the effect of a DPDM signal on the data time series $X^{(n)}(t_j)$ [as defined in \eqrefRange{X1i}{X5i}] using various combinations of the time series $H^{(n)}(t_j)$ [as defined in \eqrefRange{H1i}{H7i}], weighted by the time-dependent orientations $c_i(t_j)$.
Specifically, we inject our mock DPDM signal into the partially-processed SuperMAG time-series dataset by making the following substitution:
\begin{align}
    \mathcal{X}_j \rightarrow \mathcal{X}_j& - \pi\varepsilon_*f_*R\sqrt{2\rho_\textsc{dm}}\, \RE \lb[\sum_{i=x,y,z} c_i(t_j) \mathcal{Y}^{(i)}_{j}\rb],
    \label{eq:injection}
\end{align}
where 
\begin{align}
    \mathcal{Y}^{(x)}_j &\equiv \begin{pmatrix}
                    \left( H^{(1)}-\mathbf{1}\right)\cos(2\pi f_dt_j)-H^{(2)}\sin(2\pi f_dt_j)\\
                    -H^{(2)}\cos(2\pi f_dt_j)-H^{(1)}\sin(2\pi f_dt_j)\\
                    H^{(5)}\sin(2\pi f_dt_j)-H^{(4)}\cos(2\pi f_dt_j)\\
                    H^{(5)}\cos(2\pi f_dt_j)+\left(H^{(4)}-H^{(3)}\right)\sin(2\pi f_dt_j)\\
                    H^{(7)}\sin(2\pi f_dt_j)-H^{(6)}\cos(2\pi f_dt_j)
                \end{pmatrix},\\
    \mathcal{Y}^{(y)}_j &\equiv \begin{pmatrix}
                    H^{(2)}\cos(2\pi f_dt_j)-\left(\mathbf 1-H^{(1)}\right)\sin(2\pi f_dt_j)\\
                    H^{(1)}\cos(2\pi f_dt_j)-H^{(2)}\sin(2\pi f_dt_j)\\
                    -H^{(5)}\cos(2\pi f_dt_j)-H^{(4)}\sin(2\pi f_dt_j)\\
                    \left(H^{(3)}-H^{(4)}\right)\cos(2\pi f_dt_j)+H^{(5)}\sin(2\pi f_dt_j)\\
                    -H^{(7)}\cos(2\pi f_dt_j)-H^{(6)}\sin(2\pi f_dt_j)
                \end{pmatrix},
\end{align}
\begin{align}
    \mathcal{Y}^{(z)}_j &\equiv \begin{pmatrix}
                    0\\
                    0\\
                    H^{(6)}\\
                    -H^{(7)}\\
                    \mathbf 1-H^{(3)}
                \end{pmatrix},\\
    \mathcal{X}_j &\equiv \left.\begin{pmatrix}X^{(1)},X^{(2)},X^{(3)},X^{(4)},X^{(5)}\end{pmatrix}^{\textsc{t}}\right|_{t=t_j},
\end{align}
with ${}^{\textsc{t}}$ denoting transpose; see \eqref{XexpzApprox} and \eqrefRange{XexpxApprox}{XexpyApprox} for similar expressions for the Fourier transform of the signal.
We refer to the dataset consisting of the SuperMAG data plus this injected mock signal as the `mock-signal dataset'; the injected signal parameters are shown in \tabref{injectedParameters} along with other relevant data.

\begin{figure}[t]
\includegraphics[width=\columnwidth]{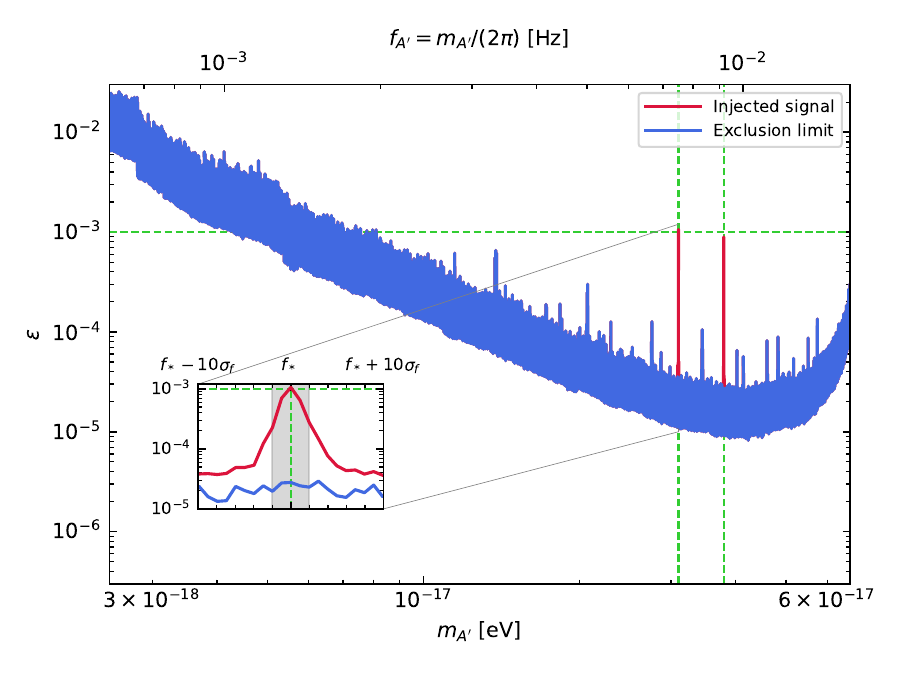}
\caption{\label{fig:resultsInjected}%
        The results, presented in red as exclusion bounds on the kinetic mixing parameter $\varepsilon$ as a function of the dark-photon mass $m_{A'}$, of our analysis pipeline as applied to mock-signal dataset consisting of an injected signal added to the SuperMAG data, as described in \secref{analysisChecksValidation}.
        Overlaid in blue, and mostly obscuring the mock-signal exclusion bounds, are the exclusion bounds obtained using the unadulterated SuperMAG dataset; see \figref{resultsExclusion}.
        In both cases, these limits account for the degradation factor $\zeta = 1.25$ discussed in \secref{analysisLimitDegradation}.
        The presence of strong peaks in the exclusion bounds from the mock-signal dataset at the injected signal frequency $f_*=7.5\times10^{-3}\,\text{Hz}$, and at its reflection across the Nyquist frequency $(1\,\text{min})^{-1}-f_*=9.2\times10^{-3}\,\text{Hz}$ (both indicated by the vertical dashed green lines) show that the analysis reconstructs the injected signal at the appropriate frequencies, but that the exclusion bounds are otherwise largely unaffected away from the vicinity of these peaks.
        The inset axes show an enlarged view of the peak at $f=f_*$, with the grey shaded region corresponding to $f_*\pm2\sigma_f$; the width of the region of the weakened exclusion bounds near $f=f_*$ is clearly consistent with our choice of $\sigma_f=10^{-6}f_* \sim \Delta f$.
        Moreover, the exclusion bounds at the frequencies where these strong peaks appear are close to the injected signal size $\varepsilon_*=10^{-3}$ (indicated by the horizontal dashed green line), as expected.
	} 
\end{figure}

We rerun the full analysis detailed in \secref[s]{analysisDetails}, \ref{sec:analysisChecksCandidates}, and \ref{sec:analysisChecksTests} on the mock-signal dataset.
The exclusion bounds resulting from this analysis are shown in red in \figref{resultsInjected}, with our exclusion bound from \figref{resultsExclusion} derived from the SuperMAG dataset superimposed in blue.
As expected, the exclusion bounds derived from the mock-signal dataset and the unadulterated SuperMAG dataset agree well everywhere except in the vicinity of the injected frequency,%
\footnote{\label{ftnt:highernoiselevel}%
    The attentive reader will note from the inset axes in \figref{resultsInjected} that even at some significant distance (as compared to $\sigma_f$) from the strong signal peaks in the mock-signal dataset (but still in its vicinity), the exclusion bounds from the two datasets do not agree exactly. 
    This is because the noise level that is used to set the exclusion bounds at any one frequency is estimated in a data-driven way using all the data, including any injected signal, in some nearby frequency range; see \secref{analysisNoise} and \appref{noiseValidation}.
    As a result, the bounds away from the region $f = f_* \pm (\text{few})\cdot\sigma_f$ that are derived using the mock-signal dataset are weakened slightly as compared to the unadulterated dataset, owing to the higher noise estimate in the former.
    A refinement of this approach would be possible, but we note from \figref{resultsInjected} that even an injected signal with a very high signal-to-noise ratio (SNR) weakens the exclusion bounds in the vicinity of that signal by a factor of only $\mathcal{O}(1)$.
    See also footnote \ref{ftnt:noiseNoSignal}.
    } %
and at its reflection across the Nyquist frequency $(1\,\text{min})^{-1}-f_*$; there are strong narrowband peaks in the exclusion bounds at those two frequencies, which are diagnostic of a signal.

We note that the exclusion bound which is set at the injected frequency actually appears slightly weaker than the value of the kinetic mixing parameter used to construct the signal; cf.~\tabref{injectedParameters} and \figref{resultsInjected}.
This is of course the expected behavior for an upper limit in the presence of a signal and additive noise.
Note however that the degradation factor of $\zeta = 1.25$ that was discussed in \secref{analysisLimitDegradation} proves crucial in obtaining this result; indeed, one can clearly see in the inset plot of \figref{resultsInjected} that injected signal power has appeared in more than one DFT bin, in line with our arguments in \secref{analysisLimitDegradation}.
The limit set at the Nyquist reflection is slightly stronger than the injected signal value of $\varepsilon_*$, but this is traceable to the fact that the Nyquist reflection of the signal appears not at an exact DFT frequency bin, and so more signal power leaks to neighboring bins than is the case for the signal at the injected frequency; we have not degraded the limits so significantly as to account for this effect, as our limits are to be formally interpreted as correct only for injected signals that lie at exact DFT frequencies.

As a separate check, we also verified that when injecting into our analysis pipeline an idealized exactly monochromatic signal (at an exact DFT frequency, so that spectral leakage effects can be ignored; see, e.g., \citeR[s]{Harris:1978wdg,Fedderke:2020yfy}) with $\bm{A}'$ aligned to the Earth's rotational axis, the limit on $\varepsilon$ correctly appears slightly weaker than the injected signal size, \emph{even without the $\zeta = 1.25$ degradation factor applied}.

Taken together, these checks confirm that the analysis operates correctly for the exactly monochromatic signal it is formally constructed to search for without any \emph{post hoc} correction, and that the amplitude of the \emph{post hoc} correction applied in \secref{analysisLimitDegradation} for a real signal with the appropriate frequency-space width is of an appropriate magnitude.

We also point out that our validation here has been phrased entirely in terms of exclusion bounds; formally, we should perform parameter estimation on the mock signal to estimate the recovered frequency and kinetic mixing parameter.
However, since the injected signal is assumed to have a large SNR in our tests, that level of detail in this validation analysis would be unwarranted.

Additionally, we reran the resampling analysis detailed in~\secref[s]{analysisChecksCandidates} and \ref{sec:analysisChecksTests} on the mock-signal dataset to ensure that we both identified candidate peaks at the injected signal frequency, and did \emph{not} reject them on the basis of the robustness tests.
All 30 of the already-discussed candidate peaks listed in~\tabref{resampling} appear again when we analyse the mock-signal dataset, with the same $p_0$-values as in the original analysis.
We also identify 15 additional na\"ive signal candidate peaks: 
two of these are clustered at $f_*+f_d$, five at $f_*$, and one at $f_*-f_d$; the remaining seven peaks appear clustered around the reflections of these peaks across the Nyquist frequency with two at $(1\,\text{min})^{-1}-(f_*+f_d)$ and five at $(1\,\text{min})^{-1}-f_*$.
Our resampling analysis correctly rejects ($p_{\text{full}}<0.01$) all five of the na\"ive signal candidates at $f_*\pm f_d$ and $(1\,\text{min})^{-1}-(f_*\pm f_d)$, along with one at $f_*$.
All the other na\"ive signal candidates that appear around $f_*$ and its Nyquist reflection at $(1\,\text{min})^{-1}-f_*$ are not rejected, indicating that the resampling analysis correctly does not rule out potential signals with the correct spatial and temporal properties.
We stress that this non-rejection of multiple na\"ive signal candidates around $f_*$ (and around its reflection through the Nyquist frequency) should not be read to indicate that our pipeline identified some unexpected multi-peaked structure; rather, this is an expected result given that $\sigma_f \sim \Delta f$ and the SNR of the injected signal is large.
Specifically, the non-rejected candidates all appear in contiguous ranges of DFT frequencies; see the inset of \figref{resultsInjected}.
It is to be expected that, evaluated on a bin-by-bin basis, some number of consecutive DFT bins (with widths similar to the signal width parameter) in the immediate vicinity of a large injected signal would each contain sufficient signal power in their own right to qualify as candidates and not be rejected, since they have the appropriate signal properties.

The results in this subsection validate that our analysis pipeline functions as expected.

\subsection{Discussion}
\label{sec:analysisChecksDiscussion}
In this section, we examined in detail the 30 na\"ive signal candidates that we identified in the data that exceed a 95\% confidence global significance threshold.
Applying a resampling analysis to temporal and spatial subsets of the data to test for the robustness of these candidates and their consistency with the expected persistent global nature of the signal, we found that we could automatically reject 23 of the 30 candidates on the grounds of their failing by a large margin the combined temporal and spatial checks. 
Of the remaining seven candidates, one fails the temporal check severely; three weak candidates are in strong tension with the combined robustness test (but cannot be definitely ruled out owing to the unaccounted-for correlation issue discussed in \secref{analysisChecksTests}); two pass all formal robustness tests and are reasonably globally significant, but they appear respectively one DFT frequency bin above or below the Nyquist sampling frequency, and must therefore be viewed with appropriate skepticism with regard to analysis systematics; and the final candidate is statistically significant globally, and passes the combined test and the spatial test, but is in strong tension with the temporal test.
As such, we do not consider any of these na\"ive signal candidates to be robust candidates for a real DPDM signal in the data on the basis of the analysis of the one-minute SuperMAG dataset presented in this work. 
Nevertheless, it would be worthwhile for future work to perform an analysis of the higher-cadence SuperMAG data, as this would likely definitively settle some questions regarding those candidates that were not automatically rejected on the basis of the formal statistical criteria we applied here.

In this section we also presented a validation of our analysis pipeline by showing that a fake signal injected at the level of the $X^{(n)}$ variables defined at \eqref{Xn} is (a) recovered by the analysis at (b) the appropriate frequency with (c) an appropriate value of the kinetic mixing parameter (up to expected deviations; see discussion in \secref{analysisChecksValidation}), and that (d) this injected signal survived the spatial and temporal robustness checks we apply to na\"ive signal candidates.
This verifies that our analysis pipeline performs as expected, and would correctly identify and not reject a real signal in the data.

In conclusion, we find no robust statistical evidence for the existence of the DPDM signal in the SuperMAG data, and we are confident that our analysis pipeline would have identified such a signal had it been present.

\section{Conclusion}
\label{sec:conclusion}
In this work, we presented the details of our analysis of the one-minute-cadence SuperMAG geomagnetic field dataset~\cite{SuperMAGwebsite,Gjerloev:2009wsd,Gjerloev:2012sdg} for the quasi-monochromatic (fractional linewidth $\sigma_f / f \sim 10^{-6}$) global magnetic field signal of dark-photon dark matter that we recently proposed in a companion paper~\cite{Fedderke:2021rys}.
Because the size of the magnetic field signal is $B \propto \varepsilon m_{A'} R \sqrt{\rho_{\textsc{dm}}}$, suffering only a geometrical suppression by the radius of the Earth $R$, we were able to place competitive limits on the parameter space for kinetically mixed dark-photon dark matter.
These limits are shown in \figref{resultsExclusion}, and cover the mass range $2\times 10^{-18}\eV \lesssim m_{A'} \lesssim 7\times 10^{-17}\eV$ corresponding to frequencies $6\times 10^{-4}\Hz \lesssim f_{A'} \lesssim 2\times 10^{-2}\Hz$, with the upper end of this mass reach being limited by the one-minute sampling cadence of the SuperMAG data.
Our analysis made use of a Bayesian framework (using a Jeffreys prior on our kinetic mixing parameter $\varepsilon$)~\cite{Centers:2019dyn,roussy2021experimental} in order to incorporate the effects of the statistically varying local dark-photon dark-matter field amplitude, which assumes values of $A'$ such that $\langle \rho_{\textsc{dm}}\rangle \sim 0.3\,\text{GeV/cm}^3$ only on average; this is a conservative value for the DM density (recent ADMX limits~\cite{ADMX:2020hay} assume a value 50\% larger; see also \citeR{Iocco:2011jz}).

In the course of our search, we initially identified 30 na\"ive signal candidates for dark-photon dark matter that exceeded a global 95\% confidence threshold on the basis of our main analysis. 
In order to ensure that we did not miss a signal, we performed robustness cross-checks on these candidates, testing them for both the spatial and temporal coherence characteristics we expected from our signal.
On the basis of these cross-checks and other indicia, we concluded that none of these signal candidates provide robust evidence for the existence of a DPDM signal in the data.

We also verified our analysis pipeline by injecting both (a) an exactly monochromatic signal aligned along the Earth's rotational axis and appearing exactly at one of the DFT frequency values where we set limits, and (b) a signal with the correct statistical properties for a dark-photon dark-matter signal whose phase- and polarization-coherence are fixed by DM velocity dispersion (i.e., the $\sigma_f/f_*\sim 10^{-6}$ linewidth of the signal).
In the former case, we correctly recovered a limit at the expected value of the kinetic mixing parameter without any \emph{post hoc} degradation to our results.
In the latter case, without any correction factor applied, we would have recovered limits slightly stronger than the injected signal owing to a mild violation of the assumptions used in the construction of the analysis by a signal with the full coherence properties of the dark-photon dark-matter signal (e.g., our analysis assumed an exactly monochromatic signal with exact phase coherence for a full coherence time, which in reality is only approximately true as the phase and polarization of the signal actually evolves fractionally by $\mathcal{O}(1)$ over a full coherence time). 
We applied a 25\% degradation to our limits to compensate for this in a \emph{post hoc} fashion. 
The resulting degraded limits correctly fail to exclude the injected signal, as expected.
We also verified in both cases that the temporal--spatial robustness checks we used to dismiss na\"ive signal candidates in the real data, did not dismiss these injected signals.

The dark-photon dark-matter exclusion bounds we set in \figref{resultsExclusion} are complementary to existing astrophysical bounds in the same mass range that arise from gas heating in various astrophysical settings~\cite{Dubovsky:2015cca,McDermott:2019lch,Wadekar:2019xnf}, and a DM-depletion bound arising from nonresonant dark-photon--photon conversion~\cite{McDermott:2019lch}.
Importantly, as we noted in \citeR{Fedderke:2021rys}, the scaling of our limits with mass is steeper than $m_{A'}^{-1}$ owing to falling noise in the SuperMAG dataset as a function of increasing frequency; this raises the prospect, assuming that this falling noise trend continues to hold, that higher-cadence magnetic data would allow the search for this signal to access currently unconstrained dark-photon dark-matter parameter space.
SuperMAG is currently in the process of releasing one-second-cadence data, and we defer analysis of those data to future work (such an analysis would also provide a separate check on our dismissal of the na\"ive signal candidates that we identified near the Nyquist frequency in the one-minute-cadence data).

Additionally, we set our limits under the assumption of a Standard Halo Model (see, e.g., \citeR{Evans:2018bqy}) DM velocity abundance and dispersion.
If stream-like structures actually dominate the local DM abundance (see, e.g., \citeR[s]{Myeong:2017skt,myeong2018shards,Lancaster_2019,Malhan_2018,Meingast_2019,OHare:2019qxc}), then the signal would be narrower and the DM abundance increased; these effects would make a signal more easily discernible in the data.

Finally, we note that while the analysis approach we presented here exploits the power of this large dataset well, it is not necessarily optimal.
We leave to future work refinement and optimization of the analysis.

\acknowledgments
We thank Surjeet Rajendran, Dmitry Budker, and Alex Sushkov for enlightening conversations at early stages of this project. 
We also thank Ari Cukierman, Michael Coughlin, Reed Essick, Pat Meyers, and Jan Harms for useful correspondence regarding data analysis.

M.A.F.~would like to thank the Berkeley Center for Theoretical Physics at the University of California Berkeley and Lawrence Berkeley National Laboratory for their long-term hospitality during which the earliest stages of this work were completed. 

M.A.F., P.W.G.,~and S.K.~were supported by the Simons Investigator Grant No.~824870, DOE Grant No.~DE-SC0012012, NSF Grant No.~PHY-2014215, DOE HEP QuantISED Award No.~100495, and the Gordon and Betty Moore Foundation Grant No.~GBMF7946.
This work was also supported by the U.S.~Department of Energy, Office of Science, National Quantum Information Science Research Centers, Superconducting Quantum Materials and Systems Center (SQMS) under contract No.~DE-AC02-07CH11359.
D.F.J.K.~was supported by NSF Grant No.~PHY-1707875 as well as the Simons and Heising-Simons Foundations. 
S.K.~was also supported by NSF Grant No.~DGE-1656518.

Some of the computing for this project was performed on the Sherlock cluster. 
We would like to thank Stanford University and the Stanford Research Computing Center for providing computational resources and support that contributed to these research results.

We gratefully acknowledge the SuperMAG Collaboration for maintaining and providing the database of ground magnetometer data that were analyzed in this work and in \citeR{Fedderke:2021rys}, and we thank Jesper W.~Gjerloev for helpful correspondence regarding technical aspects of the SuperMAG data.
SuperMAG receives funding from NSF Grant Nos.~ATM-0646323 and AGS-1003580, and NASA Grant No.~NNX08AM32G S03.

We acknowledge those who contributed data to the SuperMAG Collaboration: 
INTERMAGNET, Alan Thomson; 
CARISMA, PI Ian Mann; 
CANMOS, Geomagnetism Unit of the Geological Survey of Canada; 
The S-RAMP Database, PI K.~Yumoto and Dr.~K.~Shiokawa; 
The SPIDR database; AARI, PI Oleg Troshichev; 
The MACCS program, PI M.~Engebretson; 
GIMA; 
MEASURE, UCLA IGPP and Florida Institute of Technology; 
SAMBA, PI Eftyhia Zesta; 
210 Chain, PI K.~Yumoto; 
SAMNET, PI Farideh Honary; 
IMAGE, PI Liisa Juusola; 
Finnish Meteorological Institute, PI Liisa Juusola; 
Sodankylä Geophysical Observatory, PI Tero Raita; 
UiT the Arctic University of Norway, Troms\o\ Geophysical Observatory, PI Magnar G.~Johnsen; 
GFZ German Research Centre For Geosciences, PI J\"urgen Matzka; 
Institute of Geophysics, Polish Academy of Sciences, PI Anne Neska and Jan Reda; 
Polar Geophysical Institute, PI Alexander Yahnin and Yarolav Sakharov; 
Geological Survey of Sweden, PI Gerhard Schwarz; 
Swedish Institute of Space Physics, PI Masatoshi Yamauchi; 
AUTUMN, PI Martin Connors; 
DTU Space, Thom Edwards and PI Anna Willer; 
South Pole and McMurdo Magnetometer, PIs Louis J.~Lanzarotti and Alan T.~Weatherwax; 
ICESTAR; 
RAPIDMAG; 
British Antarctic Survey; 
McMac, PI Dr.~Peter Chi; 
BGS, PI Dr.~Susan Macmillan; 
Pushkov Institute of Terrestrial Magnetism, Ionosphere and Radio Wave Propagation (IZMIRAN); 
MFGI, PI B.~Heilig; 
Institute of Geophysics, Polish Academy of Sciences, PI Anne Neska and Jan Reda; 
University of L’Aquila, PI M.~Vellante; 
BCMT, V.~Lesur and A.~Chambodut; 
Data obtained in cooperation with Geoscience Australia, PI Marina Costelloe; 
AALPIP, co-PIs Bob Clauer and Michael Hartinger; 
SuperMAG, PI Jesper W.~Gjerloev; 
Data obtained in cooperation with the Australian Bureau of Meteorology, PI Richard Marshall.

We thank INTERMAGNET for promoting high standards of magnetic observatory practice~\cite{INTERMAGNETwebsite}.

\appendix

\section{Fourier transform conventions}
\label{app:FTconventions}
In this appendix, we give our conventions for the continuous Fourier Transform (FT) and Discrete Fourier Transform (DFT).

The continuous FT $\tilde{F}(\omega)$ [alternatively, $\tilde{F}(f)$ with $\omega = 2\pi f$] of a continuous signal $F(t)$ is defined by
\begin{align}
F(t) &= \int_{-\infty}^\infty df \tilde{F}(f) e^{+2\pi i f t } = \int_{-\infty}^\infty \frac{d\omega}{2\pi} \tilde{F}(f) e^{+i \omega t }, \\
\tilde{F}(\omega) &= \int_{-\infty}^\infty dt F(t) e^{- i \omega t },\\
\tilde{F}(f) &= \int_{-\infty}^\infty dt F(t) e^{- 2\pi i f t }.
\label{eq:FTConventions1}
\end{align}
The DFT $\hat{F}(f_k)$ [$k=0,\ldots,N-1$] of the signal with $N$ samples in the time domain $F(t_n)$ [$n=0,\ldots,N-1$] that are equally spaced and taken with a cadence $\Delta t$ for a total duration $T\equiv N \Delta t$ is defined by
\begin{align}
F(t_n) &= \frac{1}{T} \sum_{k=0}^{N-1} \hat{F}(f_k) e^{+2\pi i k n / N },\label{eq:IDFTdefn}\\
\hat{F}(f_k)&= \frac{T}{N} \sum_{n=0}^{N-1} F(t_n) e^{ - 2\pi i k n / N },
\label{eq:DFTdefn}
\end{align}
where $t_n \equiv n \Delta t = nT/N$, and $f_k \equiv k \Delta f \equiv k/T$.
The (discrete) two-sided power spectral density (PSD) is defined in terms of the DFT:
\begin{align}
\hat{S}_F(f_k) \equiv \frac{1}{T} | \hat{F}(f_k) |^2.
\label{eq:PSDdefn}
\end{align}

Note that a monochromatic signal with frequency $f_m \equiv m/T$ ($0\leq m\leq N-1;\ m\in\mathbb{Z}$),
\begin{align}
F(t_n) = f_{A'} \cos( 2 \pi f_m t_n ); \quad n=0,\ldots,N-1,
\label{eq:monochromaticSignal}
\end{align}
has a DFT given by
\begin{align}
\hat{F}(f_k) &= \frac{T}{2} f_{A'}\, \Big[ \delta_{k,m} + \delta_{k,(N-m)\!\!\!\!\mod N} \Big],
\qquad \qquad 
\label{eq:monochromaticSignalFT}
\end{align}
where $k = 0,\ldots,N-1$; the corresponding two-sided PSD is
\begin{align}
\hat{S}_F(f_k) = \frac{T}{4} |f_{A'}|^2 \, \Big[ \delta_{k,m} + \delta_{k,(N-m)\!\!\!\!\mod N} \Big]^2.
\label{eq:monochromaticSignalPSD}
\end{align}

\section{Vector spherical harmonics}
\label{app:vectorSphericalHarmonics}

\begin{figure*}[t]
\includegraphics[width=\textwidth]{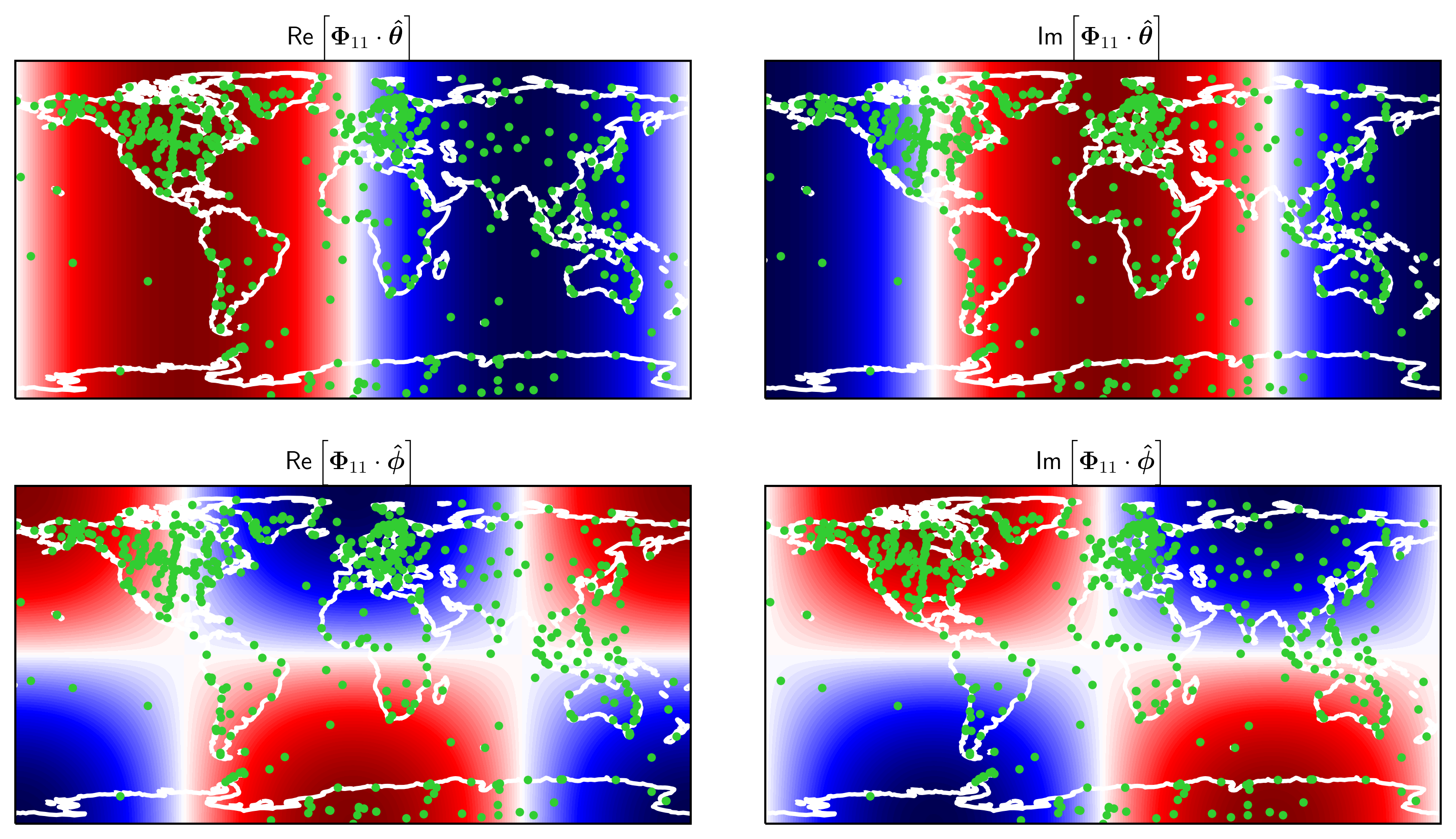}\\[1ex]
\includegraphics[width=0.5\textwidth]{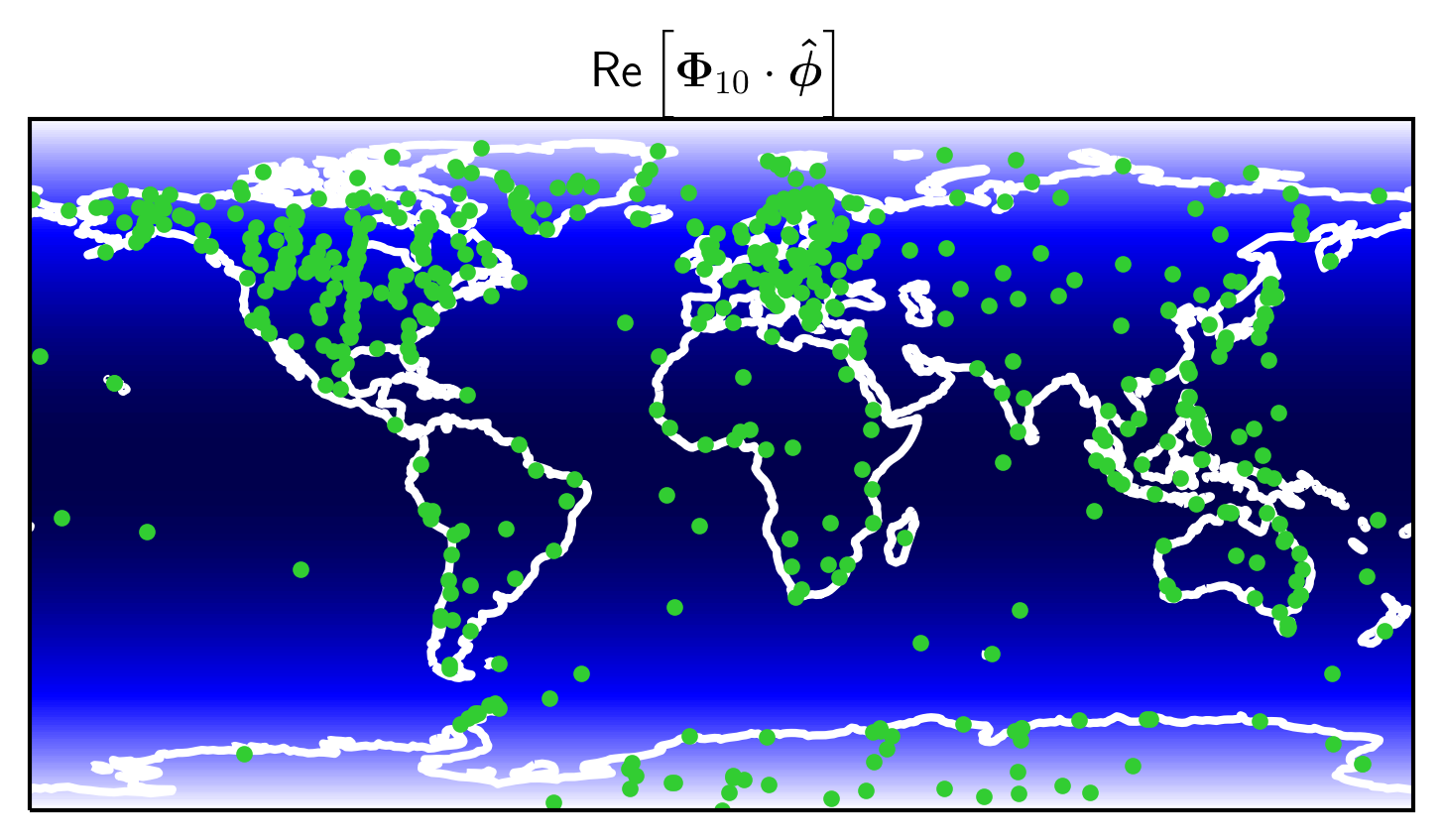}
\caption{\label{fig:VSHplots}%
    Shaded contour plots of the real and imaginary parts of all the nonzero $\bm{\hat{\theta}}$- and $\bm{\hat{\phi}}$-components of the vector spherical harmonics $\bm{\Phi}_{11}$ and $\bm{\Phi}_{10}$; the cognate plots for $\bm{\Phi}_{1,-1}$ can be read from those of $\bm{\Phi}_{11}$ using \eqref{phaseConventionPointer}.
    Red (blue) indicates positive (negative) values, with the color range for each plot independently normalized to span the range of values plotted.
    Overlaid are the outlines of the Earth's continents (white), and the locations of the SuperMAG stations (green points); see also \figref{SuperMAGstations}.
   }
\end{figure*}

This appendix, which serves only to define the conventions used in this work and in \citeR{Fedderke:2021rys}, is reproduced from \citeR{Fedderke:2021rys} with minor modifications for the convenience of the reader.

The vector spherical harmonics are defined in terms of the scalar spherical harmonic $Y_{\ell m}$ by the relations
\begin{align}
\bm{Y}_{\ell m} &= Y_{\ell m}\rhat, &
\bm{\Psi}_{\ell m} &= r\bm{\nabla} Y_{\ell m}, &
\bm{\Phi}_{\ell m} &= \bm{r}\times\bm{\nabla} Y_{\ell m},
\end{align}
where $\rhat$ is the unit vector in the direction of $\bm{r}$.
Thus $\bm{Y}_{\ell m}$ points radially, while $\bm{\Psi}_{\ell m}$ and $\bm{\Phi}_{\ell m}$ point tangentially.
Some of their relevant properties (and our phase conventions) are
\begin{align}
\bm{Y}_{\ell,-m}&=(-1)^m\bm{Y}_{\ell m}^*,\\
\bm{\Psi}_{\ell,-m}&=(-1)^m\bm{\Psi}_{\ell m}^*,\\
\bm{\Phi}_{\ell,-m}&=(-1)^m\bm{\Phi}_{\ell m}^*, \label{eq:phaseConventionPointer}\\
\bm{Y}_{\ell m}\cdot\bm{\Psi}_{\ell m}&=\bm{Y}_{\ell m}\cdot\bm{\Phi}_{\ell m}=\bm{\Psi}_{\ell m}\cdot\bm{\Phi}_{\ell m}=0,
\end{align}
\begin{align}
\int d\Omega\,\bm{Y}_{\ell m}\cdot \bm{Y}_{\ell'm'}^*&=\delta_{\ell\ell'}\delta_{mm'},\\
\int d\Omega\,\bm{\Psi}_{\ell m}\cdot\bm{\Psi}_{\ell'm'}^*&=\int d\Omega~\Phi_{\ell m}\cdot\Phi_{\ell'm'}^*\nonumber \\
	&=\ell(\ell+1)\delta_{\ell\ell'}\delta_{mm'},\\
\int d\Omega\,\bm{Y}_{\ell m}\cdot\bm{\Psi}_{\ell'm'}^*&=\int d\Omega\,\bm{Y}_{\ell m}\cdot\bm{\Phi}_{\ell'm'}^*\nonumber \\
	&=\int d\Omega\,\bm{\Psi}_{\ell m}\cdot\bm{\Phi}_{\ell'm'}^*=0.\label{eq:orthogonality}
\end{align}
The explicit expressions for the spherical harmonics which are relevant to this work [see \eqref{signal}] are
\begin{align}
\bm{\Phi}_{1,-1}(\bm{r}) &=\sqrt{\frac3{8\pi}}e^{-i\phi}(i\thetahat+\cos\theta\phihat), \label{eq:Phi1m1}\\
\bm{\Phi}_{10}(\bm{r})&=-\sqrt{\frac3{4\pi}}\sin\theta\phihat,\label{eq:Phi10}\\
\bm{\Phi}_{11}(\bm{r})&=\sqrt{\frac3{8\pi}}e^{i\phi}(i\thetahat-\cos\theta\phihat)\label{eq:Phi1p1},
\end{align}
where $\thetahat$ and $\phihat$ are unit vectors in the directions of increasing $\theta$ and $\phi$.

Note that, as written here, the spherical co-ordinate $\phi$ coincides with the definition of longitude; however, the spherical co-ordinate $\theta$ is not the latitude: $\theta$ increases from $\theta = 0$ at the Geographic North Pole (latitude $+90^\circ$), to $\theta = \pi/2$ on the Equator (latitude $0^\circ$), to $\theta = \pi$ at the Geographic South Pole (latitude $-90^\circ$).

\figref{VSHplots} shows the real and imaginary components of the nonzero $\bm{\hat{\theta}}$- and $\bm{\hat{\phi}}$-components of $\bm{\Phi}_{11}$ and $\bm{\Phi}_{10}$.

\begin{widetext}
\section{Other contributions to \texorpdfstring{$\langle\vec X_k\rangle$}{Xksig}}
\label{app:Xk}

The full expressions for the contributions to $\langle\vec X_k\rangle$ coming from the $x$- and $y$-polarizations are
\begin{align}
\langle\vec X_k\rangle_{\bm{B}=\bm{B}^{x}_{R}}\approx\varepsilon\vec\mu_{xk}\equiv\pi\varepsilon f_{A'}R\sqrt{\frac{\rho_{\textsc{dm}}}8}\begin{pmatrix}\tilde{\mathbf 1}_k(f_d-\hat f_d)-\tilde H_k^{(1)}(f_d-\hat f_d)+i\tilde H_k^{(2)}(f_d-\hat f_d)\\\tilde H_k^{(2)}(f_d-\hat f_d)+i\tilde H_k^{(1)}(f_d-\hat f_d)\\\tilde H_k^{(4)}(f_d-\hat f_d)-i\tilde H_k^{(5)}(f_d-\hat f_d)\\-\tilde H_k^{(5)}(f_d-\hat f_d)+i\tilde H_k^{(3)}(f_d-\hat f_d)-i\tilde H_k^{(4)}(f_d-\hat f_d)\\\tilde H_k^{(6)}(f_d-\hat f_d)-i\tilde H_k^{(7)}(f_d-\hat f_d)\\2\cdot\RE[\tilde 1_k(f_d)-\tilde H_k^{(1)}(f_d)]+2\cdot\IM[\tilde H_k^{(2)}(f_d)]\\2\cdot\RE[\tilde H_k^{(2)}(f_d)]+2\cdot\IM[\tilde H_k^{(1)}(f_d)]\\2\cdot\RE[\tilde H_k^{(4)}(f_d)]-2\cdot\IM[\tilde H_k^{(5)}(f_d)]\\-2\cdot\RE[\tilde H_k^{(5)}(f_d)]+2\cdot\IM[\tilde H_k^{(3)}(f_d)-\tilde H_k^{(4)}(f_d)]\\2\cdot\RE[\tilde H_k^{(6)}(f_d)]-2\cdot\IM[\tilde H_k^{(7)}(f_d)]\\\tilde{\mathbf 1}_k(\hat f_d-f_d)-\tilde H_k^{(1)}(\hat f_d-f_d)-i\tilde H_k^{(2)}(\hat f_d-f_d)\\\tilde H_k^{(2)}(\hat f_d-f_d)-i\tilde H_k^{(1)}(\hat f_d-f_d)\\\tilde H_k^{(4)}(\hat f_d-f_d)+i\tilde H_k^{(5)}(\hat f_d-f_d)\\-\tilde H_k^{(5)}(\hat f_d-f_d)-i\tilde H_k^{(3)}(\hat f_d-f_d)+i\tilde H_k^{(4)}(\hat f_d-f_d)\\\tilde H_k^{(6)}(\hat f_d-f_d)+i\tilde H_k^{(7)}(\hat f_d-f_d)\end{pmatrix},\label{eq:XexpxApprox}\\
\langle\vec X_k\rangle_{\bm{B}=\bm{B}^{y}_{R}}\approx\varepsilon\vec\mu_{yk}\equiv-\pi\varepsilon f_{A'}R\sqrt{\frac{\rho_{\textsc{dm}}}8}\begin{pmatrix}\tilde H_k^{(2)}(f_d-\hat f_d)-i\tilde{\mathbf 1}_k(f_d-\hat f_d)+i\tilde H_k^{(1)}(f_d-\hat f_d)\\\tilde H_k^{(1)}(f_d-\hat f_d)-i\tilde H_k^{(2)}(f_d-\hat f_d)\\-\tilde H_k^{(5)}(f_d-\hat f_d)-i\tilde H_k^{(4)}(f_d-\hat f_d)\\\tilde H_k^{(3)}(f_d-\hat f_d)-\tilde H_k^{(4)}(f_d-\hat f_d)+i\tilde H_k^{(5)}(f_d-\hat f_d)\\-\tilde H_k^{(7)}(f_d-\hat f_d)-i\tilde H_k^{(6)}(f_d-\hat f_d)\\2\cdot\RE[\tilde H_k^{(2)}(f_d)]-2\cdot\IM[\tilde 1_k(f_d)-\tilde H_k^{(1)}(f_d)]\\2\cdot\RE[\tilde H_k^{(1)}(f_d)]-2\cdot\IM[\tilde H_k^{(2)}(f_d)]\\-2\cdot\RE[\tilde H_k^{(5)}(f_d)]-2\cdot\IM[\tilde H_k^{(4)}(f_d)]\\2\cdot\RE[\tilde H_k^{(3)}(f_d)-\tilde H_k^{(4)}(f_d)]+2\cdot\IM[\tilde H_k^{(5)}(f_d)]\\-2\cdot\RE[\tilde H_k^{(7)}(f_d)]-2\cdot\IM[\tilde H_k^{(6)}(f_d)]\\\tilde H_k^{(2)}(\hat f_d-f_d)+i\tilde{\mathbf 1}_k(\hat f_d-f_d)-i\tilde H_k^{(1)}(\hat f_d-f_d)\\\tilde H_k^{(1)}(\hat f_d-f_d)+i\tilde H_k^{(2)}(\hat f_d-f_d)\\-\tilde H_k^{(5)}(\hat f_d-f_d)+i\tilde H_k^{(4)}(\hat f_d-f_d)\\\tilde H_k^{(3)}(\hat f_d-f_d)-\tilde H_k^{(4)}(\hat f_d-f_d)-i\tilde H_k^{(5)}(\hat f_d-f_d)\\-\tilde H_k^{(7)}(\hat f_d-f_d)+i\tilde H_k^{(6)}(\hat f_d-f_d)\end{pmatrix}.\label{eq:XexpyApprox}
\end{align}
We have again neglected subdominant Fourier contributions; see the discussion below \eqref{XexpzApprox}. 
\end{widetext}

\section{Likelihood details}
\label{app:likelihoodDetails}
In this appendix, we give some additional technical details and derivations of the likelihood function and priors that we utilized in \secref{analysisBayesian}.

\subsection{Marginalized likelihood}
\label{app:marginalizedLikelihood}
In this subsection, we supply more details of the derivation of the marginalized likelihood in \eqref{likelihood}.
For notational simplicity, we define $\bm{d}_k=\bm{a}_k+i\bm{b}_k$, and denote by $z_{ik}$, $d_{ik}$, $a_{ik}$, and $b_{ik}$ the components of $\bm{Z}_k$, $\bm{d}_k$, $\bm{a}_k$, and $\bm{b}_k$, respectively.
Moreover, we denote by $s_{ik}$ the diagonal elements of $S_k$ (i.e., the singular values of $N_k$).
From \eqref[s]{LLZ}, (\ref{eq:LLd}), and (\ref{eq:marginalLikelihood}), it is clear that the marginalized combined likelihood function factorizes over coherence times $k$, 
\begin{align}
\LL\lb(  \varepsilon \big| \{ \bm{Z}_k \} \rb) \equiv \prod_k \LL_k\lb( \varepsilon | \bm{Z}_k \rb),
\end{align}
where, noting that $\bm{d}_k$ (and thus $\bm{a}_k$ and $\bm{b}_k$) is a 3-vector,\linebreak\pagebreak
\begin{widetext}
{\noindent}we have
\begin{align}
&\LL_k\lb(\varepsilon\big|\bm{Z}_k\rb)\nonumber\\
&=\int d^3 a_kd^3 b_k~\LL_k\lb(\varepsilon,\bm{d_k}\big|\bm{Z}_k\rb)\LL_k\lb(\bm{d_k}\rb) \\
 &=\int d^3a_k d^3b_k\exp\lb[-\sum_i\lb(\lb|z_{ik}-\varepsilon s_{ik}d_{ik}\rb|^2+3|d_{ik}|^2\rb)\rb]\\
 &=\prod_i\int da_{ik}db_{ik}\exp\Big[-(3+\varepsilon^2s_{ik}^2)(a_{ik}^2+b_{ik}^2)+2\varepsilon s_{ik}\RE[z_{ik}]a_{ik}+2\varepsilon s_{ik}\IM[z_{ik}]b_{ik}-|z_{ik}|^2\Big]\\
 &=\prod_i\int da_{ik}db_{ik}\exp\lb[
 	\begin{array}{l}
		-(3+\varepsilon^2s_{ik}^2)\lb(a_{ik}-\dfrac{\varepsilon s_{ik}\RE[z_{ik}]}{3+\varepsilon^2s_{ik}^2}\rb)^2-(3+\varepsilon^2s_{ik}^2)\lb(b_{ik}-\dfrac{\varepsilon s_{ik}\IM[z_{ik}]}{3+\varepsilon^2s_{ik}^2}\rb)^2
		 - \dfrac{3|z_{ik}|^2}{(3+\varepsilon^2s_{ik}^2)}
	\end{array} \rb] \label{eq:completedSquare}\\[1ex] 
 &\propto\prod_i\frac1{3+\varepsilon^2s_{ik}^2}\exp\lb(-\frac{3|z_{ik}|^2}{3+\varepsilon^2s_{ik}^2}\rb),
\end{align}
\end{widetext}
where at \eqref{completedSquare} we completed the square in the exponent to obtain Gaussian integrals, and simplified.
Up to an arbitrary normalization, \eqref{likelihood} follows.

\subsection{Jeffreys prior}
\label{app:jeffreysPrior}
In this subsection, we derive the Jeffreys prior, \eqref{prior}.

The Fisher information matrix, $\mathcal{I}$ is defined as~\cite{Cowan:2018swr}
\begin{align}
    \mathcal{I}_{i,j}(\vec \Theta) \equiv \text{E}\lb[ \lb.\lb(\D\Theta_i\log\LL\rb)\lb(\D\Theta_j\log\LL\rb) \rb| \vec\Theta \rb],
\end{align}
where $\vec \Theta$ is the parameter vector, $\text{E}\big[\, \cdots | \vec\Theta \big]$ is the expectation value over data realizations drawn assuming the values of the parameters $\vec \Theta$, and $\LL = \LL\lb(\vec \Theta\big|\{x\}\rb)$ is the likelihood considered as a function of the model parameters given the data realization $\{x \}$.

The Jeffreys prior is the unique reparametrization-invariant prior, and is defined in terms of $\mathcal{I}$~\cite{Cowan:2018swr}:
\begin{align}
    p(\vec \Theta) \propto \lb[ \det \mathcal{I}(\vec \Theta) \rb]^{1/2}. 
\end{align}
For the case of a one-dimensional parameter vector, as is our case after marginalizing over the $\bm{d}_k$, this simplifies:
\begin{align}
p(\varepsilon) &\propto\sqrt{I(\varepsilon)} \equiv\sqrt{\text{E}\lb[ \lb.\lb(\D\varepsilon\log\LL\rb)^2 \rb| \varepsilon \rb]};
\end{align}
for notational simplicity, we leave the `$|\varepsilon$' implicit in what follows.
Therefore,
\begin{align}
\lb[p(\varepsilon)\rb]^2
			&\propto \text{E}\lb[\lb(\D\varepsilon\sum_{i,k}\lb(-\frac{3|z_{ik}|^2}{3+\varepsilon ^2s_{ik}^2}-\ln(3+\varepsilon^2s_{ik}^2)\rb)\rb)^2 \rb]\\
			&=\text{E}\lb[\lb(\sum_{i,k}\frac{2\varepsilon s_{ik}^2\lb(3|z_{ik}|^2-3-\varepsilon^2s_{ik}^2\rb)}{\lb(3+\varepsilon ^2s_{ik}^2\rb)^2}\rb)^2\rb],\label{eq:Jeffreys}
\end{align}
Interpreting the likelihood $\LL$ given by \eqref{likelihood} via its definition as the probability density function of the data given the model parameter $\varepsilon$ (see discussion at footnote~\ref{ftnt:otherArgument}), we see that (when viewed as random variables rather than as the specific data realizations we have) the real and imaginary parts of the $z_{ik}$ are all independent, zero-mean normally distributed variables satisfying $\langle|z_{ik}|^2\rangle=1+\varepsilon^2s_{ik}^2/3$.
Therefore the expectation values of all the cross terms (i.e., those with differing $i$ and $k$) in \eqref{Jeffreys} vanish since they factor into two quantities, each with expectation value zero.
The only remaining terms are thus those where $i$ and $k$ are the same for both factors.
Therefore,
\begin{align}
\lb[p(\varepsilon)\rb]^2
			&=\sum_{i,k} \text{E}\lb[\lb(\frac{2\varepsilon s_{ik}^2\lb(3|z_{ik}|^2-3-\varepsilon^2s_{ik}^2\rb)}{\lb(3+\varepsilon ^2s_{ik}^2\rb)^2}\rb)^2\rb]\\
			&=\sum_{i,k}\frac{4\varepsilon^2s_{ik}^4}{\lb(3+\varepsilon ^2s_{ik}^2\rb)^2},
\end{align}
where we used%
\footnote{\label{ftnt:not3}%
    Note that this result is distinct from the one-dimensional real-variable result $E[x^4] = 3E[x^2]^2$ that would be expected for a single zero-mean normally distributed real $x$.
    } %
$\text{E}\lb[|z_{ik}|^4\rb] = 2 \text{E}\lb[|z_{ik}|^2\rb]^2$.
The expression for the Jeffreys prior at \eqref{prior} follows.

\section{Noise validation}
\label{app:noiseValidation}
In this section, we validate some of our assumptions about the noise in the SuperMAG data, and our analysis of that dataset.
Specifically, we validate three assumptions: (1) the noise can be treated as constant over the course of one calendar year, (2) our choice of $\tau_\text{min}$ is sufficiently large, and (3) the variables $z_{ik}$ are sufficiently Gaussian.

\subsection{Noise variation within a calendar year}
\label{app:calendarYear}
Our analysis in \secref{analysisNoise} computed noise spectra $S^a_{mn}(f_p)$ associated with particular calendar years, under the assumption that the noise remained statistically stationary within a calendar year.
We evaluate this assumption by computing the same quantity $S^a_{mn}(f_p)$ on timescales shorter than a full year and comparing the results to the full-year estimate.
In particular, we divide a given year evenly into four quarters and recompute $S^a_{mn}(f_p)$ as in \secref{analysisNoise} using the data from each quarter (again taking $\tau_\text{min}=16384$ min).
Due to the finite number of samples used in \eqref{noiseEstimate}, the estimate $S^a_{mn}(f_p)$ that we compute has a large variance from one frequency to the next.
For the purposes of this comparison it is useful to examine and compare the moving average of $S^a_{mn}(f_p)$ taken over a range of frequency bins in a sliding window centered on each frequency.
That is, we compute
\begin{align}
    \bar S^a_{mn}(f_p)=\frac1{2w+1}\sum_{q=p-w}^{p+w}S^a_{mn}(f_q),
    \label{eq:noiseAve}
\end{align}
where we take the window half-width to be $w=512$, corresponding to a top-hat sliding window with width $5.2\times10^{-4}$\Hz.
Likewise it is useful to compute the standard deviation of the values of $S^a_{mn}(f_p)$ within the sliding window, as a statistic to quantify the spread:
\begin{align}
    \sigma^a_{mn}(f_p)=\sqrt{\frac1{2w+1}\sum_{q=p-w}^{p+w}\Big|S^a_{mn}(f_q)-\bar S^a_{mn}(f_q)\Big|^2}.
    \label{eq:noiseStd}
\end{align}

\begin{figure*}[t]
\includegraphics[width=0.99\textwidth]{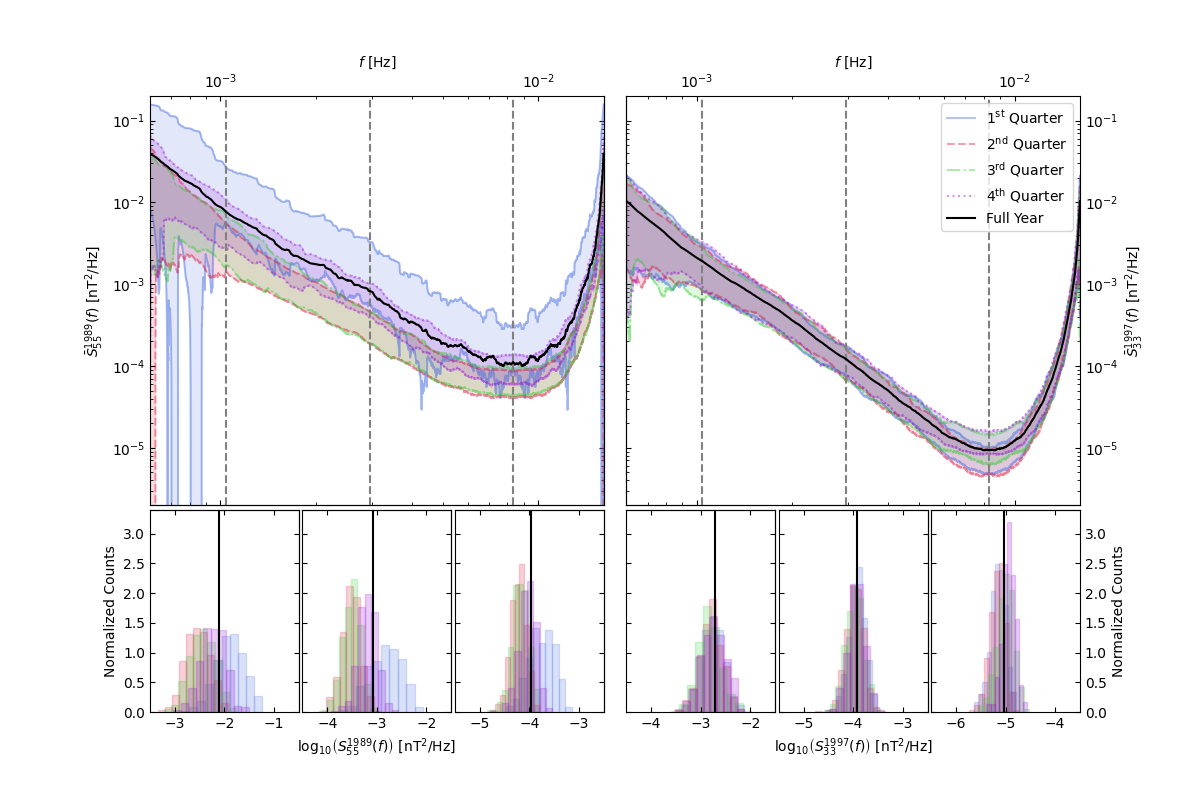}
\caption{\label{fig:subyear}%
        Noise stationarity validation; on-diagonal elements of $S^a_{mn}(f)$.
        \textsc{Upper panels:}
        The solid black line shows representative components of the full-year averaged [see \eqref{noiseAve}] noise auto-power spectra $\bar{S}^a_{mn}$ $(m=n)$ for some selected representative years $a$, while the various shaded colored bands give the values of $\bar S^a_{mn}(f)\pm\sigma^a_{mn}(f)$ that are computed using data only from one of each of the four quarters within that year $a$ (see legend).
        The left-hand side shows a case where there is some mild tension between the quarter-by-quarter noise determinations; the right-hand side shows a case where the four quarter-by-quarter determinations agree excellently.
        \textsc{Lower panels:}
        For three selected frequencies (vertical grey dashed lines marked in the upper panels), we show histograms (colored bars; see legend) of the values of $S^a_{mn}(f_q)$ for $f_q$ falling within the averaging window used to determine the quarter-by-quarter values of $\bar{S}^a_{mn}(f)$ [see discussion around \eqref{noiseAve}], along with the full-year average $\bar{S}^a_{mn}(f)$ (vertical black line).
        Each histogram panel is displayed immediately below the relevant vertical grey dashed line in the upper panel which marks the frequency to which it corresponds (i.e., in order from left to right, the histograms correspond to the same three frequencies marked, in order from left to right, by the vertical grey lines in the upper panel).
    	} 
\end{figure*}

\begin{figure*}[p]
\includegraphics[width=0.83\textwidth]{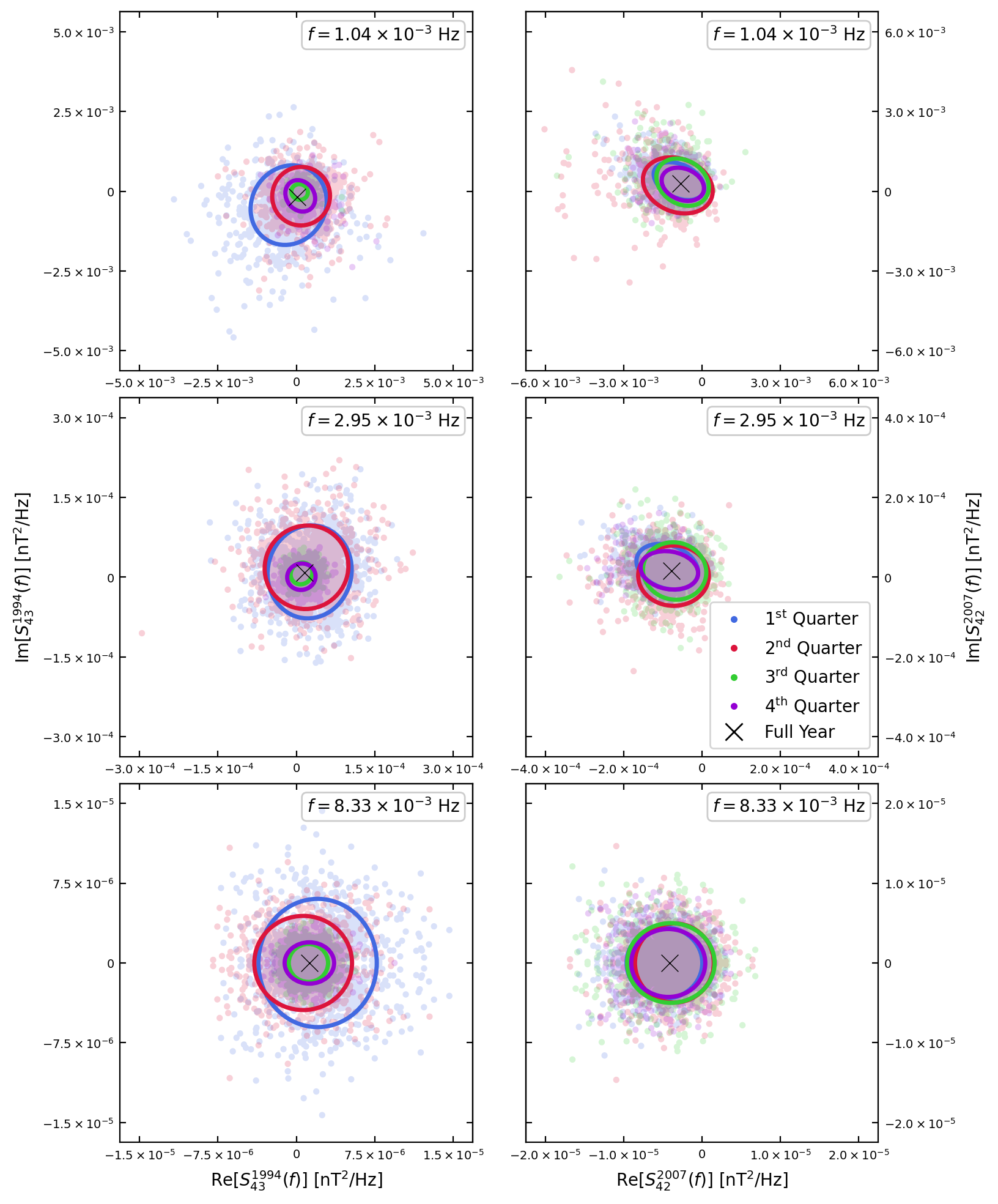}
\caption{\label{fig:subyearOffDiag}%
        Noise stationarity validation; off-diagonal elements of $S^a_{mn}(f)$.
        Because the off-diagonal components of the Hermitian matrices $S^a_{mn}(f)$ are complex, we cannot show the quarter-by-quarter agreement as a function of frequency as simply for the off-diagonal elements as we did in \figref{subyear} for the real, on-diagonal components. 
        In this figure, at three selected representative frequencies (the same ones indicated by the vertical grey dashed lines in \figref{subyear}), we select some representative off-diagonal components $(m,n)$ for some representative years $a$, and show scatter plots in the complex plane of the values of $S^a_{mn}(f_q)$ for the $f_q$ that lie within the corresponding averaging windows used to determine the quarterly values of $\bar{S}^a_{mn}(f)$ [see discussion around \eqref{noiseAve}] (shaded colored circular points; see legend).
        Note that the density of points, and not the depth of shading, indicates the clustering of values within each quarter (with some loss of resolution in the denser regions); we have fixed the depth of shading for each quarter to be independent of the density of points in order to make the differences in the clustering of points from quarter to quarter clearer. 
        Also shown are the 68\% coverage ellipses for two-dimensional Gaussian fits to the scattered points for each quarter (like-colored solid ellipses; see legend), along with the full-year average value of $\bar{S}^a_{mn}(f)$ (black cross).
        The left-hand column shows a case where there is some mild tension (within factors of $\sim 2$--$3$) between the intra-year statistical stationary of the noise assumed in our analysis, and the realised noise (i.e., the fitted 68\% coverage ellipses for different quarters vary somewhat), while the right-hand column shows a case where intra-year noise stationarity is realised well.
    	} 
\end{figure*}

In \figref[s]{subyear} and \ref{fig:subyearOffDiag}, we compare the quarterly estimates of $S^a_{mn}(f_p)$ to the full-year moving average $\bar S^a_{mn}(f_p)$, for diagonal ($m=n$) and off-diagonal ($m\neq n$) elements, respectively.
Specifically, in the top panels of \figref{subyear}, we show shaded bands that span the range of values $\bar S^a_{mn}(f_p)\pm\sigma^a_{mn}(f_p)$ for each of the four quarters in a chosen year, along with the corresponding full-year average result $\bar S^a_{mn}(f_p)$ in solid black; in the bottom panels, we show a histogram of the quarterly estimates $S^a_{mn}(f_p)$ that appear in \eqref{noiseAve} for a few representative frequencies.
In \figref{subyearOffDiag}, on the other hand, we show scatter plots of the quarterly estimates $S^a_{mn}(f_p)$ that appear in \eqref{noiseAve}, along with their corresponding 68\% coverage ellipses; the full-year average $\bar S^a_{mn}(f_p)$ is marked by a black cross.
The variety of years, components $m,n$, and frequencies $f_p$ displayed in \figref{subyearOffDiag} are broadly representative.
It is clear that for a wide range of frequencies and component choices (both diagonal and off-diagonal), the full-year average is consistent with the distribution of quarterly estimates, indicating (within the precision of the shorter-timescale estimates) that the assumption of statistical stationarity is satisfied.
We do however note that there is variation in the degree to which the individual quarterly results are consistent with each other within a year [e.g., the 1989 results for the $(m,n)=(5,5)$ component show some mild tension between the first and third quarters; whereas, e.g., the 2007 results for the $(m,n)=(4,2)$ component are in better agreement].
It is possible that a more sophisticated analysis than that presented here could account for this.

\subsection{Choice of \texorpdfstring{$\tau_\text{min}$}{tau-min}}
\label{app:tauMin}

\begin{figure*}[p]
\includegraphics[width=\textwidth]{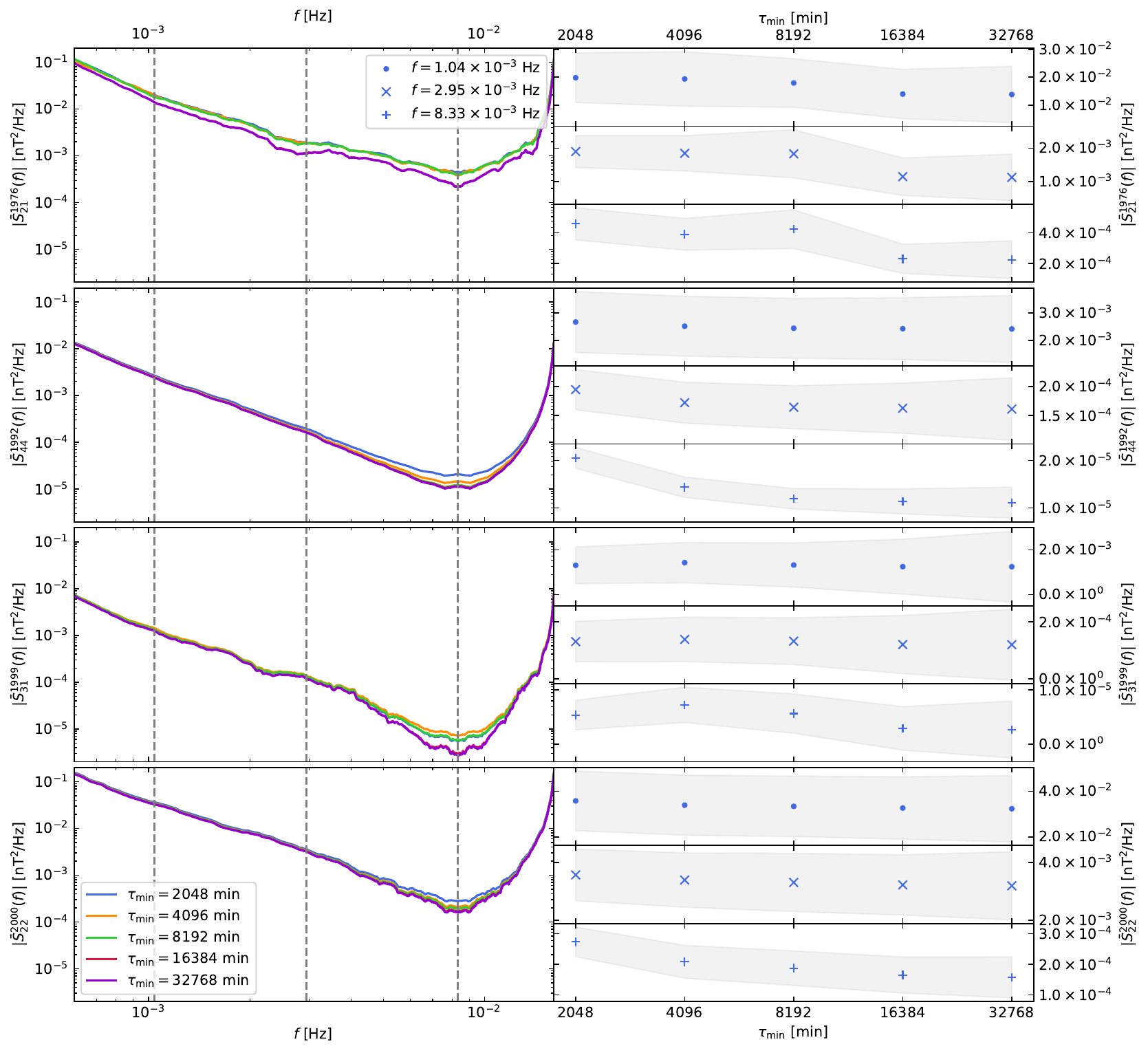}
\caption{\label{fig:tauMin}%
        Dependence of noise estimates on the parameter $\tau_\text{min}$.
        \textsc{Left column:}
        Frequency-space averaged noise spectra $|\bar S^a_{mn}(f)|$ for $\tau_\text{min}=2^j$\,min for choices $j=11,\ldots,15$, shown for multiple representative choices of year $a$ and components $(m,n)$.
        The colors as annotated in the legend on the lower left distinguish the various cases.
        Note that the sliding window used to compute the frequency-space average is varied ($w=2^{j-5}$ for the integers $j$ defined above), so that the sliding window width used to compute the frequency-space averages is maintained at $5.2\times10^{-4}$\Hz~in all cases.
        Note also that, in this plot, where relevant and in contrast to \figref{subyearOffDiag}, we show the absolute value of the complex off-diagonal $\bar S^a_{mn}(f)$. 
        \textsc{Right column:}
        Dependence of $|\bar S^a_{mn}(f)|$ on $\tau_\text{min}$ for three representative fixed frequencies: $f=1.04\times10^{-3}$\Hz, $f=2.95\times10^{-3}$\Hz, and $f=8.33\times10^{-3}$\Hz~(the locations of these frequencies are also indicated as dashed vertical grey lines in the left panels).
        The grey bands represent the spread in the values in the sliding windows, $\sigma^a_{mn}(f)$; see \eqref{noiseStd}.
        It is clear that, in most cases, for $\tau_\text{min}\geq16384$\,min, the dependence of $|\bar S^a_{mn}(f)|$ on $\tau_\text{min}$ is much smaller than the spread $\sigma^a_{mn}(f)$.
        Note that the top row shows an anomalous off-diagonal component from an early year ($a=1976$), whose behavior is likely strongly influenced by the small number of active stations.
        Additionally, the panels in the third row show a result for an off-diagonal component [$(m,n)=(3,1)$], which is somewhat smaller in normalization than the other on-diagonal components from the same year; residual variation of these results with $\tau_\text{min}$ will thus have little effect on our results.
    	} 
\end{figure*}

Our analysis in \secref{analysisNoise} has one arbitrary parameter: $\tau_\text{min}$, the lower bound for the duration of the chunks of data that are used in our noise analysis.
There is a trade-off in the selection of this parameter: it should be as small as possible to allow for more independent chunks, and thus a better overall statistical estimate of the noise.
On the other hand, taking the chunks length too short would increase the relative impact of possible correlations between data points near the edges of consecutive chunks, which could systematically bias our noise estimate as the chunks would no longer be sufficiently statistically independent.
It is therefore important to understand the dependence of $S^a_{mn}(f_p)$ on $\tau_\text{min}$, and to select $\tau_\text{min}$ large enough that the dependence of $S^a_{mn}(f_p)$ on that parameter becomes subdominant to the uncertainty on the estimate of $S^a_{mn}(f_p)$ itself.

To quantify this, we re-compute $S^a_{mn}(f_p)$ as in \secref{analysisNoise}, but using several different choices of $\tau_\text{min}$.
For each choice of $\tau_\text{min}$, we compute the frequency-space moving average spectrum, $\bar S^a_{mn}(f_p)$, as in \eqref{noiseAve}.
The results are shown in \figref{tauMin}: the panels on the left of \figref{tauMin} show the full spectra $\bar S^a_{mn}(f_p)$ obtained using values of $\tau_\text{min} = 2^j$\,min for various integers $j=11,\ldots,15$, again for various representative choices of the year $a$ and components $m,n$.
The panels on the right in \figref{tauMin} show the dependence of $\bar S^a_{mn}(f_p)$ on $\tau_\text{min}$ for the same years and components at three particular representative frequencies chosen in the low-, mid-, and high-frequency regions of our range of interest, along with their associated standard deviations (shaded bands), $\sigma^a_{mn}(f_p)$ [defined in \eqref{noiseStd}].

The figure demonstrates that for well-behaved components (typically the larger, diagonal $m=n$ components in most years), the dependence of $\bar S^a_{mn}(f_p)$ on $\tau_\text{min}$ is smaller than the spread $\sigma^a_{mn}(f_p)$ in the values of $S^a_{mn}(f_p)$, for $\tau_\text{min}=16384$\,min or greater.
The less well-behaved components that vary somewhat more with changing $\tau_\text{min}$ are typically off-diagonal (i.e., $m\neq n$) components, which tend to be smaller by a factor of $\mathcal{O}(3$--$10)$ than the on-diagonal components; these thus have less impact on our results, regardless of the choice of $\tau_\text{min}$.
Generically, the picture is that, for most years, a choice of $\tau_\text{min}=16384$\,min is reasonable, and gives results that do not strongly depend on $\tau_\text{min}$; this choice also allows sufficiently many independent chunks to obtain a noise estimate within a $\mathcal{O}(10\%)$, which is sufficient for the level of precision in our analysis. 

Nevertheless, we do note that for certain specific years, there are exceptions to this generic picture.
For instance, \figref{tauMin} shows an example of a less well-behaved  off-diagonal component [$(m,n)=(2,1)$] from an early year ($a=1976$), when much fewer stations were active as compared to later years; cf.~\figref{station_count}.
This result clearly shows a sharp change in behavior between $\tau_\text{min}=8192$\,min and $\tau_\text{min}=16384$\,min that might be worrisome.
However, anomalous behavior in years with very few stations running would not be entirely unexpected, as a single station turning on/off would have a larger impact on the overall results than in years with many stations running.
While we still include results from such `anomalous' years in our analysis, we note that---precisely because they have much fewer stations running---they will be naturally down-weighted in our analysis owing to the larger noise obtained with fewer stations running, and will thus have little influence on our final results.

\subsection{Gaussianity of variables}
\label{app:gaussianity}
In \secref{analysisBayesian}, we constructed a likelihood function for our Bayesian analysis under the assumption of Gaussianity on the real and imaginary components of the $z_{ik}$; see, e.g., \eqref{likelihood}.
We evaluate the validity of this assumption by computing the four-point function of $z_{ik}$ and comparing it to the expected result assuming Gaussianity: $\langle |z_{ik}|^4 \rangle_{\text{Gaussian}} = 2 \langle |z_{ik}|^2\rangle^2$.
We do this at each frequency by treating the $z_{ik}$ that we compute for each value of $i$ and $k$ as independent samples of \eqref{likelihood} in the absence of a signal ($\varepsilon = 0$), also assuming that values of $z_{ik}$ at different frequencies are independent.%
\footnote{\label{ftnt:gaussianityNoSignal}%
        Note again that even if any true signal were present, it would have to be very large to invalidate this treatment, as we average over 50001 neighboring frequency bins; cf.~footnote \ref{ftnt:noiseNoSignal}.
        }
We again take moving averages in frequency-space of these independent samples and study the result as a function of frequency.
That is, we compute
\begin{align}
    \langle|z_{ik}|^m\rangle(f_p)=\frac1{2w+1}\sum_{q=p-w}^{p+w}\frac1{3K(f_q)}\sum_{i,k}|z_{ik}(f_q)|^m,
\end{align}
{\noindent}for $m=2,4$, where $f_p$ runs over the full set $\{f_{ni}\}$ described in \secref{analysisFreqChoice}, $z_{ik}(f_p)$ are our analysis variables calculated for $f_{A'}=f_p$, and $K(f_p)$ is the number of subseries [cf.~the definition of $K$ in \secref{analysisTimeSeriesSubsets}] used in the analysis for $f_{A'}=f_p$ (which will vary for $f_{ni}$ of different $n$).
Here, we use $w=25000$.

\begin{figure}[!t]
\vspace{-0.3cm}
\includegraphics[width=\columnwidth]{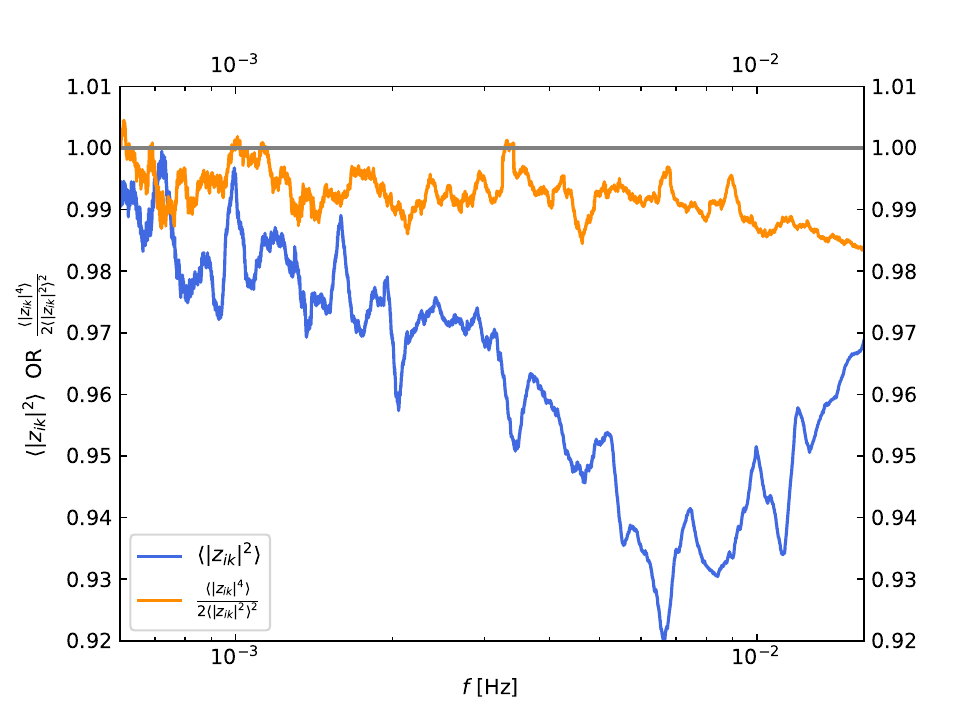}
\caption{\label{fig:Z4point}%
        Validation of Gaussianity.
        The (lower) blue line shows $\langle|z_{ik}|^2\rangle$ and the (upper) orange line shows the ratio $\langle|z_{ik}|^4\rangle / ( 2\langle|z_{ik}|^2\rangle^2)$, both as a function of frequency.  
        Both of these quantities are expected to be equal to 1 (horizontal grey line) per our analysis assumptions; cf.~\eqref{likelihood}.
        This is realised (note that the vertical axis covers only a small range of values) for $\langle|z_{ik}|^2\rangle$ to within $\sim 8$\% and for $\langle|z_{ik}|^4\rangle / ( 2\langle|z_{ik}|^2\rangle^2)$ to within $\sim 2$\%, over the entire frequency range of interest.
    	} 
\vspace{+0.3cm}
\end{figure}

In \figref{Z4point} we show in orange the ratio of $\langle|z_{ik}|^4\rangle / 2\langle|z_{ik}|^2\rangle^2$, which is expected to be 1 in the case of exact Gaussianity, as a function of frequency.
For comparison, we also show $\langle|z_{ik}|^2\rangle$ in blue, which per \eqref{likelihood} should be 1 under our analysis assumptions.
Deviations of $\langle|z_{ik}|^2\rangle$ from 1 would potentially stem from mis-estimation of the noise $S^a_{mn}$ due to the uncertainties $\sigma^a_{mn}$ referenced in the above subsections.  
It is clear from \figref{Z4point} that the deviation from Gaussianity is smaller than the mis-estimation of the noise.
There is a small deviation of $\langle|z_{ik}|^2\rangle=1$, within the 10\% level, across the entire frequency range; on the other hand, the assumption of Gaussianity as tested in this fashion is good to within 2\%.
These levels are acceptably accurate for the purposes of our analysis.

\bibliographystyle{JHEP}
\bibliography{references.bib}

\end{document}